\documentclass[longauth]{aa}  

\usepackage{graphicx}
\usepackage{caption, subcaption}
\usepackage{xcolor}
\usepackage{txfonts}

\usepackage[]{hyperref}
\usepackage[normalem]{ulem}
\usepackage{orcidlink}
\usepackage{booktabs}

\newcounter{pta}
\renewcommand{\thepta}{\Roman{pta}} 

\DeclareRobustCommand{\defpta}[1]{%
   \refstepcounter{pta}%
   \thepta\label{#1}%
}

\newcommand{\refpta}[1]{\ref{#1}}

\newcommand{\epta}[1]{{EPTA~D.R.~1}}
\newcommand{\ipta}[1]{{IPTA~D.R.~1}}

\newcommand{\uppi}{\mathrm{\pi}}
\newcommand{\e}{\mathrm{e}}

\newcommand{\C}{\textsf{\textbf{C}}}

\newcommand{\Ctot}{\tilde{\textsf{\textbf{C}}}}

\newcommand{\M}{\textsf{\textbf{M}}}

\newcommand{\tr}{\textsf{T}}

\newcommand{\ftwo}{{\tt FORTYTWO}}
\newcommand{\beph}{{\tt BAYESEPHEM}}

\newcommand{\eprise}{{\tt ENTERPRISE}}

\newcommand{\ptmc}{{\tt PTMCMCSAMPLER}}

\usepackage[switch]{lineno} 

\begin{document}

   \title{The second data release from the European Pulsar Timing Array}
   \subtitle{III. Search for gravitational wave signals}
\def\wm{3}

\author{
    J.~Antoniadis\orcidlink{0000-0003-4453-776}\inst{\ref{forth},\ref{mpifr}\ifnum\wm>1,\refpta{epta}\fi},
    \ifnum\wm>1 P.~Arumugam\orcidlink{0000-0001-9264-8024}\inst{\ref{IITR}\ifnum\wm>1,\refpta{inpta}\fi},\fi
    \ifnum\wm>1 S.~Arumugam\orcidlink{0009-0001-3587-6622}\inst{\ref{IITH_El}\ifnum\wm>1,\refpta{inpta}\fi},\fi
    S.~Babak\orcidlink{0000-0001-7469-4250}\inst{\ref{apc}\ifnum\wm>1,\refpta{epta}\fi},
    \ifnum\wm>1 M.~Bagchi\orcidlink{0000-0001-8640-8186}\inst{\ref{IMSc},\ref{HBNI}\ifnum\wm>1,\refpta{inpta}\fi},\fi
    A.-S.~Bak~Nielsen\orcidlink{ 0000-0002-1298-9392}\inst{\ref{mpifr},\ref{unibi}\ifnum\wm>1,\refpta{epta}\fi},
    C.~G.~Bassa\orcidlink{0000-0002-1429-9010}\inst{\ref{astron}\ifnum\wm>1,\refpta{epta}\fi},
    \ifnum\wm>1 A.~Bathula\orcidlink{0000-0001-7947-6703} \inst{\ref{IISERM}\ifnum\wm>1,\refpta{inpta}\fi},\fi
    A.~Berthereau\inst{\ref{lpc2e},\ref{nancay}\ifnum\wm>1,\refpta{epta}\fi},
    M.~Bonetti\orcidlink{0000-0001-7889-6810}\inst{\ref{unimib},\ref{infn-unimib},\ref{inaf-brera}\ifnum\wm>1,\refpta{epta}\fi},
    E.~Bortolas\inst{\ref{unimib},\ref{infn-unimib},\ref{inaf-brera}\ifnum\wm>1,\refpta{epta}\fi},
    P.~R.~Brook\orcidlink{0000-0003-3053-6538}\inst{\ref{unibir}\ifnum\wm>1,\refpta{epta}\fi},
    M.~Burgay\orcidlink{0000-0002-8265-4344}\inst{\ref{inaf-oac}\ifnum\wm>1,\refpta{epta}\fi},
    R.~N.~Caballero\orcidlink{0000-0001-9084-9427}\inst{\ref{HOU}\ifnum\wm>1,\refpta{epta}\fi},
    A.~Chalumeau\orcidlink{0000-0003-2111-1001}\inst{\ref{unimib}\ifnum\wm>1,\refpta{epta}\fi}\ifnum\wm=2\thanks{aurelien.chalumeau@unimib.it}\fi,
    D.~J.~Champion\orcidlink{0000-0003-1361-7723}\inst{\ref{mpifr}\ifnum\wm>1,\refpta{epta}\fi},
    S.~Chanlaridis\orcidlink{0000-0002-9323-9728}\inst{\ref{forth}\ifnum\wm>1,\refpta{epta}\fi},
    S.~Chen\orcidlink{0000-0002-3118-5963}\inst{\ref{kiaa}\ifnum\wm>1,\refpta{epta}\fi}\ifnum\wm=3\thanks{sychen@pku.edu.cn}\fi,
    I.~Cognard\orcidlink{0000-0002-1775-9692}\inst{\ref{lpc2e},\ref{nancay}\ifnum\wm>1,\refpta{epta}\fi},
    \ifnum\wm>1 S.~Dandapat\orcidlink{0000-0003-4965-9220}\inst{\ref{TIFR}\ifnum\wm>1,\refpta{inpta}\fi},\fi
    \ifnum\wm>1 D.~Deb\orcidlink{0000-0003-4067-5283}\inst{\ref{IMSc}\ifnum\wm>1,\refpta{inpta}\fi},      \fi
    \ifnum\wm>1 S.~Desai\orcidlink{0000-0002-0466-3288}\inst{\ref{IITH_Ph}\ifnum\wm>1,\refpta{inpta}\fi},\fi
    G.~Desvignes\orcidlink{0000-0003-3922-4055}\inst{\ref{mpifr}\ifnum\wm>1,\refpta{epta}\fi},
    \ifnum\wm>1 N.~Dhanda-Batra \inst{\ref{UoD}\ifnum\wm>1,\refpta{inpta}\fi},\fi
    \ifnum\wm>1 C.~Dwivedi\orcidlink{0000-0002-8804-650X}\inst{\ref{IIST}\ifnum\wm>1,\refpta{inpta}\fi},\fi
    M.~Falxa\inst{\ref{apc},\ref{lpc2e}\ifnum\wm>1,\refpta{epta}\fi}
    R.~D.~Ferdman\inst{\ref{uea}\ifnum\wm>1,\refpta{epta}\fi},
    A.~Franchini\orcidlink{0000-0002-8400-0969}\inst{\ref{unimib},\ref{infn-unimib}\ifnum\wm>1,\refpta{epta}\fi},
    J.~R.~Gair\orcidlink{0000-0002-1671-3668}\inst{\ref{aei}\ifnum\wm>1,\refpta{epta}\fi},
    B.~Goncharov\orcidlink{0000-0003-3189-5807}\inst{\ref{gssi},\ref{lngs}\ifnum\wm>1,\refpta{epta}\fi}
    \ifnum\wm>1 A.~Gopakumar\orcidlink{0000-0003-4274-4369}\inst{\ref{TIFR}\ifnum\wm>1,\refpta{inpta}\fi},\fi
    E.~Graikou\inst{\ref{mpifr}\ifnum\wm>1,\refpta{epta}\fi},
    J.-M.~Grie{\ss}meier\orcidlink{0000-0003-3362-7996}\inst{\ref{lpc2e},\ref{nancay}\ifnum\wm>1,\refpta{epta}\fi},
    L.~Guillemot\orcidlink{0000-0002-9049-8716}\inst{\ref{lpc2e},\ref{nancay}\ifnum\wm>1,\refpta{epta}\fi},
    Y.~J.~Guo\inst{\ref{mpifr}\ifnum\wm>1,\refpta{epta}\fi}\ifnum\wm=3\thanks{yjguo@mpifr-bonn.mpg.de}\fi,
    \ifnum\wm>1 Y.~Gupta\orcidlink{0000-0001-5765-0619}\inst{\ref{NCRA}\ifnum\wm>1,\refpta{inpta}\fi},\fi
    \ifnum\wm>1 S.~Hisano\orcidlink{0000-0002-7700-3379}\inst{\ref{KU_J}\ifnum\wm>1,\refpta{inpta}\fi},\fi
    H.~Hu\orcidlink{0000-0002-3407-8071}\inst{\ref{mpifr}\ifnum\wm>1,\refpta{epta}\fi}, 
    F.~Iraci\inst{\ref{unica}\ref{inaf-oac}\ifnum\wm>1,\refpta{epta}\fi},
    D.~Izquierdo-Villalba\orcidlink{0000-0002-6143-1491}\inst{\ref{unimib},\ref{infn-unimib}\ifnum\wm>1,\refpta{epta}\fi},
    J.~Jang\orcidlink{0000-0003-4454-0204}\inst{\ref{mpifr}\ifnum\wm>1,\refpta{epta}\fi}\ifnum\wm=1\thanks{jjang@mpifr-bonn.mpg.de}\fi,
    J.~Jawor\orcidlink{0000-0003-3391-0011}\inst{\ref{mpifr}\ifnum\wm>1,\refpta{epta}\fi},
    G.~H.~Janssen\orcidlink{0000-0003-3068-3677}\inst{\ref{astron},\ref{imapp}\ifnum\wm>1,\refpta{epta}\fi},
    A.~Jessner\orcidlink{0000-0001-6152-9504}\inst{\ref{mpifr}\ifnum\wm>1,\refpta{epta}\fi},
    \ifnum\wm>1 B.~C.~Joshi\orcidlink{0000-0002-0863-7781}\inst{\ref{NCRA},\ref{IITR}\ifnum\wm>1,\refpta{inpta}\fi},\fi
    \ifnum\wm>1 F.~Kareem\orcidlink{0000-0003-2444-838X} \inst{\ref{IISERK},\ref{CESSI}\ifnum\wm>1,\refpta{inpta}\fi},\fi
    R.~Karuppusamy\orcidlink{0000-0002-5307-2919}\inst{\ref{mpifr}\ifnum\wm>1,\refpta{epta}\fi},
    E.~F.~Keane\orcidlink{0000-0002-4553-655X}\inst{\ref{tcd}\ifnum\wm>1,\refpta{epta}\fi},
    M.~J.~Keith\orcidlink{0000-0001-5567-5492}\inst{\ref{jbca}\ifnum\wm>1,\refpta{epta}\fi}\ifnum\wm=2\thanks{michael.keith@manchester.ac.uk}\fi,
    \ifnum\wm>1 D.~Kharbanda\orcidlink{0000-0001-8863-4152}\inst{\ref{IITH_Ph}\ifnum\wm>1,\refpta{inpta}\fi},\fi
    \ifnum\wm>1 T.~Kikunaga\orcidlink{0000-0002-5016-3567} \inst{\ref{KU_J}\ifnum\wm>1,\refpta{inpta}\fi},\fi
    \ifnum\wm>1 N.~Kolhe\orcidlink{0000-0003-3528-9863} \inst{\ref{XCM}\ifnum\wm>1,\refpta{inpta}\fi},\fi
    M.~Kramer\inst{\ref{mpifr},\ref{jbca}\ifnum\wm>1,\refpta{epta}\fi},
    M.~A.~Krishnakumar\orcidlink{0000-0003-4528-2745}\inst{\ref{mpifr},\ref{unibi}\ifnum\wm>1,\refpta{epta}\fi\ifnum\wm>1,\refpta{inpta}\fi},
    K.~Lackeos\orcidlink{0000-0002-6554-3722}\inst{\ref{mpifr}\ifnum\wm>1,\refpta{epta}\fi},
    K.~J.~Lee\inst{3,8,\ref{mpifr}\ifnum\wm>1,\refpta{epta}\fi},
    K.~Liu\inst{\ref{mpifr}\ifnum\wm>1,\refpta{epta}\fi}\ifnum\wm=1\thanks{kliu@mpifr-bonn.mpg.de}\fi,
    Y.~Liu\orcidlink{0000-0001-9986-9360}\inst{\ref{naoc}, \ref{unibi}\ifnum\wm>1,\refpta{epta}\fi},
    A.~G.~Lyne\inst{\ref{jbca}\ifnum\wm>1,\refpta{epta}\fi},
    J.~W.~McKee\orcidlink{0000-0002-2885-8485}\inst{\ref{milne},\ref{daim}\ifnum\wm>1,\refpta{epta}\fi},
    \ifnum\wm>1 Y.~Maan\inst{\ref{NCRA}\ifnum\wm>1,\refpta{inpta}\fi},\fi
    R.~A.~Main\inst{\ref{mpifr}\ifnum\wm>1,\refpta{epta}\fi},
    M.~B.~Mickaliger\orcidlink{0000-0001-6798-5682}\inst{\ref{jbca}\ifnum\wm>1,\refpta{epta}\fi},
    I.~C.~Ni\c{t}u\orcidlink{0000-0003-3611-3464}\inst{\ref{jbca}\ifnum\wm>1,\refpta{epta}\fi},
    \ifnum\wm>1 K.~Nobleson\orcidlink{0000-0003-2715-4504}\inst{\ref{BITS}\ifnum\wm>1,\refpta{inpta}\fi},\fi
    \ifnum\wm>1 A.~K.~Paladi\orcidlink{0000-0002-8651-9510}\inst{\ref{IISc}\ifnum\wm>1,\refpta{inpta}\fi},\fi
    A.~Parthasarathy\orcidlink{0000-0002-4140-5616}\inst{\ref{mpifr}\ifnum\wm>1,\refpta{epta}\fi}\ifnum\wm=2\thanks{aparthas@mpifr-bonn.mpg.de}\fi,
    B.~B.~P.~Perera\orcidlink{0000-0002-8509-5947}\inst{\ref{arecibo}\ifnum\wm>1,\refpta{epta}\fi},
    D.~Perrodin\orcidlink{0000-0002-1806-2483}\inst{\ref{inaf-oac}\ifnum\wm>1,\refpta{epta}\fi},
    A.~Petiteau\orcidlink{0000-0002-7371-9695}\inst{\ref{irfu},\ref{apc}\ifnum\wm>1,\refpta{epta}\fi},
    N.~K.~Porayko\inst{\ref{unimib},\ref{mpifr}\ifnum\wm>1,\refpta{epta}\fi},
    A.~Possenti\inst{\ref{inaf-oac}\ifnum\wm>1,\refpta{epta}\fi},
    \ifnum\wm>1 T.~Prabu\inst{\ref{RRI}\ifnum\wm>1,\refpta{inpta}\fi},\fi
    H.~Quelquejay~Leclere\inst{\ref{apc}\ifnum\wm>1,\refpta{epta}\fi}
    \ifnum\wm>1 P.~Rana\orcidlink{0000-0001-6184-5195}\inst{\ref{TIFR}\ifnum\wm>1,\refpta{inpta}\fi},\fi
    A.~Samajdar\orcidlink{0000-0002-0857-6018}\inst{\ref{uni-potsdam}\ifnum\wm>1,\refpta{epta}\fi},
    S.~A.~Sanidas\inst{\ref{jbca}\ifnum\wm>1,\refpta{epta}\fi},
    A.~Sesana\inst{\ref{unimib},\ref{infn-unimib},\ref{inaf-brera}\ifnum\wm>1,\refpta{epta}\fi},
    G.~Shaifullah\orcidlink{0000-0002-8452-4834}\inst{\ref{unimib},\ref{infn-unimib},\ref{inaf-oac}\ifnum\wm>1,\refpta{epta}\fi}\ifnum\wm=1\thanks{golam.shaifullah@unimib.it}\fi,
    \ifnum\wm>1 J.~Singha\orcidlink{0000-0002-1636-9414}\inst{\ref{IITR}\ifnum\wm>1,\refpta{inpta}\fi},\fi
    L.~Speri\orcidlink{0000-0002-5442-7267}\inst{\ref{aei}\ifnum\wm>1,\refpta{epta}\fi},
    R.~Spiewak\inst{\ref{jbca}\ifnum\wm>1,\refpta{epta}\fi},
    \ifnum\wm>1 A.~Srivastava\orcidlink{0000-0003-3531-7887} \inst{\ref{IITH_Ph}\ifnum\wm>1,\refpta{inpta}\fi},\fi
    B.~W.~Stappers\inst{\ref{jbca}\ifnum\wm>1,\refpta{epta}\fi},
    \ifnum\wm>1 M.~Surnis\orcidlink{0000-0002-9507-6985}\inst{\ref{IISERB}\ifnum\wm>1,\refpta{inpta}\fi},\fi
    S.~C.~Susarla\orcidlink{0000-0003-4332-8201}\inst{\ref{uog}\ifnum\wm>1,\refpta{epta}\fi},
    \ifnum\wm>1 A.~Susobhanan\orcidlink{0000-0002-2820-0931}\inst{\ref{CGCA}\ifnum\wm>1,\refpta{inpta}\fi},\fi
    \ifnum\wm>1 K.~Takahashi\orcidlink{0000-0002-3034-5769}\inst{\ref{KU_J1},\ref{KU_J2}\ifnum\wm>1,\refpta{inpta}\fi} \fi
    \ifnum\wm>1 P.~Tarafdar\orcidlink{0000-0001-6921-4195}\inst{\ref{IMSc}\ifnum\wm>1,\refpta{inpta}\fi}\fi
    G.~Theureau\orcidlink{0000-0002-3649-276X}\inst{\ref{lpc2e}, \ref{nancay}, \ref{luth}\ifnum\wm>1,\refpta{epta}\fi},
    C.~Tiburzi\inst{\ref{inaf-oac}\ifnum\wm>1,\refpta{epta}\fi},
    E.~van~der~Wateren\orcidlink{0000-0003-0382-8463}\inst{\ref{astron},\ref{imapp}\ifnum\wm>1,\refpta{epta}\fi},
    A.~Vecchio\orcidlink{0000-0002-6254-1617}\inst{\ref{unibir}\ifnum\wm>1,\refpta{epta}\fi},
    V.~Venkatraman~Krishnan\orcidlink{0000-0001-9518-9819}\inst{\ref{mpifr}\ifnum\wm>1,\refpta{epta}\fi},
    J.~P.~W.~Verbiest\orcidlink{0000-0002-4088-896X}\inst{\ref{FSI},\ref{unibi},\ref{mpifr}\ifnum\wm>1,\refpta{epta}\fi},
    J.~Wang\orcidlink{0000-0003-1933-6498}\inst{\ref{unibi}, \ref{airub}, \ref{bnuz}\ifnum\wm>1,\refpta{epta}\fi},
    L.~Wang\inst{\ref{jbca}\ifnum\wm>1,\refpta{epta}\fi} and 
    Z.~Wu\orcidlink{0000-0002-1381-7859}\inst{\ref{naoc},\ref{unibi}\ifnum\wm>1,\refpta{epta}\fi}.
    }

\institute{
{Institute of Astrophysics, FORTH, N. Plastira 100, 70013, Heraklion, Greece\label{forth}}\and 
{Max-Planck-Institut f{\"u}r Radioastronomie, Auf dem H{\"u}gel 69, 53121 Bonn, Germany\label{mpifr}}\and
\ifnum\wm>1{Department of Physics, Indian Institute of Technology Roorkee, Roorkee-247667, India\label{IITR}}\and \fi
\ifnum\wm>1{Department of Electrical Engineering, IIT Hyderabad, Kandi, Telangana 502284, India \label{IITH_El}}\and\fi
{Universit{\'e} Paris Cit{\'e}, CNRS, Astroparticule et Cosmologie, 75013 Paris, France\label{apc}}\and
\ifnum\wm>1{The Institute of Mathematical Sciences, C. I. T. Campus, Taramani, Chennai 600113, India \label{IMSc}}\and\fi
\ifnum\wm>1{Homi Bhabha National Institute, Training School Complex, Anushakti Nagar, Mumbai 400094, India \label{HBNI}}\and\fi
{Fakult{\"a}t f{\"u}r Physik, Universit{\"a}t Bielefeld, Postfach 100131, 33501 Bielefeld, Germany\label{unibi}}\and
{ASTRON, Netherlands Institute for Radio Astronomy, Oude Hoogeveensedijk 4, 7991 PD, Dwingeloo, The Netherlands\label{astron}}\and
\ifnum\wm>1{Department of Physical Sciences, Indian Institute of Science Education and Research, Mohali, Punjab 140306, India \label{IISERM}}\and\fi
{Laboratoire de Physique et Chimie de l'Environnement et de l'Espace, Universit\'e d'Orl\'eans / CNRS, 45071 Orl\'eans Cedex 02, France \label{lpc2e}}\and
{Observatoire Radioastronomique de Nan\c{c}ay, Observatoire de Paris, Universit\'e PSL, Université d'Orl\'eans, CNRS, 18330 Nan\c{c}ay, France\label{nancay}}\and
{Dipartimento di Fisica ``G. Occhialini", Universit{\'a} degli Studi di Milano-Bicocca, Piazza della Scienza 3, I-20126 Milano, Italy\label{unimib}}\and
{INFN, Sezione di Milano-Bicocca, Piazza della Scienza 3, I-20126 Milano, Italy\label{infn-unimib}}\and
{INAF - Osservatorio Astronomico di Brera, via Brera 20, I-20121 Milano, Italy\label{inaf-brera}}\and
{Institute for Gravitational Wave Astronomy and School of Physics and Astronomy, University of Birmingham, Edgbaston, Birmingham B15 2TT, UK\label{unibir}}\and
{INAF - Osservatorio Astronomico di Cagliari, via della Scienza 5, 09047 Selargius (CA), Italy\label{inaf-oac}}\and
{Hellenic Open University, School of Science and Technology, 26335 Patras, Greece\label{HOU}}\and
{Kavli Institute for Astronomy and Astrophysics, Peking University, Beijing 100871, P. R. China\label{kiaa}}\and
\ifnum\wm>1{Department of Astronomy and Astrophysics, Tata Institute of Fundamental Research, Homi Bhabha Road, Navy Nagar, Colaba, Mumbai 400005, India \label{TIFR}}\and\fi
\ifnum\wm>1{Department of Physics, IIT Hyderabad, Kandi, Telangana 502284, India \label{IITH_Ph}}\and\fi
\ifnum\wm>1{Department of Physics and Astrophysics, University of Delhi, Delhi 110007, India \label{UoD}}\and\fi
\ifnum\wm>1{Department of Earth and Space Sciences, Indian Institute of Space Science and Technology, Valiamala, Thiruvananthapuram, Kerala 695547,India \label{IIST}}\and\fi
{School of Physics, Faculty of Science, University of East Anglia, Norwich NR4 7TJ, UK\label{uea}}\and
{Max Planck Institute for Gravitational Physics (Albert Einstein Institute), Am Mu{\"u}hlenberg 1, 14476 Potsdam, Germany\label{aei}}\and
{Gran Sasso Science Institute (GSSI), I-67100 L'Aquila, Italy \label{gssi}}\and
{INFN, Laboratori Nazionali del Gran Sasso, I-67100 Assergi, Italy \label{lngs}}\and 
\ifnum\wm>1{National Centre for Radio Astrophysics, Pune University Campus, Pune 411007, India \label{NCRA}}\and\fi
\ifnum\wm>1{Kumamoto University, Graduate School of Science and Technology, Kumamoto, 860-8555, Japan \label{KU_J}}\and\fi
{Universit{\'a} di Cagliari, Dipartimento di Fisica, S.P. Monserrato-Sestu Km 0,700 - 09042 Monserrato (CA), Italy\label{unica}}\and
{Department of Astrophysics/IMAPP, Radboud University Nijmegen, P.O. Box 9010, 6500 GL Nijmegen, The Netherlands\label{imapp}}\and
\ifnum\wm>1{Department of Physical Sciences,Indian Institute of Science Education and Research Kolkata, Mohanpur, 741246, India \label{IISERK}}\and\fi
\ifnum\wm>1{Center of Excellence in Space Sciences India, Indian Institute of Science Education and Research Kolkata, 741246, India \label{CESSI}}\and \fi
{School of Physics, Trinity College Dublin, College Green, Dublin 2, D02 PN40, Ireland\label{tcd}}\and
{Jodrell Bank Centre for Astrophysics, Department of Physics and Astronomy, University of Manchester, Manchester M13 9PL, UK\label{jbca}}\and
\ifnum\wm>1{Department of Physics, St. Xavier’s College (Autonomous), Mumbai 400001, India \label{XCM}}\and\fi
{National Astronomical Observatories, Chinese Academy of Sciences, Beijing 100101, P. R. China\label{naoc}}\and
{E.A. Milne Centre for Astrophysics, University of Hull, Cottingham Road, Kingston-upon-Hull, HU6 7RX, UK\label{milne}}\and
{Centre of Excellence for Data Science, Artificial Intelligence and Modelling (DAIM), University of Hull, Cottingham Road, Kingston-upon-Hull, HU6 7RX, UK\label{daim}}\and
\ifnum\wm>1{Department of Physics, BITS Pilani Hyderabad Campus, Hyderabad 500078, Telangana, India \label{BITS}}\and\fi
\ifnum\wm>1{Joint Astronomy Programme, Indian Institute of Science, Bengaluru, Karnataka, 560012, India \label{IISc}}\and\fi
{Arecibo Observatory, HC3 Box 53995, Arecibo, PR 00612, USA\label{arecibo}}\and
{IRFU, CEA, Université Paris-Saclay, F-91191 Gif-sur-Yvette, France \label{irfu}}\and
\ifnum\wm>1{Raman Research Institute India, Bengaluru, Karnataka, 560080, India \label{RRI}}\and\fi
{Institut f\"{u}r Physik und Astronomie, Universit\"{a}t Potsdam, Haus 28, Karl-Liebknecht-Str. 24/25, 14476, Potsdam, Germany\label{uni-potsdam}}\and
\ifnum\wm>1{Department of Physics, IISER Bhopal, Bhopal Bypass Road, Bhauri, Bhopal 462066, Madhya Pradesh, India \label{IISERB}}\and\fi
{Ollscoil na Gaillimhe --- University of Galway, University Road, Galway, H91 TK33, Ireland\label{uog}}\and
\ifnum\wm>1{Center for Gravitation, Cosmology, and Astrophysics, University of Wisconsin-Milwaukee, Milwaukee, WI 53211, USA \label{CGCA}}\and\fi
\ifnum\wm>1{Division of Natural Science, Faculty of Advanced Science and Technology, Kumamoto University, 2-39-1 Kurokami, Kumamoto 860-8555, Japan \label{KU_J1}}\and\fi
\ifnum\wm>1{International Research Organization for Advanced Science and Technology, Kumamoto University, 2-39-1 Kurokami, Kumamoto 860-8555, Japan \label{KU_J2}}\and\fi
{Laboratoire Univers et Th{\'e}ories LUTh, Observatoire de Paris, Universit{\'e} PSL, CNRS, Universit{\'e} de Paris, 92190 Meudon, France\label{luth}}\and
{Florida Space Institute, University of Central Florida, 12354 Research Parkway, Partnership 1 Building, Suite 214, Orlando, 32826-0650, FL, USA\label{FSI}}\and
{Ruhr University Bochum, Faculty of Physics and Astronomy, Astronomical Institute (AIRUB), 44780 Bochum, Germany \label{airub}}\and
{Advanced Institute of Natural Sciences, Beijing Normal University, Zhuhai 519087, China \label{bnuz}}\\
\ifnum\wm>1\\ {\defpta{epta} : The European Pulsar Timing Array}\fi
\ifnum\wm>1\\ {\defpta{inpta} : The Indian Pulsar Timing Array}\fi
}

   \date{Received May 8, 2023; accepted }
\titlerunning{GWB Search}
\authorrunning{EPTA+InPTA}

  \abstract 
  {We present the results of the search for an isotropic stochastic gravitational wave background (GWB) at nanohertz frequencies using the second data release of the European Pulsar Timing Array (EPTA) for 25 millisecond pulsars and a combination with the first data release of the Indian Pulsar Timing Array (InPTA).
  A robust GWB detection is conditioned upon resolving the Hellings-Downs angular pattern in the pairwise cross-correlation of the pulsar timing residuals. 
  Additionally, the GWB is expected to yield the same (common) spectrum of temporal correlations across pulsars, which is used as a null hypothesis in the GWB search. 
  Such a common-spectrum process has already been observed in pulsar timing data.
  We analysed (i) the full 24.7-year EPTA data set, (ii) its 10.3-year subset based on modern observing systems, (iii) the combination of the full data set with the first data release of the InPTA for ten commonly timed millisecond pulsars, and (iv) the combination of the 10.3-year subset with the InPTA data. 
  These combinations allowed us to probe the contributions of instrumental noise and interstellar propagation effects.
  With the full data set, we find marginal evidence for a GWB, with a Bayes factor of four and a false alarm probability of $4\%$.
  With the 10.3-year subset, we report evidence for a GWB, with a Bayes factor of $60$ and a false alarm probability of about $0.1\%$ ($\gtrsim 3\sigma$ significance).
  The addition of the InPTA data yields results that are broadly consistent with the EPTA-only data sets, with the benefit of better noise modelling.
  Analyses were performed with different data processing pipelines to test the consistency of the results from independent software packages.
  The latest EPTA data from new generation observing systems show non-negligible evidence for the GWB. At the same time, the inferred spectrum is rather uncertain and in mild tension with the common signal measured in the full data set. 
  However, if the spectral index is fixed at 13/3, the two data sets give a similar amplitude of ($2.5\pm0.7)\times10^{-15}$ at a reference frequency of $1\,{\rm yr}^{-1}$.
  Further investigation of these issues is required for reliable astrophysical interpretations of this signal.
  By continuing our detection efforts as part of the International Pulsar Timing Array (IPTA), we expect to be able to improve the measurement of 
  spatial correlations and better characterise this signal in the coming years.}


   \maketitle
%

\section{Introduction}

The first direct gravitational wave (GW) detection \citep{abb2016} marked the beginning of a new era in the exploration of the Universe. Although terrestrial interferometers such as LIGO, Virgo, and KAGRA are sensitive to GWs at kilohertz frequencies, where stellar mass compact binary mergers leave their imprint, a variety of astrophysical and cosmological phenomena are expected to generate GWs over a much broader frequency spectrum, reaching down to the nanohertz regime and beyond.

Supermassive black holes (SMBHs) are ubiquitous in galaxies \citep{1998AJ....115.2285M,2013ARA&A..51..511K} and there is growing evidence that some of them formed when the Universe was less than a gigayear old \citep[e.g.][]{2019ApJ...884...30W,2021ApJ...907L...1W}. According to the established cold dark matter cosmological scenarios, galaxy formation proceeds in a hierarchical fashion, with small galaxies merging with each other over cosmic history to build progressively larger structures \citep{1978MNRAS.183..341W}. If these galaxies host SMBHs in their centres, in the aftermath of the merger, SMBH binaries (SMBHBs) will inevitably form \citep{1980Natur.287..307B}.
Adiabatically inspiralling SMBHBs are anticipated to be the loudest sources of GWs at nanohertz frequencies. The incoherent superposition of their emitted GW signals forms a stochastic GW background (GWB) whose amplitude and spectral index relate to the galactic merger history of the Universe and to the dynamical properties of the emitting binaries \citep{jb03,svc08,2013CQGra..30v4014S}.
Besides SMBHBs, a stochastic GWB can be produced by a number of other physical processes potentially occurring in the early Universe, including non-standard inflationary fields \citep{2016NCimR..39..399G}, first-order phase transitions \citep{2010PhRvD..82f3511C}, and cosmological defects such as a network of cosmic strings \citep{2000PhRvL..85.3761D}. A comprehensive overview of these phenomena can be found in \cite{2018CQGra..35p3001C}.

Currently, this very low-frequency GW regime can only be accessed with pulsar timing arrays \citep[PTAs,][]{fb1990}. The technique of pulsar timing relies on the exceptional rotational stability of a particular population of neutron stars, the millisecond pulsars (MSPs). The times of arrival (TOAs) of radio pulses observed at the telescope are measured precisely using maser clocks referenced to the international atomic time. A model, known as a phase-connected timing solution, is then used to account for every rotation of the pulsar for the entire series of TOAs \citep[see][for a detailed explanation]{lk04}. The pulsar timing technique has allowed high-precision measurements that have led to several significant breakthroughs, including the first indirect detection of GWs through the measured orbital shrinkage of PSR~B1913+16 \citep{tw89}. In a PTA a network of the most stable MSPs is observed regularly and the TOAs are modelled. It is within the small deviations from the model (the residuals) that nanohertz GWs can be searched for.

The idea of using MSPs to detect nanohertz GWs was first proposed by \cite{saz78} and \cite{det79}. The distortions in spacetime caused by a GW propagating over the Earth or over a pulsar lead to stochastic advances or delays in the TOAs.
Astrophysical sources produce GWs that cause larger delays over longer timescales, that is, a temporally correlated (red) signal.
However, disentangling the GW signal from other red noise sources, such as variations in the interstellar medium (ISM) or intrinsic pulsar spin noise, with a single pulsar is impossible. \cite{fb90} were the first to suggest a PTA as a method to overcome this problem. Not only would the GW signal result in a common red signal (CRS) in all pulsars, but the signal would be spatially correlated across the sky. This correlation is related to the quadrupolar nature of GWs. Although GWs passing over the individual pulsars would not be correlated, those propagating over the Earth would be. When the degree of correlation for each pair of pulsars is plotted against their angular separation, this results in the Hellings and Downs curve \citep[HD,][]{hd83}. It is this HD curve that allows us to distinguish the GWB from other potential sources of correlated signal \citep[e.g. the modelling of Earth's motion in the Solar system and local clock instabilities,][]{thk+2016}.

The European Pulsar Timing Array \citep[EPTA,][]{kc2013} was formed in 2004 to facilitate the detection of GWs. However, it uses pulsar observations taken well before its formal creation, some of which were specifically for PTA-style analysis. The EPTA data are provided by some of the largest radio telescopes in Europe: the Lovell telescope at the Jodrell Bank Observatory, the Nan\c{c}ay decimetric radio telescope, the Westerbork synthesis radio telescope, the Effelsberg 100\,m radio telescope, and the Sardinia radio telescope. These telescopes supply independent data sets at a range of observing frequencies \citep[see] []{wm1}, but since 2009 they have also worked together as the Large European Array for Pulsars (LEAP), a coherently phased interferometer with an equivalent dish diameter of up to 194\,m \citep{bjk+16}.

The earliest EPTA data date back to 1994, with most pulsars having over 15 years of data. The bandwidth available and the backends used to record the data have improved over the years. While only some coherently dedispersing backend systems were used initially, considerable upgrades were made around 2005--2010, when most telescopes switched to broadband, coherent dedispersion systems. In addition to offering a wide range of observing frequencies and high observing cadence (with weekly or even shorter spacings between successive observations), multiple telescopes also allow the data sets to be checked against each other, highlighting any local issues. This is crucial for reliability.

THE EPTA is a founding member of the International Pulsar Timing Array \citep[IPTA,][]{vbc+2009}, along with the Parkes Pulsar Timing Array (PPTA) which uses the Parkes telescope \citep{man06, hob2013}, and the North American Nanohertz Observatory for Gravitational Waves (NANOGrav), which uses data from the Arecibo observatory, Green Bank radio telescope, and Very Large Array \citep{jenet09, abb+2015a}. Recently, the Indian Pulsar Timing Array \citep[InPTA, using the Giant Meterwave Radio Telescope,][]{2018JApA...39...51J} has joined the IPTA as a full member, while the Chinese Pulsar Timing Array \citep[CPTA, using the Five Hundred Meter Spherical Telescope,][]{jyg+19}, and the MeerTIME programme \citep[using the MeerKAT telescope,][]{bja+2020} have become observer members.

The first EPTA data release (DR1) was made in 2015 \citep{dcl+2016} and was used to place an upper limit on the GWB \citep{ltm+2015}. However, during the analysis of the six best pulsars in the array, a weak CRS was observed in the data. An analysis of the same pulsars in 2021 \citep{ccg+21} allowed for a direct comparison with the earlier work and clearly showed that not only was the CRS still present in the data, but also that its properties could be significantly better constrained. This CRS is consistent with the findings of the NANOGrav 12.5-year data set \citep{abb+2020}, PPTA DR2 \citep{ppta+22}, and IPTA DR2 \citep{ipta+22}. However, six pulsars do not provide enough pairs to sufficiently sample the HD curve, the crucial signature of GWs. To this end, EPTA Data Release 2 \citep[DR2,][]{wm1} has been created using 25 MSPs optimally selected among those timed by the collaboration, following the method described in \cite{2023MNRAS.518.1802S}.

In this paper, we present the results of the search for a stochastic GWB at nanohertz frequencies in the EPTA DR2. A summary of the data set and noise models is given in Section~\ref{sec:data}. For more details, we refer interested readers to the companion papers \citep{wm1,wm2}, respectively. In Section~\ref{sec:method} we briefly review our analysis methods, which are similar to those used in the six-pulsar analysis presented in \cite{ccg+21}. Our main results, including a comparison of the full DR2 data set against a reduced data set that includes only the new generation backend systems, are presented in Section~\ref{sec:results}. In Section~\ref{sec:InPTA} we discuss the addition of InPTA data and its impact on the GWB search, and draw our conclusions in Section~\ref{sec:conclusions}.

\section{Data set and noise models}
\label{sec:data}

The DR2 includes observations of 25 pulsars selected from the DR1 source list.
These data were collected with six EPTA telescopes, including LEAP.
The DR2 data set is a combination of data from DR1 with those recorded with a new generation of data acquisition systems which offer significantly wider bandwidth and thus greater sensitivity. The DR2 data set offers a variety of time spans for different pulsars, from a minimum of 14 to a maximum of 25 years. That data set also has a broad observing frequency coverage, starting from ($\sim$300\,MHz) and extending up to ($\sim$4\,GHz). A subset of DR2, using data for six pulsars was used for the common red noise process search presented in \cite{ccg+21}. Since then, our multi-telescope data have allowed us to detect and correct an issue in the clock corrections applied to the data collected with the `NUPPI' pulsar backend \citep{ctg+13} at the Nan\c{c}ay Radio Telescope. 
More details of the EPTA DR2 data set and timing analysis results can be found in our data release paper \citep{wm1}.

For the analysis presented in this paper, we used two versions of the DR2, the full data set and a truncated version that features data collected with the new generation of pulsar backends only. These are extended by incorporating data from the first InPTA data release \citep{2022PASA...39...53T} for an overlapping set of ten pulsars.
The InPTA data set was obtained using the upgraded Giant Metrewave Radio Telescope (uGMRT) from MJD 58235 to 59496 covering about 3.5 years. It complements the EPTA data with simultaneous observations in the 300--500\,MHz and 1260--1460\,MHz bands and adds about 0.7 years to the EPTA time span.
To summarise, we analyse the following four data sets, additional details of which can be found in the EPTA data release paper \citep{wm1} and the accompanying pulsar noise analysis paper \citep{wm2}:
\begin{enumerate}
    \item \texttt{DR2full}. The complete EPTA DR2 data set, covering 24.7 years of data;
    \item \texttt{DR2new}. A reduced version of the entire data set, including only the last 10.3 years of data, collected with new generation wide-band backends;
    \item \texttt{DR2full+}. The same as \texttt{DR2full}, but with the addition of InPTA data for ten pulsars, covering 25.4 years of data; 
    \item \texttt{DR2new+}. The same as \texttt{DR2new}, but with the addition of InPTA data for ten pulsars, covering 11.0 years of data.
\end{enumerate}

The new generation backends use improved hardware for the conversion of the electric signals to digital data streams and allow for coherent dedispersion during the observations, whereas previous systems mostly operated with incoherent dedispersion. The increased processing power also enables us to use up to four times the frequency bandwidth as compared to the older, legacy backends.

Before a correlated signal can be searched for, the deterministic properties of individual pulsars need to be modelled. This includes the spin, astrometric, orbital (for binaries), and noise parameters of the pulsar \citep{wm1}. Single pulsar noise models for the data sets mentioned above have been obtained from a specific model selection scheme presented in \citet{wm2}. For all pulsars, the timing model parameters were analytically marginalised and the white noise parameters EFAC and EQUAD were set at fixed values; cf. Section \ref{sec:method}.
The TOAs are measured by averaging over a frequency range in which the pulse profile can be considered stable. For the legacy systems, the full bandwidth of about 128\,MHz was used. However, for the new generation backends with larger bandwidths, we split the observation into sub-bands, treating each sub-band independently with a template and offset. This method allowed us to reduce the number of TOAs while retaining most of the information from the observations.
With at most four TOAs per observation and due to the significantly lowered sensitivity as a result of the sub-banding we could not measure significant, time-correlated white noise. Thus the ECORR parameter, which describes the presence of such noise, was not included in the analysis.
Model selection was applied for other time-correlated signals, allowing for the selection of the most favoured 
combination among observing-frequency independent red noise (RN), dispersion measure variations (DM), 
and scattering variations (SV). These correspond to stochastic signals that induce a 
delay in the timing residuals with a chromatic index $k$ of zero, two, and four, 
respectively, which characterises the dependence on the observing radio frequency $\nu^{-k}$.
Two large events were observed in PSR~J1713+0747, one at MJD $\sim 54757$ \citep{cks+2015,zsd+2015,dcl+2016} and one at MJD $\sim 57510$ \citep{leg2018,grs+2021,2022MNRAS.509.5538C}. These were assumed to be caused by sudden changes in the scattering and dispersion variation and modelled as deterministic signals with fixed chromatic indices $k=4$ and $k=2$, 
respectively, as obtained from a Bayesian fit to the data. While both events are spanned by \texttt{DR2full}, only the second event falls within the time-span of \texttt{DR2new}. 

Each noise process is modelled as a sum over Fourier components. Following \cite{2022MNRAS.509.5538C}, we did not fix a priori the number of Fourier components
of the various processes in the noise analysis. Instead, for all combinations of RN, DM, and SV, we determined the optimal number of Fourier components that best describes the data. We did not consider models that include SV but not DM, as this is not physically motivated. We obtained customised noise models for each pulsar from a Bayes factor (BF) evaluation among the candidate models and 
performed a final analysis by refitting the timing model parameters simultaneously with these favoured noise models. This enabled further refinement of the timing model parameters and a more reliable evaluation of the white noise parameters, which are subsequently fixed in the GWB analyses. The interpretation of custom noise models and further discussion on the robustness of these results are presented in \citet{wm2}.

\section{Methods}
\label{sec:method}
The analysis methods closely follow those of \cite{ccg+21} and references therein. In the following, we summarise the key components of the analysis and provide details of additional analyses included in this work.

The PTA likelihood function for a CRS search is given by \citep{vlm+2009}
\begin{equation}
\label{eq:gwb_reduc_lik}
L_{\rm{PTA}}\propto\frac{\e^{-\frac{1}{2}\sum_{I,J,i,j}(\widetilde{\delta{}t}_{I,i})^{\tr}\Ctot^{-1}(\widetilde{\delta{}t}_{J,j})}}{\sqrt{|\Ctot|}}\,.
\end{equation}
The post-fit residuals of pulsar $I$ at observation $i$ are denoted as $\widetilde{\delta{}t}_{I,i}$ and $\Ctot = \Ctot^* + \Gamma \C_{\rm CRS}$ is the sum of the block diagonal covariance matrices for all pulsars\footnote{To avoid $\Ctot$ becoming singular, it is regularised as $|\Ctot^*| = |\C^*|\times|\M^{*T}\C^{*-1}\M^*|$, where $\M^*$ is the design matrix of all pulsars.}
 and the overlap reduction function (ORF) $\Gamma$ multiplied by the covariance matrix for the correlated common red signal $\C_{\rm CRS}$. The timing model parameters are analytically marginalised over; see \cite{vl2013, vv2014} for more details.

The covariance matrix for each pulsar contains information on the white and red noise components RN, DM, and SV. Measurement uncertainties on the TOAs can be calibrated with a pair of white noise parameters, EFAC and EQUAD, for each telescope, receiver and backend combination to modify the initial estimate from the instrument and TOA extraction method,
\begin{equation}
    \widetilde{\sigma}_{ij}^2 = (\sigma_{ij} \times {\rm EFAC})^2 + {\rm EQUAD}^2
\end{equation}

The red noise power spectra were modelled with a power law
\begin{equation}
    S_{RN} = \frac{A^2}{12\pi^2} \left(\frac{f}{\rm{1yr}}\right)^{-\gamma}\, \frac{\rm{yr^3}}{T},
\end{equation}
representing long-term variations in the ToAs which are independent of the radio frequency of the observations. This model was used for both, individual pulsars as well as for any putative common red noise.

Propagation of the radio signals through the interstellar medium adds delays that depend on the frequency of the radio photons. Following \cite{2022MNRAS.509.5538C} we considered two types of processes; DM variations and scattering of the photons by electrons encountered along the line of sight between the pulsar and Earth. These were also modelled with power laws
\begin{equation}
    S_{k} = \frac{A^2}{12\pi^2} \frac{K}{\nu^{-k}}  \left(\frac{f}{\rm{1yr}}\right)^{-\gamma}\, \frac{\rm{yr^3}}{T},
\end{equation}
where $K$ is the DM constant at a reference frequency of 1\,MHz, $k$ is the chromatic index of two or four for DM and SV, respectively, and $\nu$ is the radio frequency of the propagating photons.

The frequencies $f$ of the Fourier basis were chosen to be $f_{n}=n/T\, (n=1,..., N)$, where $T$ is the time interval between the first and last observations, and $N$ is the number of frequency bins considered. This number was customised for each noise process in each individual pulsar, as described in the companion noise analysis paper \citep{wm2}. 

For the common red signals, we used two methods to determine the optimal number of frequency bins: 
1) we fitted a broken power law to estimate the frequency where the red noise became dominant over the white noise \citep{abb+2020}; and 2) we constructed a free-spectrum model, where the power at each frequency was modelled with an independent parameter \citep{lah+2013,abb+2020}.

For the detection of the GWB, the characteristic spatial correlation described by the HD curve 
\begin{equation}
\label{eq:HDcurve}
\Gamma_{\textrm{GWB}}(\zeta_{IJ})=\frac{3}{2}x_{IJ}\ln{x_{IJ}}-\frac{x_{IJ}}{4}+\frac{1}{2}+\frac{1}{2}\delta{x_{IJ}}\,,
\end{equation}
is the key criterion. Here, $\zeta_{IJ}$ is the spatial angular separation between pulsars $I$ and $J$, $x_{IJ}=[1-\cos(\zeta_{IJ})]/2$, and $\delta(x_{IJ})$ is the Kronecker delta function.

We employed three types of model to search for generic spatial correlations in the data to compare against the expected HD correlation from a GWB:
\begin{enumerate}
\item a binned correlation function \citep{tg2013}, where we weighted the evidence for the pulsar pair correlations in ten bins of angular separations between pulsars; 
\item a Chebyshev polynomial decomposition to the third order \citep{ltm+2015,ccg+21}
\begin{equation}
\label{eq:cheby1}
\Gamma(\zeta_{IJ}) \approx c_1 + c_2 y_{IJ} + c_3(2y_{IJ}^2-1)+c_4(4y_{IJ}^3-3y_{IJ})\,,
\end{equation}
where $y_{IJ}=(\zeta_{IJ}-\uppi/2)/(\uppi/2)$ and $c_i$ are the Chebyshev polynomial parameters whose priors are uniform in the range $[-1,1]$. The cross-correlation is normalised so that $\Gamma(x) \in [-1,1]$. This decomposition can be used to compare against constraints from previous EPTA data sets.

\item a Legendre polynomial decomposition to fifth order \citep{grt+2014}

\begin{equation}
\label{eq:leg1}
\Gamma(\zeta_{IJ}) \approx \sum_{i=0}^{5}l_i P_i (\cos \zeta_{IJ})\,,
\end{equation}
where $P_i$ are the Legendre polynomial functions of order $i$ and $l_i$ are the Legendre polynomial parameters whose priors are uniform in the range $[-1,1]$. The parameters can be interpreted as the amount of power in the monopole $i=0$, dipole $i=1$, quadrupole $i=2$, and higher modes. A pure GWB would have no monopole or dipole contributions (i.e. $l_0=l_1=0$) in this decomposition with all power at $i \geq 2$.
\end{enumerate}

The off-diagonal elements of the covariance matrix encode the information on cross-pulsar correlated common signals. Apart from the quadrupole HD, we tested for the presence of a monopole (associated with, e.g. clock time errors) and a dipole (associated with, e.g. systematics in the model of the position of the Earth, the Solar system ephemeris, SSE) term. Unless otherwise stated, all analyses were performed with a fixed SSE model, DE440, produced by \cite{2021AJ....161..105P}. To check for possible SSE systematics, we performed additional analyses using the \beph{} model \citep{abb+2018,2020ApJ...893..112V}.

Using only the diagonal terms of the covariance matrix allows for fast computational analysis and corresponds to a model without any spatial correlations, which we refer to as common uncorrelated red noise (CURN). It is also possible to use only the cross-terms to search for a common signal in a split-likelihood analysis \citep{abb+2020,ipta+22}.
If the posterior distribution of the uncorrelated model has a substantial number of samples that are within the support of the correlated model, it is possible to employ the reweighting formalism, which was introduced in \cite{Hourihane:2022ner}, to approximate the posterior of the correlated model. The reweighting process involves sampling from the posterior distribution of the uncorrelated model (CURN) and then adjusting the weights of the obtained samples to reflect their likelihood under the correlated model. This technique enables the posterior of a correlated model to be obtained efficiently, the Bayes factor between the two models to be obtained, and the quality of the reweighted samples to be quantified.

\begin{table}
    \caption{Prior ranges for the parameters of the power laws used in the analysis: amplitude, $A$, and spectral index, $\gamma$. 
    Subscripts RN, DM, and CRS denote the
    red noise, DM noise, and common red signal, respectively.
    }
    \def\arraystretch{1.5}
    \begin{tabular}{c||c|c}
    \hline
        Parameter & Prior Type & Range\\
    \hline
    \hline
        $A_{\rm{RN}}$, $A_{\rm{DM}}$, $A_{\rm{CRS}}$ & log-Uniform & $[10^{-18} - 10^{-10}]$ \\
        $\gamma_{\rm{RN}}$,~$\gamma_{\rm{DM}}$,~$\gamma_{\rm{CRS}}$ & Uniform & $[0 - 7]$ \\
    \hline
    \end{tabular}
    \label{tab:priors}
\end{table}

\subsection{Bayesian analysis}
\label{sec:bayes}

We estimated the parameters by evaluating the posterior probability, which is proportional to the likelihood given by Equation~\ref{eq:gwb_reduc_lik} multiplied by the prior.
The inverse of the proportionality coefficient is referred to as Bayesian evidence and it is equal to the integral of the likelihood times the prior over the prior range. When searching for a GWB, we fixed the white noise parameters of each pulsar to the maximum likelihood values produced by the single pulsar noise analysis \citep{wm2} and we simultaneously evaluated the RN, DM, and SV of each individual pulsar and the CRS. The prior ranges adopted for these parameters is given in Table~\ref{tab:priors}.
 Bayesian evidence was used for model selection, where the Bayes factor for one hypothesis over the other is equal to the ratio of the two evidence values corresponding to these hypotheses.
The posterior evaluation was done mainly with \ptmc{} with other samplers used for cross-checking: \texttt{m3c2} \citep{Falxa:2022yrm}, \texttt{Eryn} \citep{Karnesis:2023ras}, and \citep{nessai}.

In this work, we performed model selection in two ways.
First, directly, by introducing a hyperparameter that switches between likelihoods corresponding to the two models. The ratio of the fraction of samples using one model to the fraction using the other model is the Bayes factor. This method is known as the product-space sampling method~\citep{cc1995,hhh+2016}. The second method involves resampling from the CURN model; it is mentioned above and described in detail in \cite{Hourihane:2022ner}.

\begin{figure*}
\centering
\includegraphics[width=\textwidth]{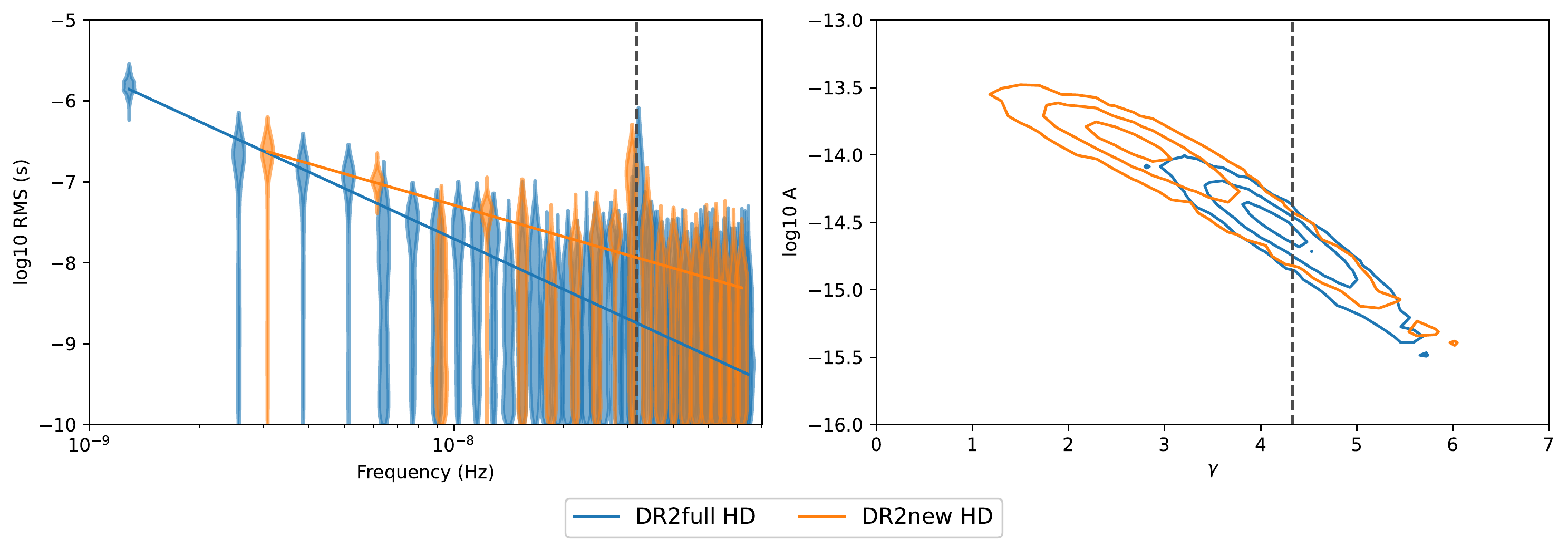}
\caption{
Spectral properties of a CRS assuming HD correlations. The left panel shows the free spectrum, the independent measurement of common power at each frequency bin, for the two versions of the EPTA-only data set. The right panel shows the 1/2/3$\sigma$ contour of the 2D posterior distribution of amplitude and spectral index when modelling the spectrum with a power law. In both panels, results for \texttt{DR2full} are in blue, while those of \texttt{DR2new} are in orange. The solid lines in the left panel are the power law best-fits to the GWB (see main text for the parameters of the fit), while the vertical dashed line indicates the position of $f=1\,{\rm yr}^{-1}$. The vertical dashed line in the right panel denotes $\gamma=13/3$.}
\label{fig:fs_pl}
\end{figure*}

\begin{table*}
\caption{90\% credible regions for the power law parameters constraints in the different Bayesian analyses with DE440 for both \texttt{DR2full} and \texttt{DR2new}. The analyses included the search for common uncorrelated red noise (CURN), gravitational wave background (GWB), and a common correlated signal with overlap reduction function (ORF) modelled with different methods (binned ORF, Chebyshev ORF, and Legendre ORF).}
\begin{center}
\def\arraystretch{1.5}
\begin{tabular*}{0.7\textwidth}{c||cc|cc}
\hline
 & \multicolumn{2}{c}{\texttt{DR2full}} & \multicolumn{2}{c}{\texttt{DR2new}} \\
Software + Model & $\log_{10} A_{\rm CRS}$ & $\gamma_{\rm CRS}$ & $\log_{10} A_{\rm CRS}$ & $\gamma_{\rm CRS}$ \\
\hline \hline
\eprise{} + CURN             & $-14.53_{-0.44}^{+0.29}$ & $4.13_{-0.59}^{+0.80}$ & $-14.00_{-0.77}^{+0.28}$ & $2.91_{-0.87}^{+1.72}$  \\
\hline
\ftwo{} + CURN               & $-14.52_{-0.40}^{+0.30}$ & $4.12_{-0.60}^{+0.74}$ & $-14.00_{-0.66}^{+0.27}$ & $2.91_{-0.85}^{+1.51}$  \\
\hline
\eprise{} + GWB              & $-14.54_{-0.41}^{+0.28}$ & $4.19_{-0.63}^{+0.73}$ & $-13.94_{-0.48}^{+0.23}$ & $2.71_{-0.71}^{+1.18}$  \\
\hline
\ftwo{} + GWB                & $-14.53_{-0.40}^{+0.30}$ & $4.16_{-0.66}^{+0.74}$ & $-13.94_{-0.55}^{+0.24}$ & $2.71_{-0.75}^{+1.30}$  \\
\hline
\eprise{} + Binned ORF       & $-14.47_{-0.35}^{+0.27}$ & $4.10_{-0.56}^{+0.64}$ & $-13.89_{-0.32}^{+0.22}$ & $2.63_{-0.71}^{+0.86}$  \\
\hline
\ftwo{} + Binned ORF         & $-14.49_{-0.39}^{+0.29}$ & $4.11_{-0.62}^{+0.72}$ & $-13.87_{-0.37}^{+0.22}$ & $2.58_{-0.74}^{+0.98}$  \\
\hline
\eprise{} + Chebyshev ORF    & $-14.50_{-0.40}^{+0.32}$ & $4.17_{-0.72}^{+0.73}$ & $-13.87_{-0.31}^{+0.22}$ & $2.57_{-0.76}^{+0.86}$  \\
\hline
\eprise{} + Legendre ORF     & $-14.51_{-0.40}^{+0.30}$ & $4.19_{-0.63}^{+0.74}$ & $-13.89_{-0.35}^{+0.23}$ & $2.59_{-0.72}^{+0.98}$  \\
\hline
\end{tabular*}
\label{tab:pl}
\end{center}
\end{table*}

\subsection{Frequentist analysis}

We also used the frequentist optimal statistic (OS) framework, developed by \cite{abc+2009}, \cite{dfg+2013} and \cite{ccx+2015} with the noise marginalisation introduced by \cite{2018PhRvD..98d4003V}, to compare against the Bayesian results.
The Bayesian output of a CURN analysis was used as the input for the OS. This allowed for high computational efficiency, as the posterior distributions of pulsar noise were directly used to compute the statistics. With the OS we can compute the amount of correlated power for each pulsar pair. Comparing this correlation against different models gives signal-to-noise (S/N) estimates for different types of spatial correlations.

\subsection{Software packages}

As in \cite{ccg+21}, we used both \eprise{}\footnote{\url{https://gitlab.in2p3.fr/epta/enterprise} and \url{https://gitlab.in2p3.fr/epta/enterprise_extensions}} \citep{enterprise,enterprise_extensions} and \ftwo{}\footnote{\url{https://github.com/caballero-astro/fortytwo}} \citep{cll+2016} for the PTA likelihood computation and cross-check the main results with both pipelines. Some of the more specific analyses were performed only with \eprise{} for computational cost efficacy since we have demonstrated in \cite{ccg+21} that the two pipelines produce broadly consistent results.

\begin{figure*}[htp]
\centering
     
     \begin{subfigure}[b]{0.45\textwidth}
         \centering
         \includegraphics[width=\textwidth]{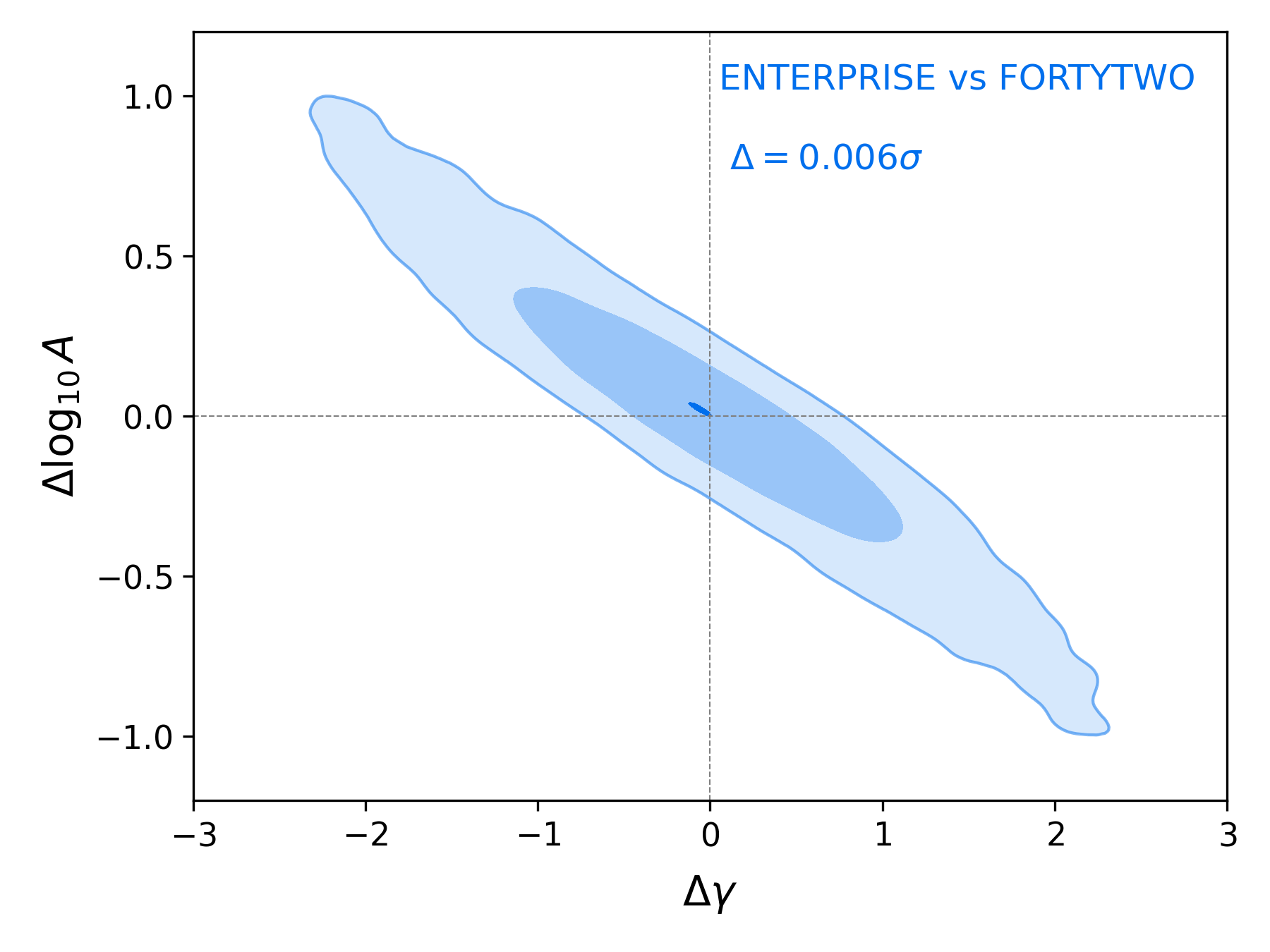}
         \label{}
     \end{subfigure}
     \hspace{0.05cm}
     \begin{subfigure}[b]{0.45\textwidth}
         \centering
         \includegraphics[width=\textwidth]{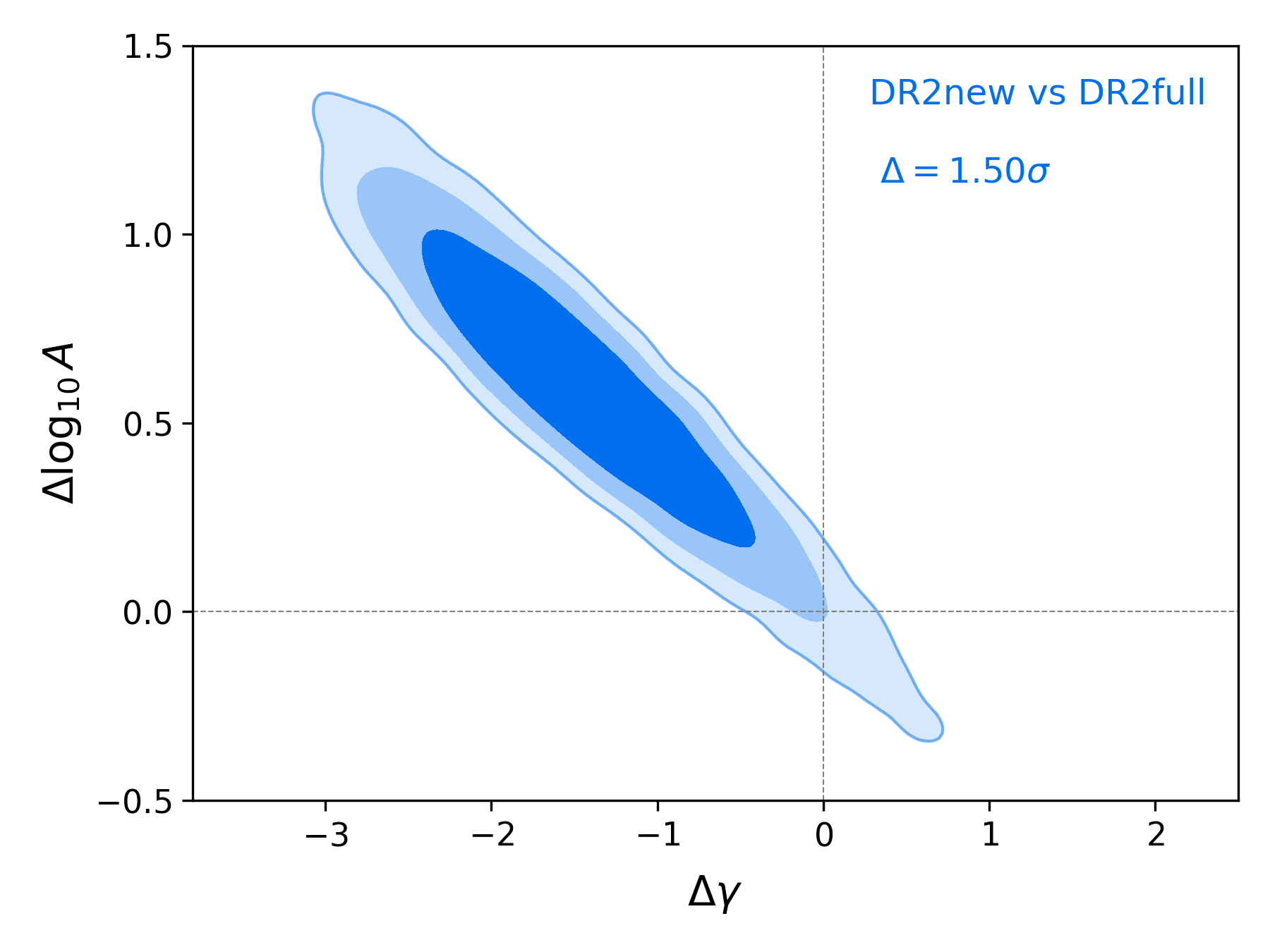}
         \label{}
     \end{subfigure}
        \caption{
       Difference distributions between two posterior distributions originating from GWB processes. The left panel depicts the difference distribution between \texttt{DR2new} and \texttt{DR2full} data sets while employing the \texttt{ENTERPRISE} package. In comparison, the right panel shows the tension contour between \eprise{} and \ftwo{} software packages when we employ the \texttt{DR2new} data set. The plots contain three contours: $1\sigma$, $2\sigma$, and the $\Delta$ contours that correspond to the value of the computed tension.}
        \label{tension-plots}
\end{figure*}

\begin{table*}
\caption{
  $Z$-score (in number of $\sigma$) produced by the \texttt{tensiometer} package, detailed in \cite{raveri2021non} when comparing posteriors produced by 
 various data sets and software packages.
 The second column compares the posteriors between the \texttt{DR2new} and \texttt{DR2full} data set 
 while employing \eprise{} and \ftwo{} (in brackets).
 On the contrary, the third column compares the posteriors given by the
 \eprise{} and \ftwo{} software packages running on the \texttt{DR2new} and \texttt{DR2full} data sets (in brackets). 
}

\def\arraystretch{1.5}
\centering
\begin{tabular}{c||c|c}
 \hline
 & Data set comparison & Software package comparison \\
\cmidrule(rl){2-2} \cmidrule(rl){3-3}
 & \texttt{DR2new} vs \texttt{DR2full} & \eprise{} vs \ftwo{} \\
\cmidrule(rl){2-2} \cmidrule(rl){3-3}
 & {\eprise{} (\ftwo{})}  & {\texttt{DR2new} (\texttt{DR2full})}  \\
\hline \hline
CURN		& 1.06 (1.15)	&  0.0063 (0.0274)	 \\
GWB		&	1.50 (1.49)	& 0.006 (0.0229)\\				
Binned ORF		&  1.69 (1.68) &	0.002 (0.0325) \\
 \hline
\end{tabular}
\label{tension-table}
\end{table*}

\section{Search results on the EPTA data sets \texttt{DR2full} and \texttt{DR2new}}
\label{sec:results}

We first present results for the analysis carried out with EPTA-only versions of the data set, namely \texttt{DR2full} and \texttt{DR2new}. Results for the EPTA$+$InPTA data set analysis are presented in Section~\ref{sec:InPTA}.

For simplicity and efficiency, our general analysis setup uses the DE440 Solar system ephemeris fit. The starting values for the marginalisation of the timing model are taken from the timing analysis \citep{wm1}. Pulsar noise models and observing system white noise parameters are taken from the single pulsar noise analysis \citep{wm2}.

Previous PTA analyses \citep[e.g.][]{abb+2020,ipta+22} have shown the importance of choosing the optimal number of frequencies to model any putative common signal and that most of the power of a common red signal can be found in the lowest frequency bins. We choose the width of the frequency bin to be $1/T$, where $T$ is the time span of the data set. For the power law the lowest 24 (9) frequency bins are used to model the CRS in the \texttt{DR2full} (\texttt{DR2new}) data set, which corresponds to a maximum frequency $f_{\rm max}=24/24.7$ yr$^{-1}$($9/10.3$ yr$^{-1}$) respectively.
We subsequently used these limiting frequencies 
for the remaining analyses (unless otherwise specified). This choice has been verified with a broken power law analysis, which shows that using a larger number of frequency bins does not impact the recovery of the parameters of the CRS.

We show the posterior distributions of the free-spectrum model for the first 50 (20) frequency bins for the \texttt{DR2full} (\texttt{DR2new}) data set in Figure \ref{fig:fs_pl}. The most noticeable difference is in the lowest constrained frequency bin. Extending the \texttt{DR2new} only best fit power law to lower frequencies would result in a lower CRS in the 1/24.7 yr$^{-1}$ bin, compared to what is measured in \texttt{DR2full}. Analogously, fitting a power law to the \texttt{DR2full} data set excluding the lowest frequency bin could give constraints that are more consistent with the new data set. This could be due to the non-stationarity of the processes involved, deviations from the power law model or some additional unmodelled noise in the full data set. Further investigations are needed to understand better this difference and if and how it could be mitigated.

\subsection{Spectral parameter constraints}

The right panel of Figure \ref{fig:fs_pl} shows the posteriors of the recovered parameters for a power law model for a HD correlated process.
The recovered power law parameters with \texttt{DR2new} are logarithmic amplitude $\log_{10} A = -13.94_{-0.48}^{+0.23}$ and spectral index $\gamma = 2.71_{-0.73}^{+1.18}$ (90\% credible regions). The spectral index is shallower than the expected mean value of $13/3$ from a population of circular SMBHBs, which still lies within the 3$\sigma$ credible region \citep[also see][]{msc+2021}. With \texttt{DR2full} the recovered logarithmic amplitude $\log_{10} A = -14.54_{-0.41}^{+0.28}$ and spectral index $\gamma = 4.19_{-0.63}^{+0.73}$ is closer to $13/3$. The two data sets give consistent results in the sense that the two posteriors overlap and lie on the same $A-\gamma$ degeneracy line that corresponds to fixing the total HD-correlated power. Therefore, the HD-correlated power measured in \texttt{DR2full} and \texttt{DR2new} is comparable, although the spectral shape is not well constrained and appears to be different in the two data sets. In support of this statement, when fixing the spectral index to $13/3$, the background amplitude inferred from the two data sets is consistent, with a value of $\log_{10}A=-14.61_{-0.12}^{+0.11}$.

\begin{figure}
\includegraphics[width=0.9\linewidth]{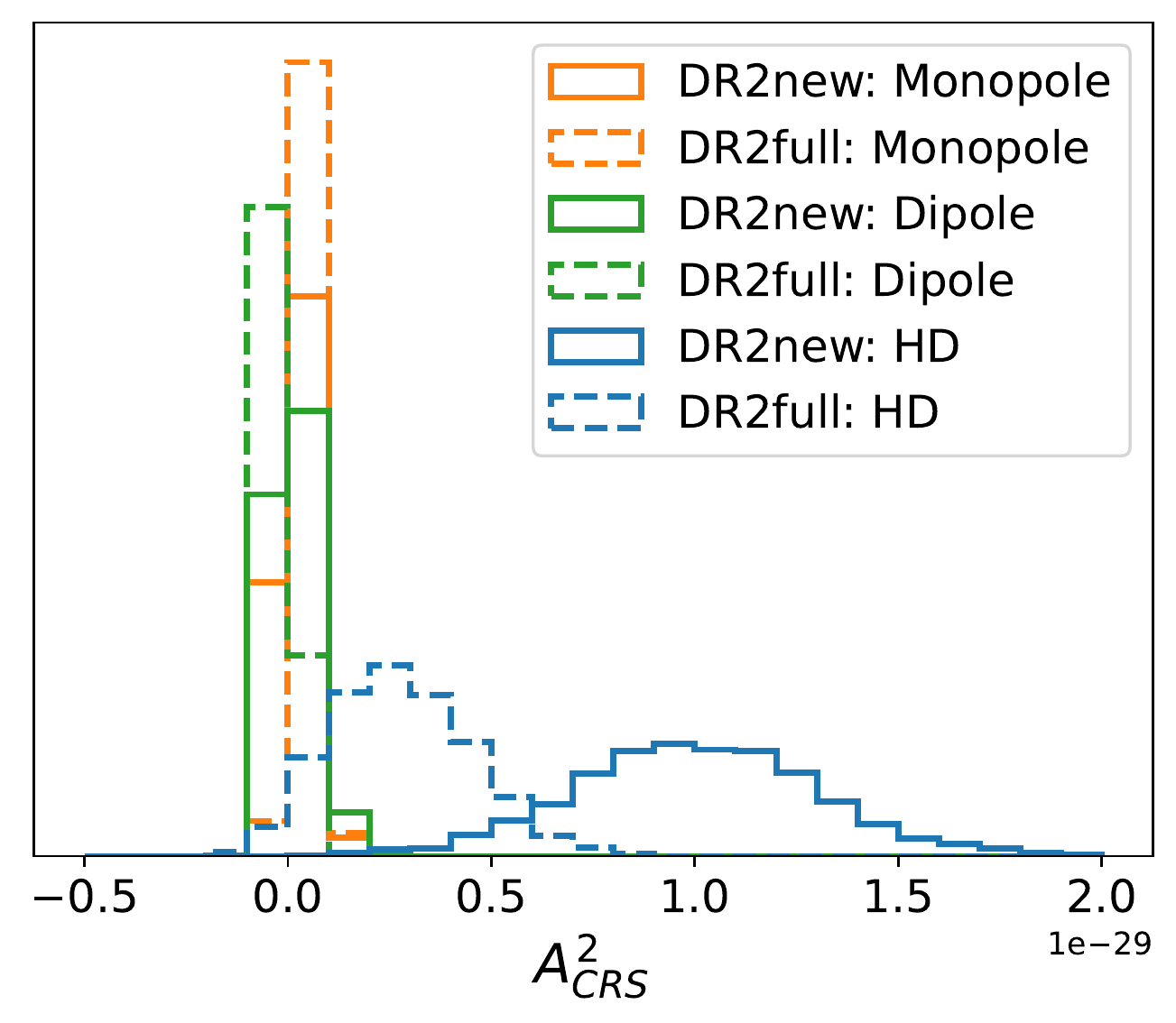}
\includegraphics[width=0.9\linewidth]{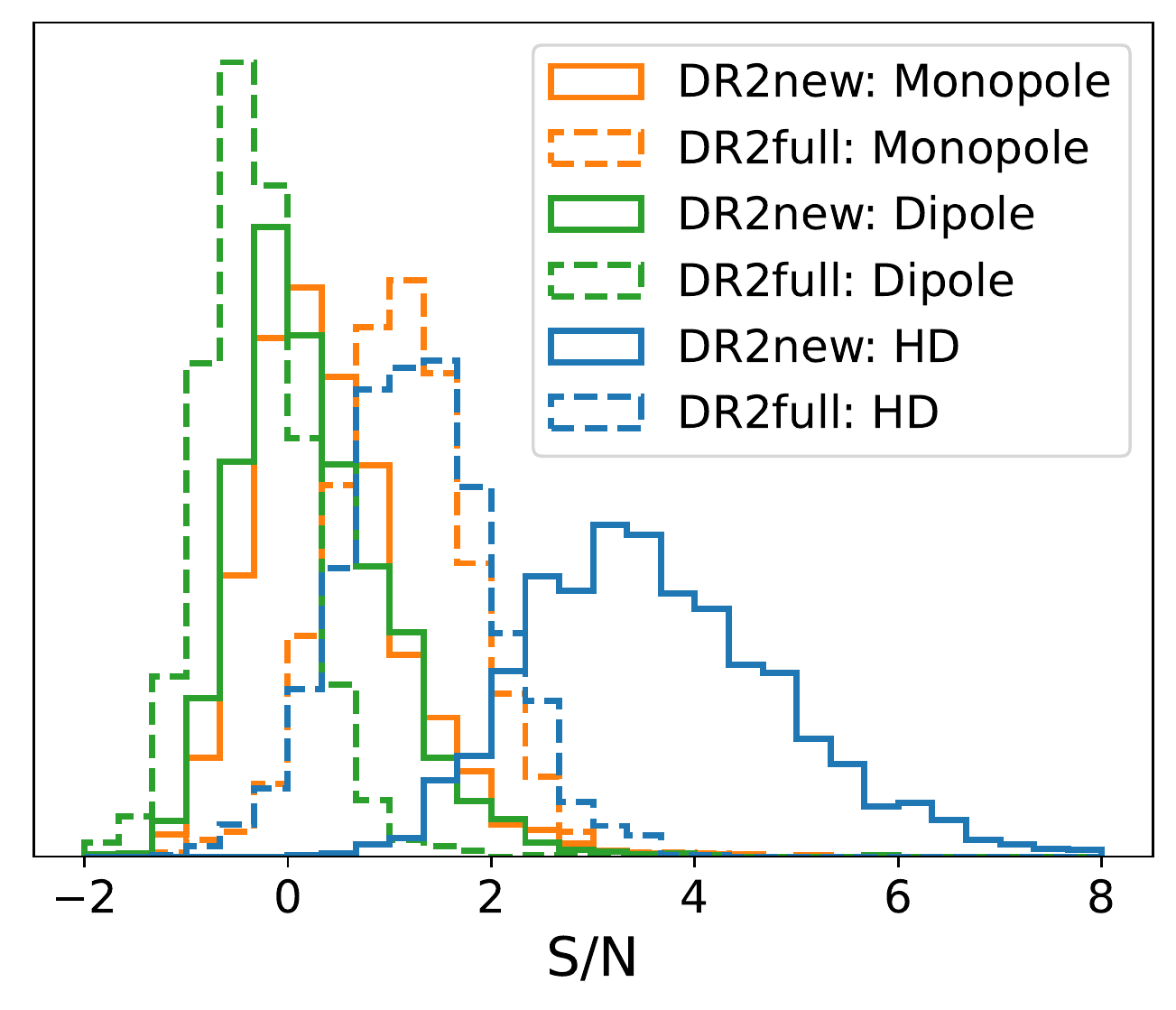}
\caption{Amplitude and S/N for a common red signal with $\gamma=13/3$ for the optimal statistic. The top and bottom panels show the noise-marginalised distributions of the squared amplitude $A_{\rm CRS}^2$ and S/N, respectively, for a common signal with different correlation patterns, with HD in blue, monopole in orange, and dipole in green. Solid lines are results from \texttt{DR2new} and the dashed lines are results from \texttt{DR2full}.}
\label{fig:os_amp_sn}
\end{figure}

\begin{figure}
\includegraphics[width=0.9\linewidth]{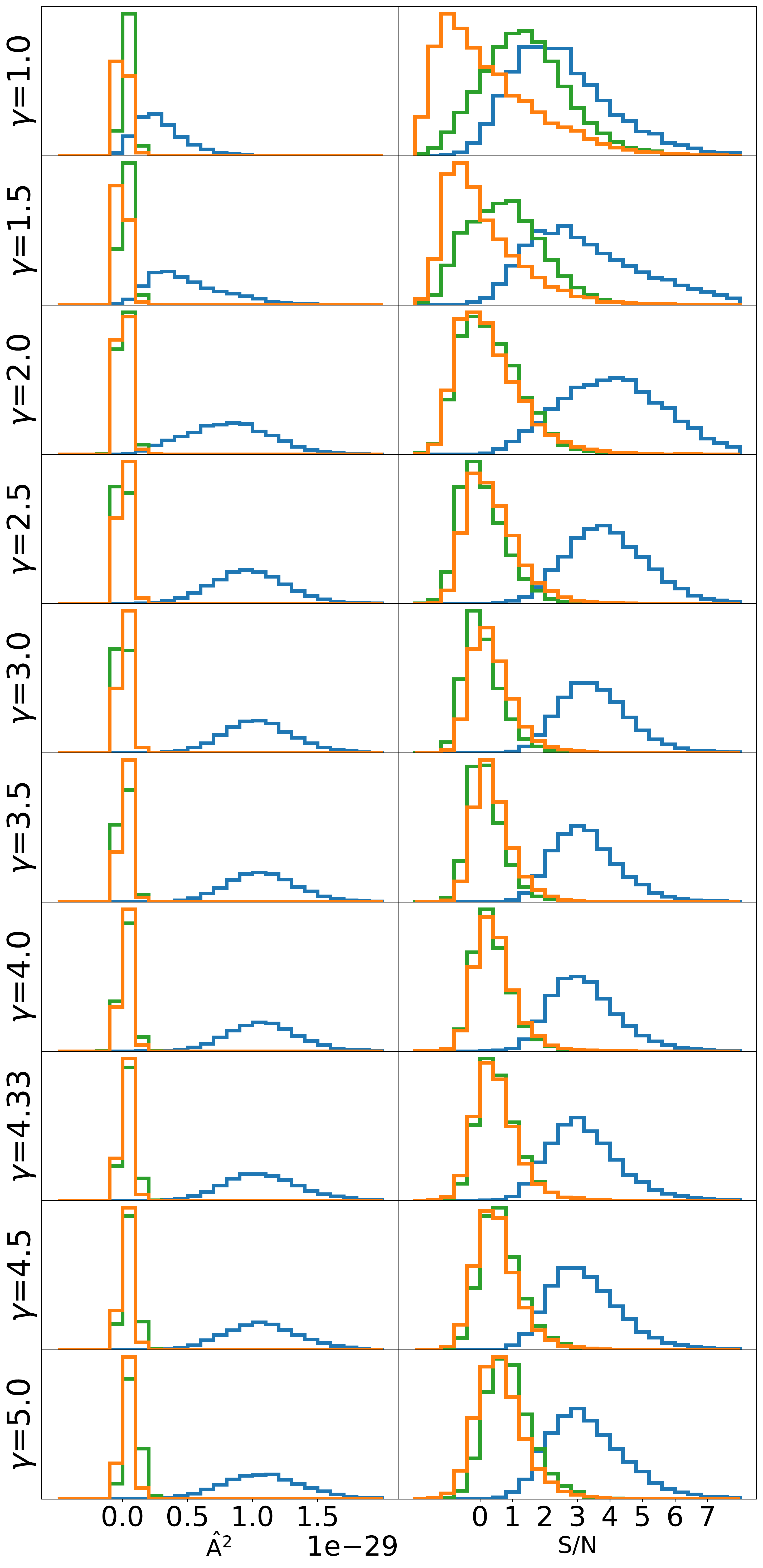}
\caption{Optimal statistic amplitude (left) and S/N (right) for a range of spectral indices for a common red signal using the \texttt{DR2new} data set. Results for the HD model are shown in blue, the dipole model in green, and the monopole model in orange.}
\label{fig:os_varied_index}
\end{figure}

\begin{table*}
\caption{
Median optimal statistic amplitudes and S/N for a single component fit for the monopole, dipole, or HD correlation fixing the spectral index of the common signal to $13/3$. The uncertainties indicate the 90\% credible region. 
}
\begin{center}
\def\arraystretch{1.5}
\begin{tabular*}{0.65\linewidth}{c||c|c|c|c}
\hline
 & \texttt{DR2full} & \texttt{DR2full+} & \texttt{DR2new} & \texttt{DR2new+} \\
\hline \hline
$A_{\rm MP}^2$ & $4.1^{+5.0}_{-3.9} \times 10^{-31}$ & $3.7^{+5.1}_{-4.1} \times 10^{-31}$ & $1.5^{+6.7}_{-4.3} \times 10^{-31}$ & $4.0^{+7.0}_{-4.6} \times 10^{-31}$ \\
$A_{\rm DP}^2$ & $-1.9^{+4.6}_{-4.1} \times 10^{-31}$ & $-0.8^{+5.1}_{-4.3} \times 10^{-31}$ & $0.8^{+9.3}_{-5.8} \times 10^{-31}$ & $3.9^{+9.1}_{-6.5} \times 10^{-31}$ \\
$A_{\rm HD}^2$ & $2.7^{+3.0}_{-2.5} \times 10^{-30}$ & $2.9^{+2.9}_{-2.4} \times 10^{-30}$ & $10.0^{+5.1}_{-4.9} \times 10^{-30}$ & $11.0^{+4.6}_{-4.4} \times 10^{-30}$ \\
S/N$_{\rm MP}$ & $1.1^{+1.1}_{-1.0}$ & $1.0^{+1.1}_{-1.1}$ & $0.3^{+1.4}_{-0.9}$ & $0.8^{+1.7}_{-1.0}$ \\
S/N$_{\rm DP}$ & $-0.4^{+0.9}_{-0.8}$ & $-0.2^{+1.0}_{-0.9}$ & $0.1^{+1.5}_{-0.9}$ & $0.6^{+1.5}_{-0.9}$ \\
S/N$_{\rm HD}$ & $1.3^{+1.3}_{-1.2}$ & $1.4^{+1.2}_{-1.1}$ & $3.5^{+2.4}_{-1.7}$ & $4.1^{+2.7}_{-1.7}$ \\
\hline
\end{tabular*}
\label{tab:os}
\end{center}
\end{table*}

\begin{table*}
\caption{
Model selection for different inter-pulsar correlation models for a common red signal (CRS). We present Bayes factors (${\rm  BF}$), for different CRS models against the CURN model. We assume the DE440 SSE fit and use the PSRN+CURN model as the reference model.
The model component acronyms are:
(i) PSRN = individual Pulsar noise, (ii) CURN = common uncorrelated red noise,
(iii) GWB = gravitational wave background with quadrupolar (HD) angular correlation, 
(iv) CLK = common signal with monopolar spatial correlation, as expected from a clock error,
(v) EPH = common signal with dipolar spatial correlation, as expected from SSE errors.
}
\begin{center}
\def\arraystretch{1.5}
\begin{tabular*}{0.95\linewidth}{c|c||cc|c|cc|c}
\hline
 &  & \multicolumn{2}{c}{\texttt{DR2full}} & \texttt{DR2full+} & \multicolumn{2}{c}{\texttt{DR2new}} & \texttt{DR2new+} \\

ID & Model & \eprise{} & \ftwo{}  & \eprise{} & \eprise{} & \ftwo{} & \eprise{} \\
\hline \hline
1 & PSRN + CURN          & -- & -- & -- & -- & -- & -- \\
\hline
2 & PSRN + GWB           & $4$ & $5$ & $4$ & $60$ & $62$ & $65$ \\
\hline
3 & PSRN + CLK           & $<0.01$ & $<0.01$ &  $<0.01$ & $0.2$ & $1.2$ & $0.3$ \\
\hline
4 & PSRN + EPH           & $<0.01$ & $\sim10^{-4}$  &  $<0.01$ & $0.2$ & $0.2$ & $1.3$ \\
\hline
5 & PSRN + CURN + CLK    & $2$ & $1$ &  $2.7$ & $0.8$ & $2$ & $1.6$ \\
\hline
6 & PSRN + CURN + EPH    & $1$ & $0.1$ &  $1$ & $1$ & $1$ & $1.6$ \\
\hline
7 & PSRN + GWB + CURN    & $3$ & $3$ &  $4$ & $27$ & $13$ & $25$ \\
\hline
8 & PSRN + GWB + CLK     & $5$ & $12$ &  $7$ & $28$ & $35$ & $57$ \\
\hline
9 & PSRN + GWB + EPH     & $3$ & $3$ &  $3.6$ & $33$ & $29$ &$43$ \\
\hline
\end{tabular*}
\label{tab:bf_dr2}
\end{center}
\end{table*}

Table \ref{tab:pl} gives the 90\% credible regions of the recovered power law parameters for different analyses and models. Once the data set is fixed, the CRS (CURN or GWB) parameter constraints are consistent between different software packages, namely \eprise{} and \ftwo{}, as well as different models. However, as already indicated by the free-spectrum analysis, there is a systematic difference in the CRS recovery between \texttt{DR2full} and \texttt{DR2new}.

We quantify the differences between the power law posterior distributions that arise from  using  different software packages and data sets 
by adapting the \texttt{tensiometer} package, outlined in \cite{raveri2021non} and 
summarised in \cite{wm2}.
This package essentially provides the probability density function of the parameter differences which can be integrated to obtain the mean probability for the presence of parameter shifts (see equation 4 in \cite{raveri2021non}). The resulting probability for a parameter shift can be converted into an effective number of $\sigma$s using the standard normal distribution.
In short, the package produces a score that can be interpreted as `within how many $\sigma$' two distributions are consistent \citep[see also][for more details]{RaveriHu}. The results of this analysis in Table~\ref{tension-table} indicate that the differences are minimal when comparing posteriors between different analysis software packages (\eprise{} vs \ftwo{}) regardless of the data set (either \texttt{DR2full} or \texttt{DR2new}).
However, when comparing GWB posteriors between different data sets (\texttt{DR2full} vs \texttt{DR2new}), there are tensions of  
$\sim 1\sigma$ for CURN, $\sim 1.4 \sigma$ for HD, and $\sim 1.6 \sigma$ for Binned ORF, regardless of the software package used.

Figure~\ref{tension-plots} shows, in the left panel, the two-dimensional posterior difference distribution between the \eprise{} and \ftwo{} posteriors obtained for the \texttt{DR2new} data set, again showing consistency of the results provided by the two independent analysis packages. On the contrary, the corresponding distribution for the difference in the posteriors 
associated with \texttt{DR2full} and \texttt{DR2new}, shown in the right panel, highlights the significant difference between the two data sets, more detailed comparisons can be found at the following URL\footnote{\url{https://github.com/subhajitphy/Posterior\_comparisons}}.

The parameter constraints from the Bayesian pipelines can be compared with the results of OS estimates. We first fix the spectral index to $\gamma=13/3$ and compute the OS amplitude and S/N for a CRS with monopole, dipole, or HD correlation. A summary of our findings is given in Table \ref{tab:os}, for the three correlation patterns in the four different data sets.
The best-fit amplitudes for the HD correlation from the OS can be compared with the Bayesian value found when slicing the posterior at $\gamma=13/3$, which is $A_{\rm HD}^2=6.0^{+4.0}_{-3.0} \times 10^{-30}$. We notice that this value sits halfway between the OS amplitude estimate for the two data sets, with $A_{\rm HD}^2$ of $2.7^{+3.0}_{-2.5} \times 10^{-30}$ for \texttt{DR2full} and $10.0^{+5.1}_{-4.9} \times 10^{-30}$ for \texttt{DR2new}. Both estimates overlap with the Bayesian value within their 90\% credible region. The median value for the OS S/N estimate for a HD-correlated process increases from 1.3 in \texttt{DR2full} to 3.5 for \texttt{DR2new}. 
The $A^2$ and S/N distributions of the correlated processes as estimated by the OS are shown in Figure~\ref{fig:os_amp_sn}, which further highlight the HD correlated signal emerging in \texttt{DR2new}.

\begin{figure*}
\centering
\includegraphics[width=\textwidth]{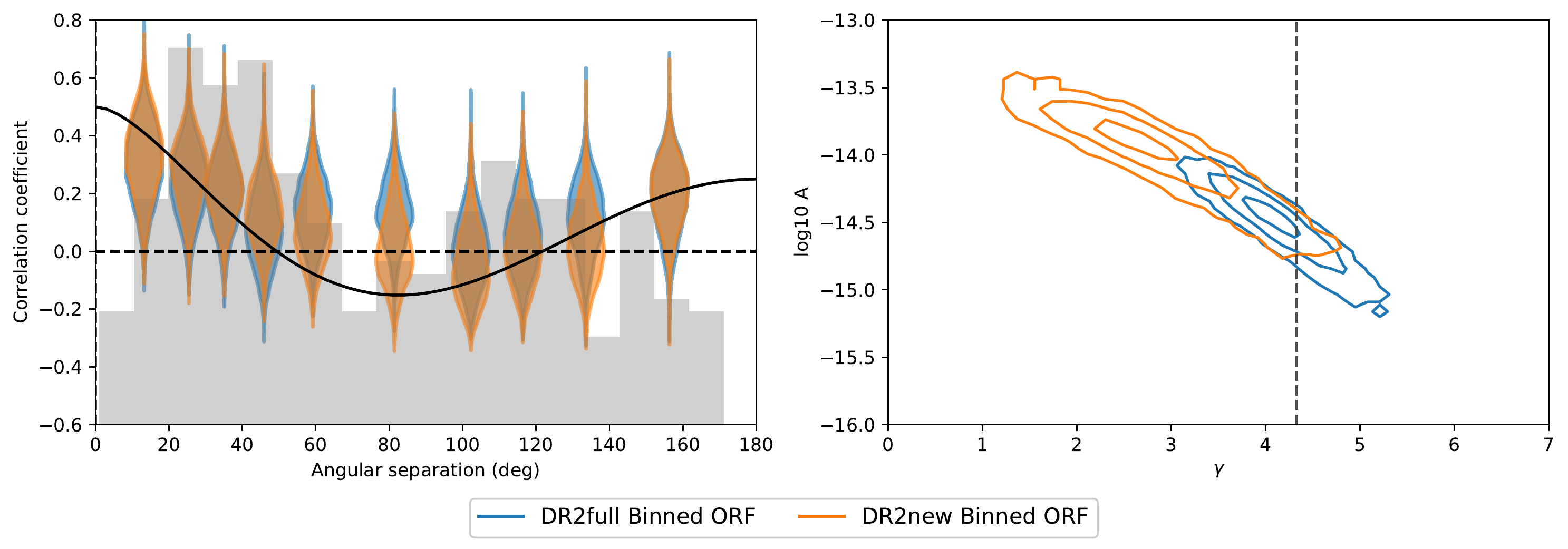}
\caption{
Binned overlap reduction function. Blue is for \texttt{DR2full} while orange is for \texttt{DR2new}. The left panel shows violins of the posterior of the correlation coefficients averaged at ten bins of angular separations with 30 pulsar pairs each. The black line is the HD curve based on theoretical expectation of a GWB signal. The grey histogram is the arbitrarily normalised distribution of the number of pulsar pairs at different angular separations. The right panel is the corresponding 2D posterior for the amplitude and spectral index of the common correlated signal, showing 1/2/3 $\sigma$ contours.}
\label{fig:bin_orf_pl}
\end{figure*}

\begin{figure}
\centering
\includegraphics[width=0.9\linewidth]{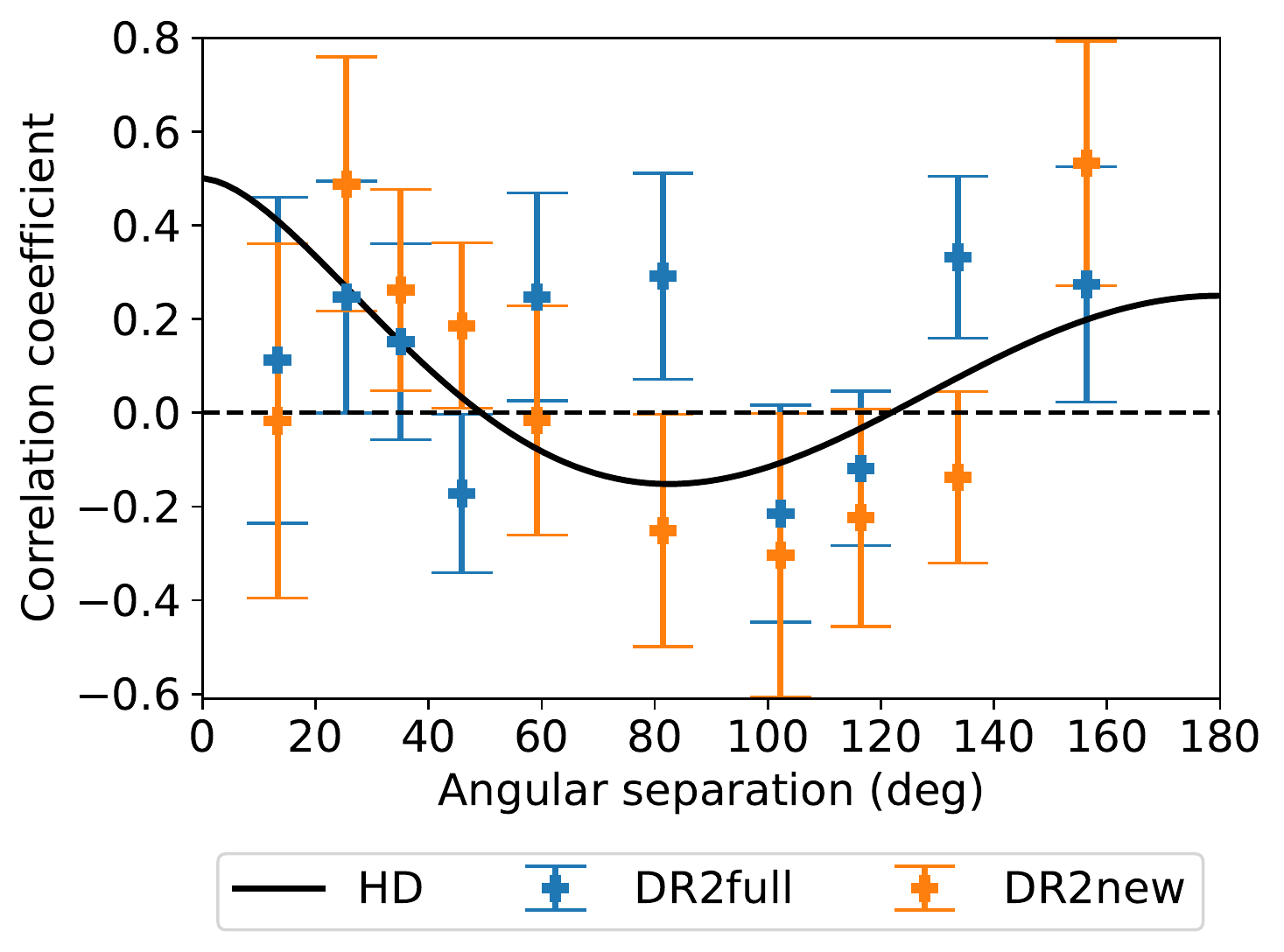}
\caption{Constraints on the overlap reduction function from the optimal statistic. Blue and orange points indicate the results for \texttt{DR2full} and \texttt{DR2new} respectively. The correlation coefficients for each pair of pulsars are weighted and averaged following the description in \cite{2022arXiv220807230A} and grouped in the same way as those in Figure \ref{fig:bin_orf_pl} for comparison. The HD correlation is plotted as a black line for reference.}
\label{fig:os_orf}
\end{figure}

Although we fixed $\gamma=13/3$ in the previous analysis, the OS can also be computed for a common red process with an arbitrary spectral slope.
Figure~\ref{fig:os_varied_index} shows how the OS amplitude and S/N of the \texttt{DR2new} data subset change as we vary the spectral index $\gamma$ of the CRS model. We increased $\gamma$ from 1.0 to 5.0 in steps of 0.5 and also
included $\gamma=13/3$ to show the expected spectral index of
a stationary ensemble of inspiralling SMBHBs. We evaluate the S/N for the monopole, dipole, and HD correlations for each $\gamma$.
The median of the HD S/N appears to peak around a $\gamma$ of 2.0, broadly consistent with the shallow posterior found in the Bayesian analysis (cf. the right panel of Figure \ref{fig:fs_pl}). The spread of the histograms, however, means that S/N values are self-consistent across the whole range of $\gamma$.

\subsection{Spatial correlation constraints}
\label{sec:spatial}

After checking for spectral properties, we reconstruct the spatial correlation of the common red signal. The results of the Bayesian search for the correlations with ten binned free parameters and a common red signal power law are shown in Figure \ref{fig:bin_orf_pl}. The bins are chosen so that each of them contains 30 pulsar pairs. The grey-shaded histogram represents the distribution of pulsar pairs as a function of separation. Since the pulsar distribution is concentrated in the galactic plane, we have more pairs at small angular separations compared to an array of pulsars uniformly distributed across the sky. However, broad coverage of all angles is still achieved with the 25 pulsars chosen using the ranking procedure of \citet{2023MNRAS.518.1802S}. When comparing the \texttt{DR2full} ORF constraints with those of \texttt{DR2new}, one can see that the latter appears much more consistent with the expected HD correlation. In particular, the bins around 60, 80 and 135 degrees (i.e. the fifth, sixth, and ninth bins) have more positive correlation coefficients in \texttt{DR2full}. These appear to be responsible for the signal in \texttt{DR2full} being consistent with a CURN and a monopole, as also implied by the OS amplitude and S/N for a monopole correlation reported in Table~\ref{tab:os}. In contrast, \texttt{DR2new} is very consistent with a HD-correlated process.
We also use Chebyshev and Legendre decompositions for the ORF in the Bayesian analysis and find ORF and power law constraints that are consistent with the binned free parameter analysis presented here; see Appendix~\ref{App:A}.

For comparison, the spatial correlations computed with the OS marginalised over the pulsar noise parameters are shown in Figure \ref{fig:os_orf}, where the correlation coefficients have been obtained by scaling to the median amplitude at fixed $\gamma = 13/3$, as given in Table \ref{tab:os}. For each noise realisation, only the median values of the pulsar pair correlation are used. While Bayesian analysis averages the correlation within each bin, the OS uses each pulsar pair independently and fits the best correlation across all pairs. For comparison and visual purposes, we choose the same binning and avoid showing 300 individual pulsar pairs. Although the two methods give broadly comparable ORF constraints, several differences can be found. Firstly, the first bin with the pulsar pairs with the closest separations deviates away from the HD and is consistent with no correlation. Second, the fourth and seventh bins drop significantly into negative correlations. These dips are most prominent in \texttt{DR2full}, while \texttt{DR2new} follows the HD curve more closely. Consistent with the Bayesian evaluation, the OS reconstruction also shows prominent positive correlations for the fifth, sixth, and ninth bins in the \texttt{DR2full} data set, making the overall curve inconsistent with HD. 

To quantify how likely the data set is actually showing evidence for a GWB with HD correlation, we compute Bayes factors comparing different spatial correlations: Hellings-Downs (HD) correlations that arise from a GWB, monopole correlations that could be produced by clock errors (CLK) and dipole correlations that could be due to SSE systematics (EPH). Firstly, the presence of a common uncorrelated red noise (CURN) is strongly favoured against a model that includes only individual pulsar noises (PSRN). Compared to $3.3$ from the 6PSR data set, the log10 Bayes factors are $\sim5$ for \texttt{DR2full} and $\sim3$ for \texttt{DR2new}. Our main comparison was thus against the PSRN+CURN model, with respect to which we referred all Bayes factors. These are summarised in Table \ref{tab:bf_dr2}. Both the PSRN+CLK and PSRN+EPH models are heavily disfavoured against the PSRN+CURN model in both \texttt{DR2full} and \texttt{DR2new} (row IDs 3 and 4 in Table~\ref{tab:bf_dr2}). The evidence for these two correlations as additional processes to the CURN is inconclusive (row IDs 5 and 6). Conversely, while \texttt{DR2full} shows very little evidence for a GWB compared to the CURN, \texttt{DR2new} has a significant Bayes factor in favour of HD (row ID 2). Since this is a significant result, we have recomputed the Bayes factor using several alternative samplers and methods, as described in Section \ref{sec:bayes}, obtaining Bayes factors of 66, 56, 62. In this data set, we also find that BFs for models including an additional CLK or EPH or CURN are about a factor of two smaller compared to the model including a GWB alone (row IDs 7, 8 and 9). On the contrary, the analogue BFs for \texttt{DR2full} are inconclusive with an indication for an additional monopole process (row ID 8).

These BFs can be compared to the S/N from the OS in Table \ref{tab:os} and Figure \ref{fig:os_amp_sn}. The \texttt{DR2new} data set yields a median S/N $\approx 3.5$ for the HD correlation, while it is about $1.3$ in \texttt{DR2full}. These S/N estimates act as semi-independent confirmation of the BFs from Table \ref{tab:bf_dr2}. Consistent with the slightly higher BF for an additional monopole in \texttt{DR2full} the S/N is $\approx 1.2$. For \texttt{DR2new} the S/N for a monopole drops to be consistent with zero. Lastly, no significant signal for a dipole correlation is found in either of the two data sets.

\subsection{Significance tests}

To quantitatively estimate the significance of the hypothesis that a GWB signal with HD correlation is present in the data, the null hypothesis distribution need to be constructed.
Many repetitions of an experiment need to be performed in order to define a strict p-value. This is, unfortunately, not possible for PTAs. Thus, we can only attempt to find a good proxy to estimate the true statistical p-value for the null hypothesis. In the following, we refer to the estimated value from our proxy methods as p-values for simplicity.
The respective distributions can be constructed in two different ways, by introducing random phase shifts in the Fourier basis of the common red noise process \citep{tlb+2017} or by moving the positions of the pulsars in the sky via a random scramble \citep{2016PhRvD..93j4047C}. The aim of both methods is to effectively destroy the distinctive cross-pulsar correlations, unique to the GWB signal, while retaining the individual pulsar noise characteristics.
One should emphasise that both methods should be robust against any mismodelled features in the data set, therefore they, in general, provide more conservative estimates of the significance in comparison to the possibly oversimplified noise simulation bootstrapping.

\begin{figure}
\includegraphics[width=0.9\linewidth]{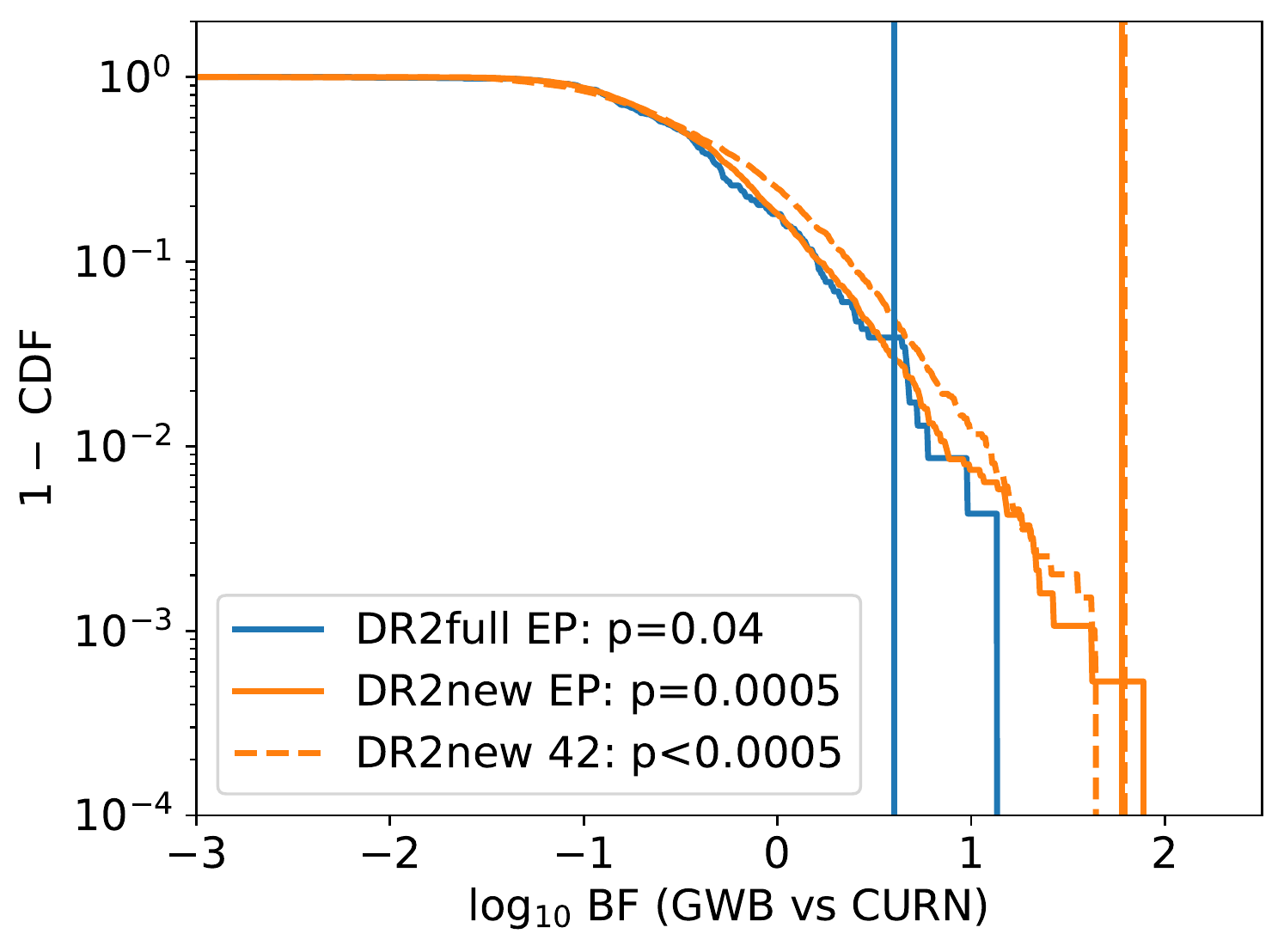}
\includegraphics[width=0.9\linewidth]{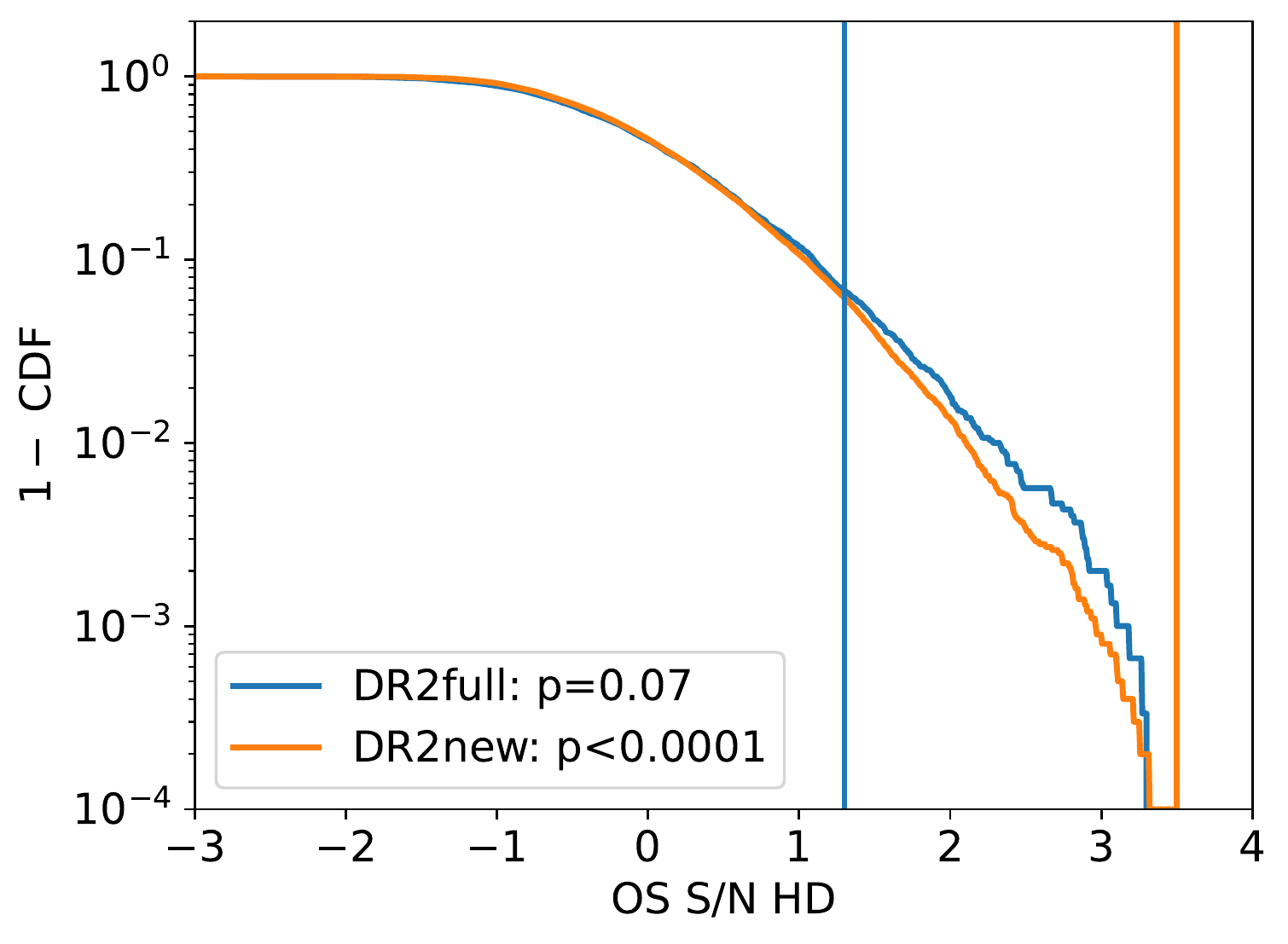}
\caption{$1-$cumulative density function (CDF) from phase shifts. The top panel is for the Bayes factors for PSRN+GWB versus PSRN+CURN using \eprise{} (EP) and \ftwo{} (42). It should be noted that due to the computational cost, we calculated only a limited number of phase-shifted BFs. This could explain the differences in the 1$-$CDFs. The OS S/N for the HD correlation with \eprise{} is shown in the bottom panel. Blue lines are for \texttt{DR2full} while orange lines are for \texttt{DR2new}. Vertical lines are the measured Bayes factor for PSRN+GWB versus PSRN+CURN reported in Table~\ref{tab:bf_dr2} or the OS S/N$_{\rm HD}$ reported in Table~\ref{tab:os} respectively. The estimated p-values for each method are given in the legends.}
\label{fig:fap_pshift}
\end{figure}

The distributions of BFs under the null hypothesis (PSRN + CURN) were constructed for \texttt{DR2full} and \texttt{DR2new} using about 200 and 2000 phase shifts, respectively and are displayed in the upper panel of Figure \ref{fig:fap_pshift}. The \texttt{DR2full} measured BF from Table \ref{tab:bf_dr2} lies within the 2$\sigma$ range of the null hypothesis distribution with a p-value of $0.04$. The p-value for the BF derived with the \texttt{DR2new} data set reaches a statistically interesting value of $0.0005$, which corresponds to the 3$\sigma$ level of significance ('evidence'). The analysis was performed using both \eprise{} and \ftwo{} and shows consistent results between the two software packages. This significance test was repeated for the OS S/N values for the HD correlation and results are shown in the bottom panel of Figure \ref{fig:fap_pshift}. For \texttt{DR2full} a p-value of $0.07$ is found. None of the 10000 realisations produced a S/N that is comparable to what has been found in \texttt{DR2new}. Therefore, only an upper limit can be set for the p-value $<0.0001$, which corresponds to a significance of $>3.5\sigma$.

\begin{figure}
\includegraphics[width=0.9\linewidth]{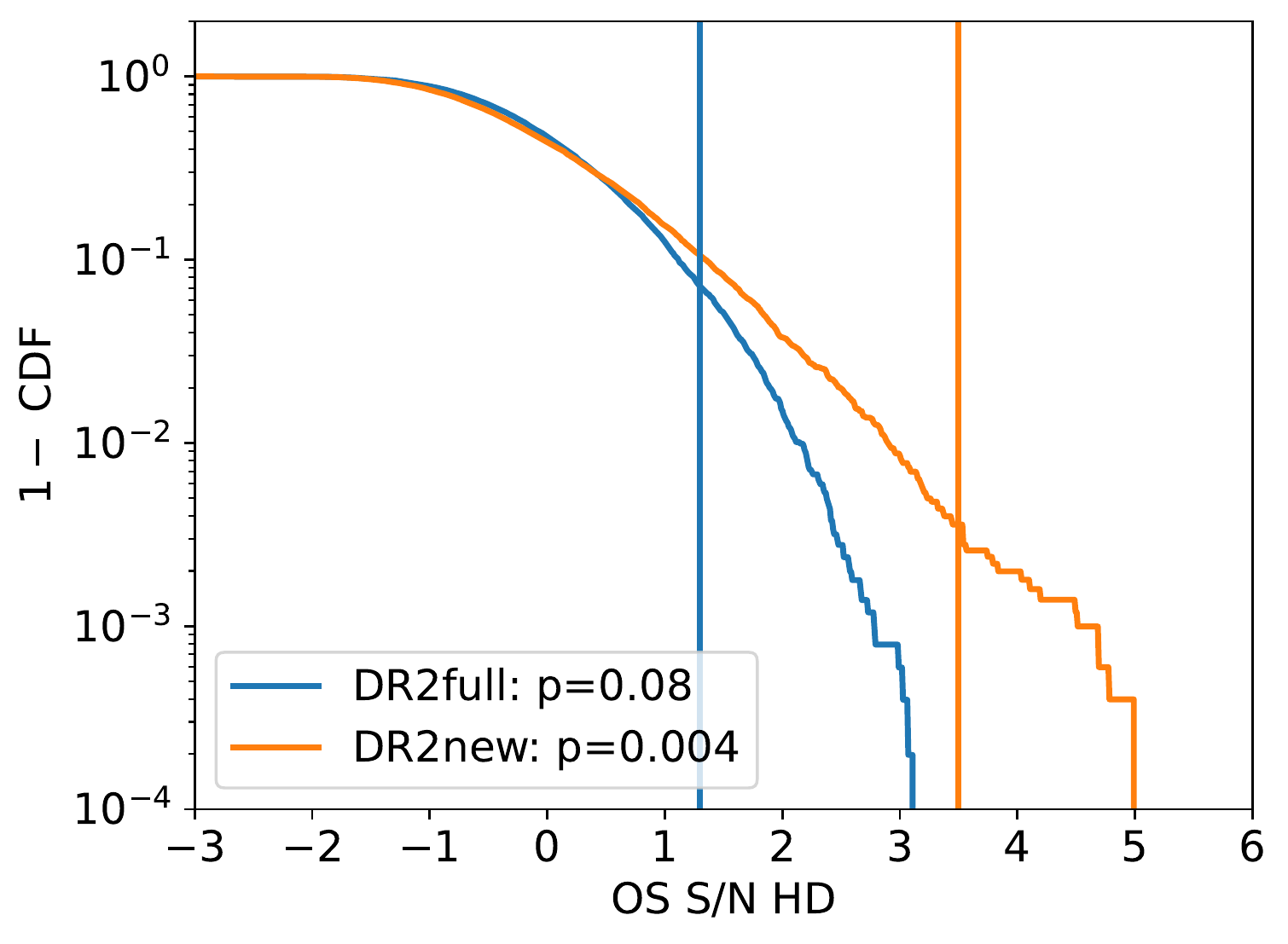}
\includegraphics[width=0.9\linewidth]{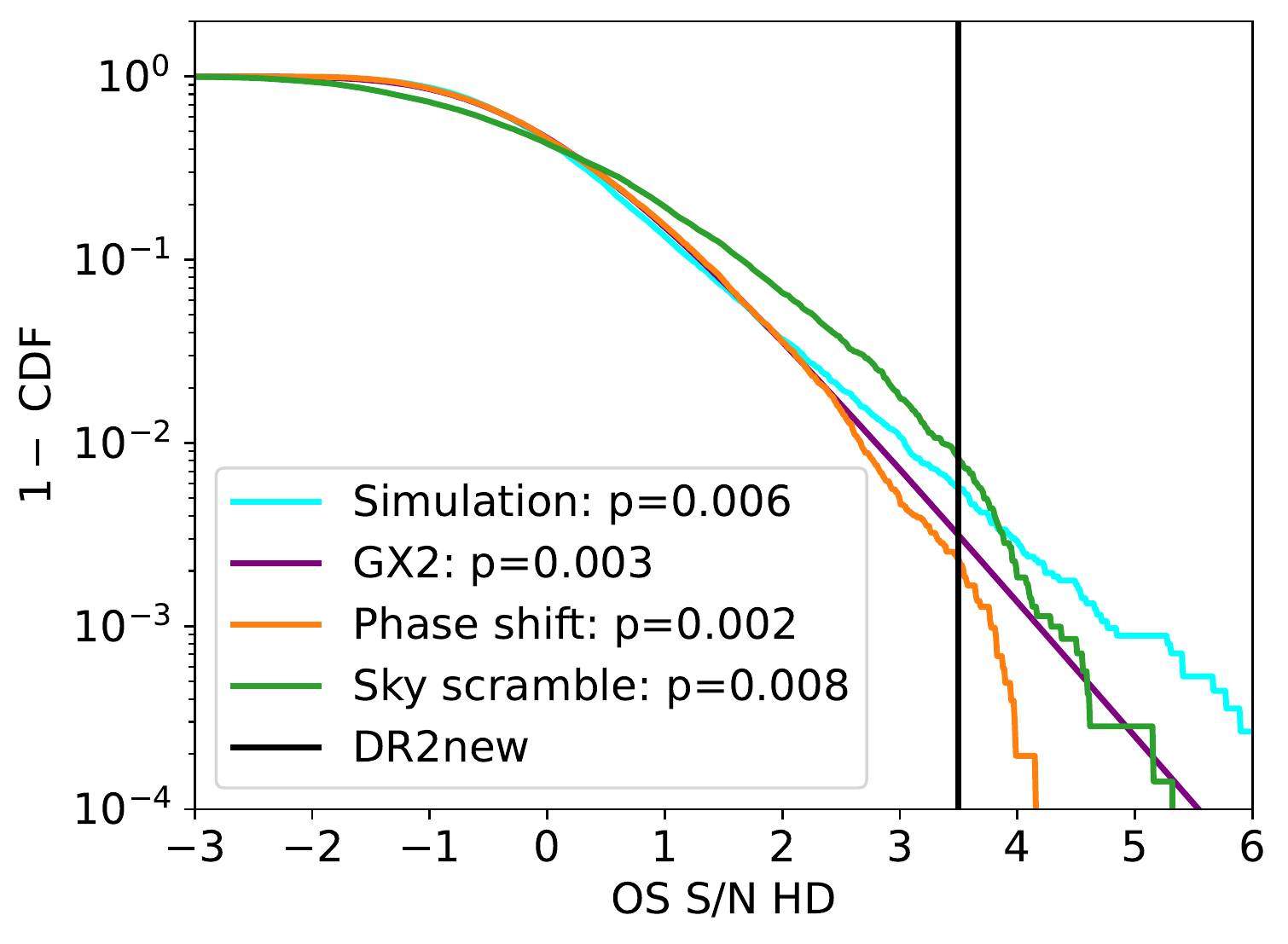}
\caption{$1-$cumulative density function (CDF) from sky-scrambled optimal statistic HD S/N distributions and a comparison between different methods. In the top panel, the blue histograms are for \texttt{DR2full} while the orange histograms are for \texttt{DR2new}. Vertical lines are the measured S/N$_{\rm HD}$ values reported in Table~\ref{tab:os}. The large discrepancy between the two data sets could be an indication that some remaining signal is still present after the scrambling, especially in the strong S/N regime. More checks need to be performed to assess the validity of this method. The bottom panel compares a simulated null distribution in cyan, the generalised $\chi^2$ (GX2) distribution from \citep{2023arXiv230501116H} in purple and the two proxy methods of phase shifting in orange and sky scrambling in green. At the measured value from the \texttt{DR2new}, the two methods differ by a factor of a few. The estimated p-values for each method are given in the legends.}
\label{fig:fap_scramble}
\end{figure}

Figure \ref{fig:fap_scramble} shows the null distribution obtained with sky scrambles in the OS analysis in the top panel. A matching threshold of 0.2 for any two sky scrambles was imposed to produce about 5000 samples. A large difference particularly in the high S/N tail of the density functions can be found between \texttt{DR2full} and \texttt{DR2new}. The p-value for \texttt{DR2full} of $0.08$ is comparable to that obtained with the phase shifts. This could indicate that in the low S/N regime, both methods produce reliable null distributions. In the high S/N regime, however, with \texttt{DR2new} the sky scramble p-value of $0.004$ is not consistent with the phase shift method.

The bottom panel of Figure \ref{fig:fap_scramble} compares p-values from simulations, theoretical computation and the two methods. A null distribution was generated using a set of realistic simulations resembling the statistical properties of the real \texttt{DR2new} data set and with the injected CURN only. The noise parameters as well as the amplitude and slope of the CURN for the null distribution were taken as random samples from the posteriors of the CURN search with \texttt{DR2new}. Additionally, we compare these p-values with those obtained from the theoretical null distribution described by generalised $\chi^2$ (GX2) distributions and derived in \cite{2023arXiv230501116H}. The distribution is computed by fixing the noise parameters to the median values of the posteriors\footnote{This difference in how the parameters for the pulsar noise are chosen between the simulated and theoretical null distributions can explain the more conservative  p-value for the former. As the randomly drawn parameters for the simulations could allow for more large S/Ns to be found by chance.}.

Both null distributions are compared to the proxy distributions obtained via phase shifting and sky scrambling. For consistency, instead of using the real data set, a target data set was simulated using the same procedure as the simulations for the null distribution, but with GWB injected instead of CURN. The lowest p-value of 0.002 is obtained using phase shifts. The most conservative number is obtained when using all sky scrambles without introducing any threshold or weighting with the p-value of 0.008. At S/N$_{\rm HD} = 3.5$ the p-values of the simulated (0.006) and theoretical (0.003) null distributions lie between those from phase shifting and sky scrambling, see Figure \ref{fig:fap_scramble}. We have tested introducing thresholds on the match between the true and scrambled pulsar positions and amongst the scrambles themselves and found that in general smaller thresholds lead to a decrease in the proxy null distribution. Similar results are obtained when adding weights to account for the different contributions from each pulsar. From all the above we can conclude that the p-value for the S/N found with \texttt{DR2new} should be $\sim 0.004$, which was also obtained with sky scrambling at a threshold of 0.2 and no weights. The inconsistencies between the p-values obtained with the real and simulated target data sets can be due to the incompleteness of our simulations (e.g. exponential dips for J1713$+$0747 are not included and a simpler noise model with a fixed number of frequency components was used).

For an optimal sky scrambling orthogonality may need to be ensured between different realisations. Additionally, a realistic PTA does not have equally good pulsars, which should be taken into account when assessing the match between different scrambles. This can limit the maximum number of possible sky scrambles or their effectiveness in breaking correlated processes \citep{2023arXiv230504464D}.
On the other hand, \cite{2023arXiv230501116H} have shown that sky scrambling can lead to null distributions that are consistent with the theoretical prediction.
Further studies are required to determine whether any method can be a good conservative proxy for PTA experiments to accurately estimate the p-value and significance of a detected signal.

\subsection{Consistency tests}

\subsubsection{Comparing the power in the auto-correlation and cross-correlation terms.}

For a true GWB both the auto-correlation and cross-correlation-terms should constrain the same process. Since the cross-correlated power is proportional to the square of the expectation value of the ORF, that is $\Gamma^2$,  which is always $\ll 1$, it is expected that the power in the auto-correlation terms -- equivalent to the CURN -- is the dominant contributor to the signal. However, we stress that the cross-correlation terms contain the `smoking gun' that can provide conclusive evidence for the presence of a GWB in the data.

\begin{figure}
\centering
\includegraphics[width=0.9\linewidth]{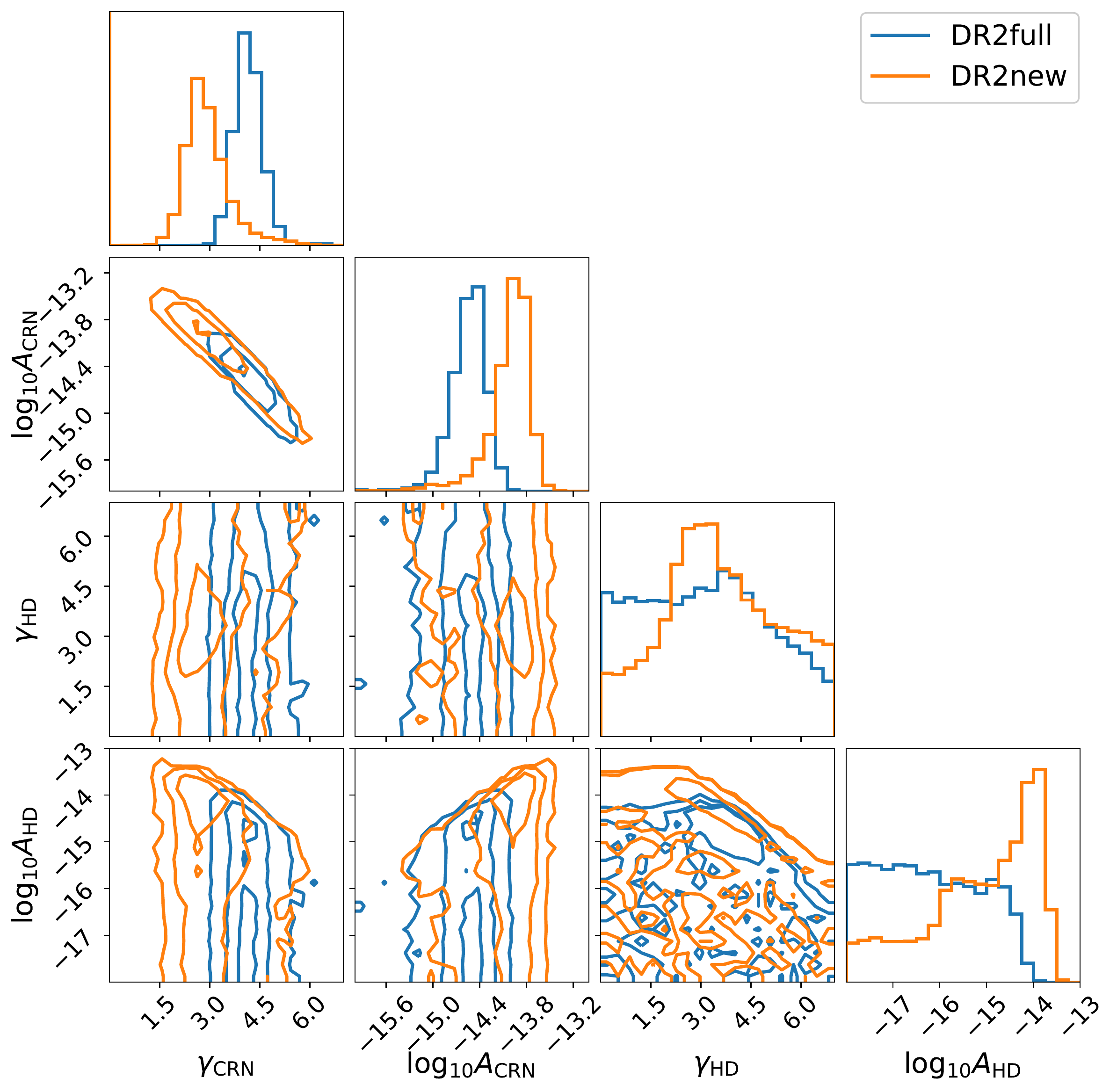}
\caption{Power law posterior constraints from a split analysis using the auto-correlation terms (CURN) and the cross-correlation terms (assuming HD correlation) separately. The left two columns show the recovered power law from the auto-correlation terms and the right two columns show the cross-correlation terms. Blue distributions are for \texttt{DR2full} while orange distributions are for \texttt{DR2new}.}
\label{fig:zero_diag}
\end{figure}

We apply the split likelihood technique \citep{abb+2020,ipta+22} to both \texttt{DR2full} and \texttt{DR2new}. Figure \ref{fig:zero_diag} shows the resulting posterior contours. While the auto-correlation-terms-only-analyses recover the CURN with consistent amplitude and slope for both data sets, noticeable differences can be seen for the power law parameters that are constrained using the cross-correlation terms only and assuming the HD correlation. For the \texttt{DR2full} data set, the cross-correlation terms contain virtually no power and thus only an upper limit is found for an HD-correlated signal. On the contrary, consistent with the much larger evidence for a GWB, the cross-correlation terms in \texttt{DR2new} produce a peak in amplitude. Since a long tail still remains, one cannot rule out the possibility of a zero amplitude.

\subsubsection{Solar System Ephemeris systematics}
The effects of SSE systematics on the PTA GWB search have been studied in \cite{thk+2016,gll+2019} and models that can help mitigate the dipolar correlated signal induced by SSE systematics have been added to the tools for PTA analysis.
We employ the widely used \beph{} model \citep{abb+2018,2020ApJ...893..112V} on the EPTA-only data sets.

\begin{figure*}
\centering
\includegraphics[width=\textwidth]{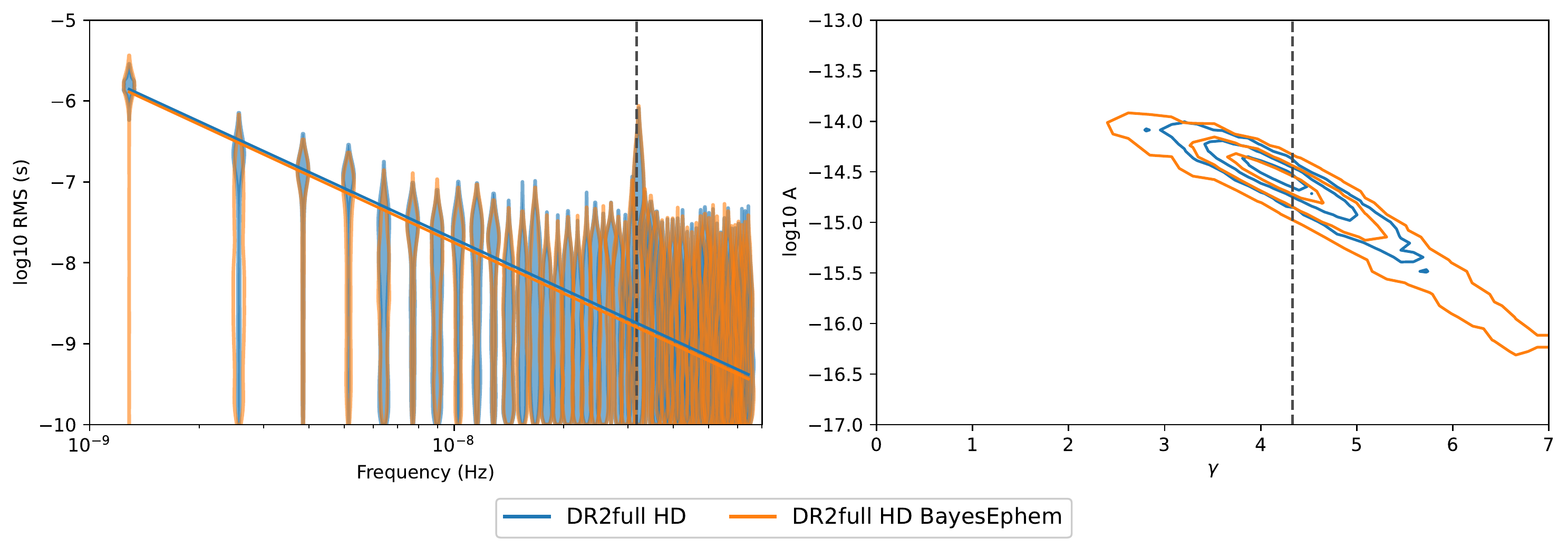}
\includegraphics[width=\textwidth]{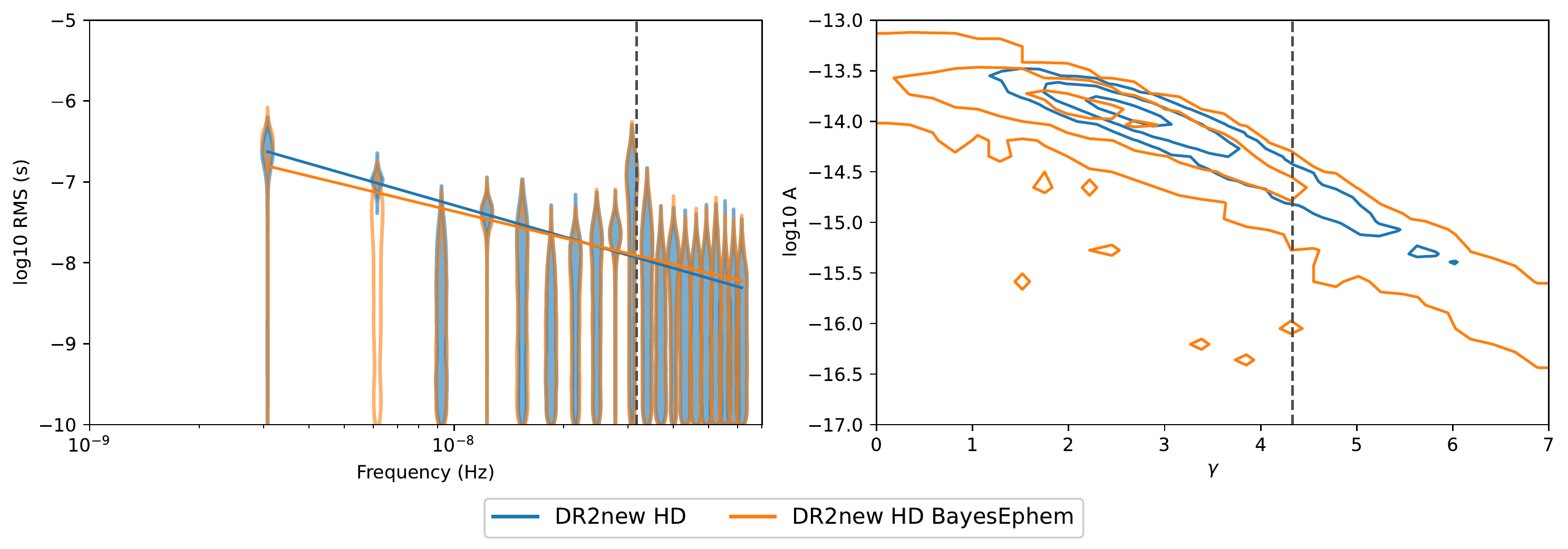}
\caption{Spectral properties of a GWB signal in the style of Figure \ref{fig:fs_pl} for the \texttt{DR2full} (top) and \texttt{DR2new} (bottom) comparing the analysis without (blue) and with (orange) \beph{}.}
\label{fig:fs_pl_be}
\end{figure*}

Figure \ref{fig:fs_pl_be} compares the addition of \beph{} to each data set against the use of the DE440 fit alone. Allowing SSE systematics to be present and absorbed by a model is a more conservative approach. In general, the left panels show that certain frequency bins have lower power compared to the DE440 analysis. This in turn broadens the power law posteriors. Comparing the \texttt{DR2full} data set against the \texttt{DR2new} data set, the longer time span of \texttt{DR2full} helps to produce results that are more independent of the SSE fitting used, while the short time span of \texttt{DR2new} strongly limits its ability to separate SSE systematics from other common signals. In fact 10.3 years is close to the Jovian orbital period of about 12 years, which could lead to a strong degeneracy.

Adding \beph{} also affects the BF for the different signal models. 
Table \ref{tab:bf_dr2_be} shows a selection of the same models as Table \ref{tab:bf_dr2}. The most relevant effect in the search for a GWB is that the BF in favour of the HD correlation in \texttt{DR2new} is reduced from about 60 to 17, a factor of about 3. Since \beph{} is known to partially absorb power from the GWB, this reduction follows the expectation, although the exact amount is difficult to predict \citep{2020ApJ...893..112V}.

\begin{table}
\caption{
Bayes factors for different CRS models against the CURN model adding \beph{} to all models with the \eprise{} pipeline.
The model component acronyms are as in Table \ref{tab:bf_dr2}
}
\begin{center}
\def\arraystretch{1.5}
\begin{tabular*}{0.8\linewidth}{c|c||c|c}
\hline
ID & Model & \texttt{DR2full} & \texttt{DR2new} \\
\hline \hline
1 & PSRN + CURN          & -- & -- \\
\hline
2 & PSRN + GWB           & $1.5$ & $17$ \\
\hline
3 & PSRN + CLK           & $0.5$ & $2$ \\
\hline
4 & PSRN + EPH           & $<0.01$ & $0.4$ \\
\hline
\end{tabular*}
\label{tab:bf_dr2_be}
\end{center}
\end{table}

\begin{figure*}
\centering
\includegraphics[width=0.43\linewidth]{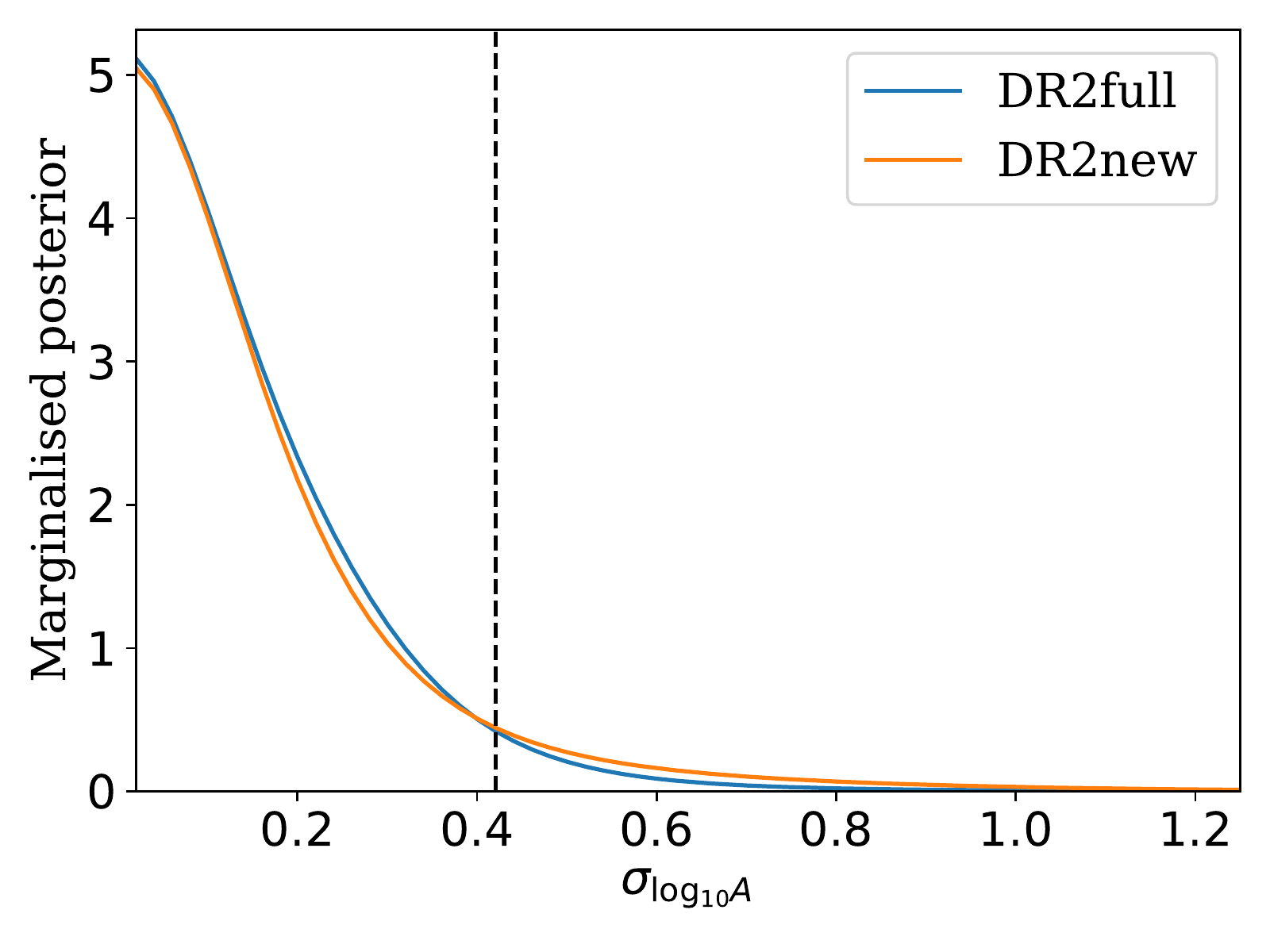}
\includegraphics[width=0.46\linewidth]{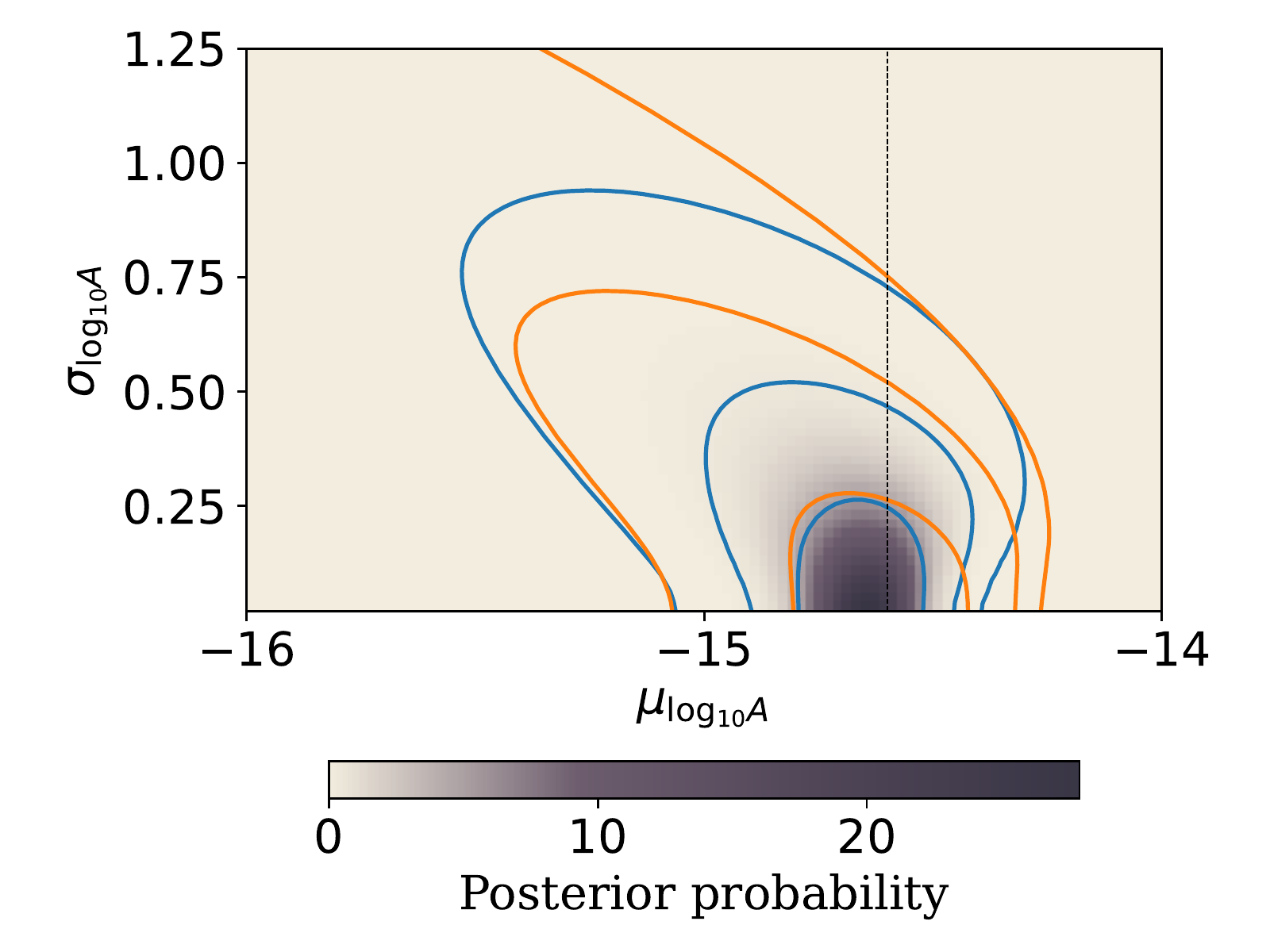}
\includegraphics[width=0.9\textwidth]{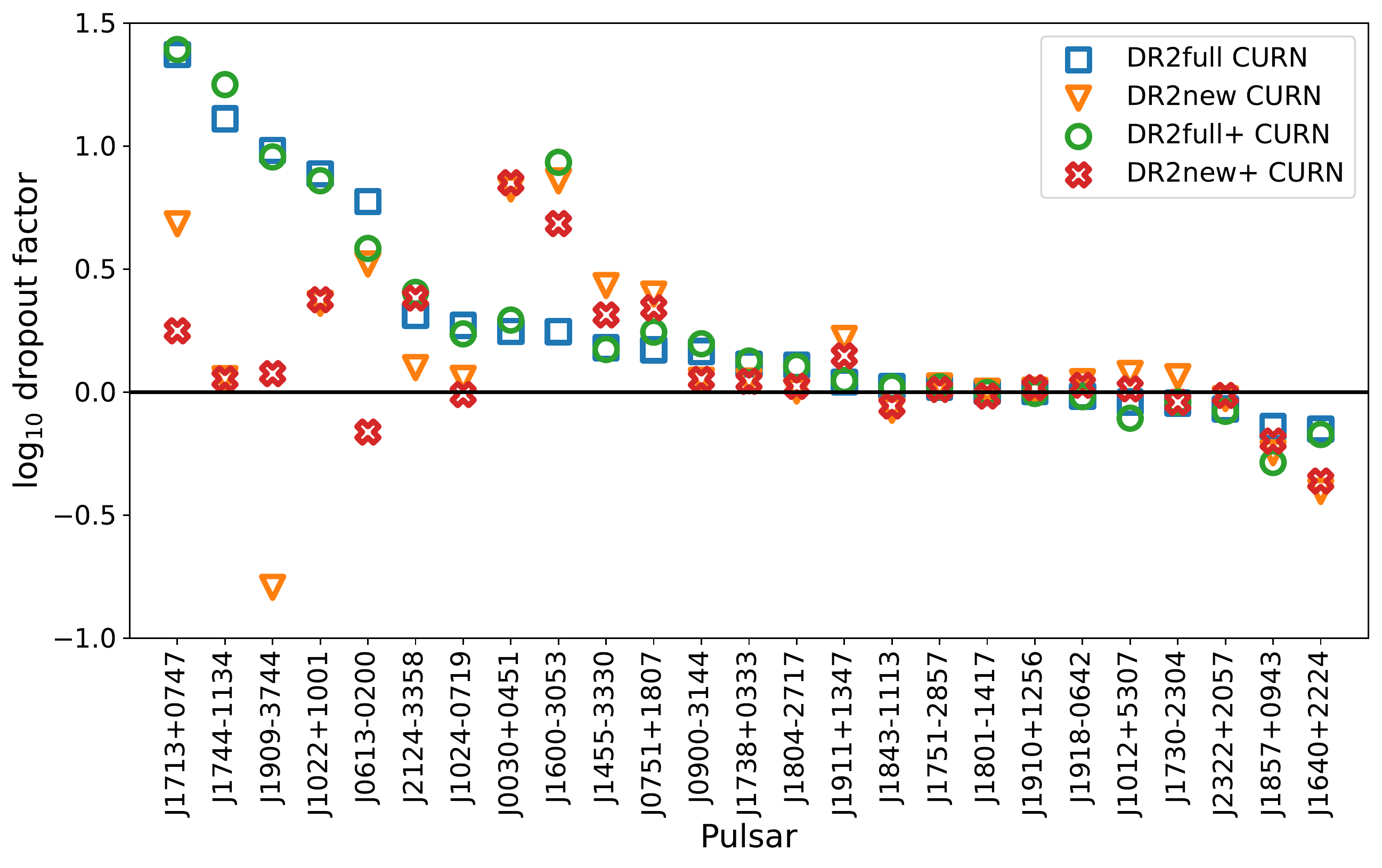}
\caption{Tests of the CURN model. The top two panels show that pulsar spectra of temporal correlations attributed to CURN are indeed consistent with representing the same spectrum~\citep{gts+2022}.
The blue lines show the measurement for \texttt{DR2full}, whereas the orange lines show the measurement for \texttt{DR2new}.
The top left panel is the measurement of the standard deviation of the CURN amplitudes across pulsar spectra, $\sigma_{\log_{10}A}$, marginalised over the mean of the CURN amplitudes, $\mu_{\log_{10}A}$, as well as pulsar-intrinsic noise parameters.
The dashed vertical line denotes an upper limit at 95\% credibility.
The top right panel shows both the mean and the standard deviation.
The inferred mean value is consistent with the measurement of $\log_{10}A_{\text{CURN}}$ denoted by a dashed vertical line.
The bottom panel shows the logarithm of the dropout factors~\citep{abb+2020} which suggest pulsars' contribution to the CURN model.
Positive values represent support for the CURN model, whereas negative values point towards inconsistency of pulsar data with the CURN model.
The differences in the dropout factors between the data sets could be due to differences in the recovered CURN or the pulsar red noise spectral properties.
}
\label{fig:cvsqc}
\end{figure*}

\subsubsection{Distinguishing a common red spectrum from similar individual pulsar noise spectra}

It was shown in \cite{gsr+2021} that the CURN hypothesis may become strongly favoured over the PSRN hypothesis even when spectra of temporal correlations across pulsars are similar and yet not strictly common, as implied by the CURN model.
This is because the standard uniform prior distributions for power law parameters of pulsar-intrinsic red noise do not match the observed distributions of these parameters.
The issue was addressed in \cite{gts+2022}, where the authors introduced a hierarchical model that governs the distribution of power law amplitudes across pulsars in the CURN.
As the distribution width is consistent with zero, this indicates the likely presence of a common-spectrum stochastic process in the data.
Dropout analysis, \citep[e.g.][]{abb+2020} also enables the mitigation of this issue by identifying pulsar outliers. 
The effect of the range of simulated pulsar noise parameters on the spurious evidence for CURN is demonstrated in Figure 6 in \cite{zhs+2022}.

To test each pulsar's consistency with the CURN, we measured both the dropout factors and the distribution of power law amplitudes of the CURN spectrum in the pulsars. The results are shown in Figure \ref{fig:cvsqc}.
The two top panels show the measurements of $\sigma_{\log_{10}A}$ and $\mu_{\log_{10}A}$, the standard deviation and the mean of the power law amplitude of CURN measured in the pulsars, following \cite{gts+2022}.
As expected, the mean $\mu_{\log_{10}A}$ is consistent with the measurement of $\log_{10}A_{\text{CURN}}$.
At the same time the standard deviation is consistent with zero, confirming the presence of common temporal correlations in the data.
Based on the measured dropout factors, we find that in both data sets only a few pulsars have intrinsic red noise that seems to be inconsistent with the CURN. The majority of pulsars display indifferent behaviour, leaving around five pulsars to contribute most to the constraints of the CURN power law. However, the two data sets have CURNs with very different posterior contours. This can have an impact on the difference between certain pulsars agreeing more or less with the CURN. J1909$-$0747 and J1744$-$0744, for example, have very steep intrinsic red noise, thus they are more consistent with the steep CURN from \texttt{DR2full} compared to the shallow CURN from \texttt{DR2new}. More discussion on the intrinsic pulsar noise properties can be found in \cite{wm2}.

\subsection{Continuous GW signal search}

In addition to the GWB search, we have also searched our data for the presence of a continuous GW (CGW) signal from an individually resolvable SMBHB in a circular orbit. This subsection presents a preliminary analysis and a detailed investigation (which includes simulations and significance estimation) will be given in a separate paper. The main aim of this section is twofold: (i) to look for an alternative explanation of the observed common signal and (ii) to understand how the inclusion of the CGW signal in the model affects the main findings of our analysis. 

First, we performed an analysis using the \texttt{DR2full} data set. The addition of the CGW signal to the custom pulsar noise and CURN is not informative about the presence of a CGW, with a Bayes factor $BF^{\textrm{CGW+CURN}}_{\textrm{CURN}} = 0.5$ (model containing CGW and CURN over the model with CURN only). 

\begin{figure}
\centering
\includegraphics[width=0.98\linewidth]{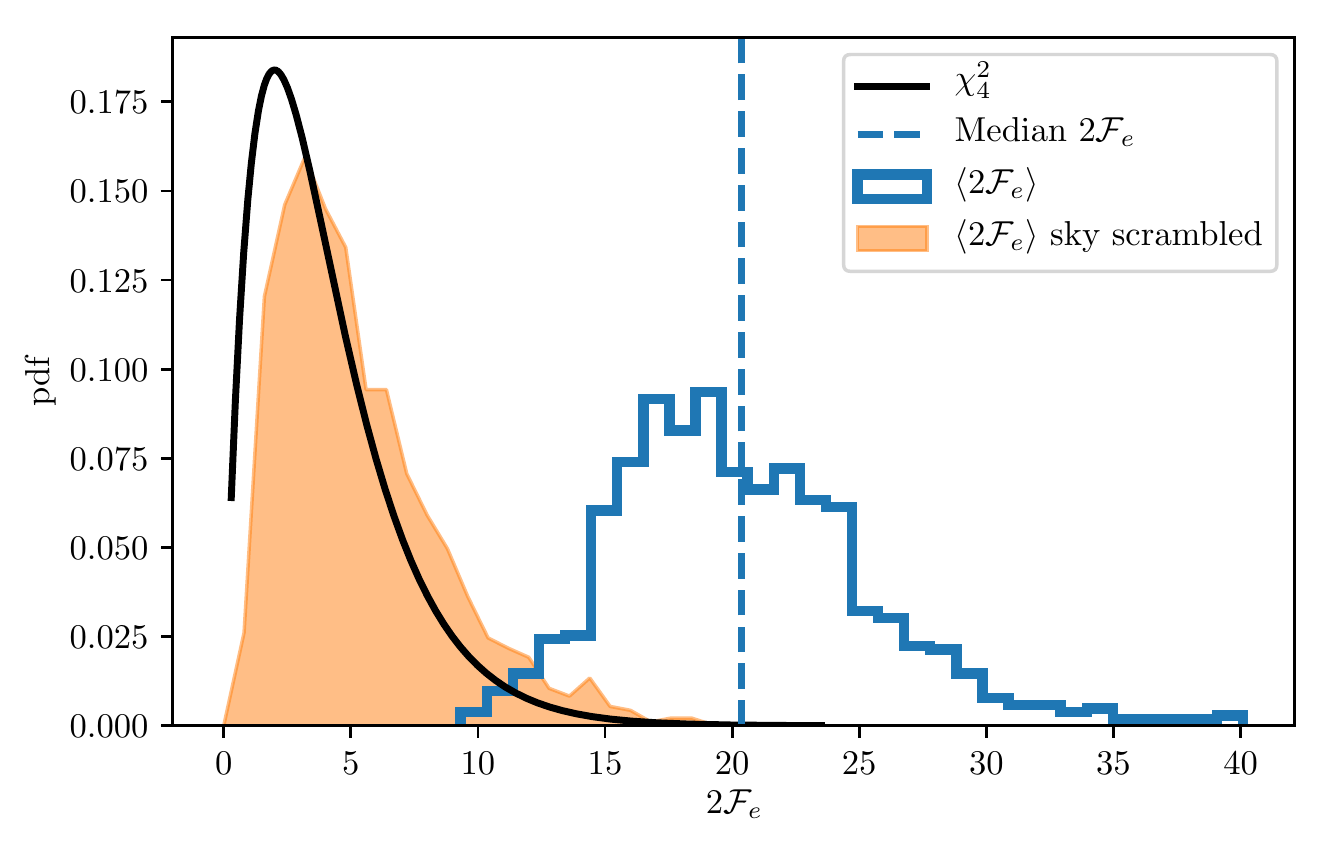}
\caption{$F_e$-statistic for a CGW search marginalised over the noise uncertainties (blue). The $F_e$ obtained from the analysis of the data with scrambled sky position of pulsars is shown in orange and it is compared with theoretical ($\chi^2_4$) distribution in black. 
}
\label{fig:Festat}
\end{figure}

Next, we move on to the analysis of the \texttt{DR2new} data set. We started by considering a simple model of a CGW source superimposed on PSRN. We found substantial support for the presence of a CGW: 
the BF of the model CGW+PSRN over the null model, PSRN, is 200 with the Earth term only and 260 if we include the pulsar term in the search. For this search we used the sampler and method described in \cite{Becsy:2022zbu}. 
The candidate source is localised in sky position and frequency. 
We have also computed the $F_e$ statistic 
\citep[a frequentist approach; see][]{Babak:2011mr, Ellis:2012zv}. The results are shown in Figure \ref{fig:Festat}. The $F_e$ statistic can be marginalised over the individual pulsar noise parameters in a similar manner as the OS. It is shown in blue with the mean value corresponding to S/N$\approx 4.5$. We computed the corresponding p-value ($\approx 0.1$\% ) by scrambling the pulsar sky positions, assuming 0.2 match as the orthogonality criterion. The $F_e$-statistic takes only the Earth-term into account and scrambling the pulsar's position destroys the coherence of the CGW, while preserving the noise properties. The $F_e$ of sky scrambles is shown in orange and is compared to the theoretically expected (for Gaussian noise) central $\chi^2$ distribution with four degrees of freedom.   

\begin{figure}
\centering
\includegraphics[width=0.8\linewidth]{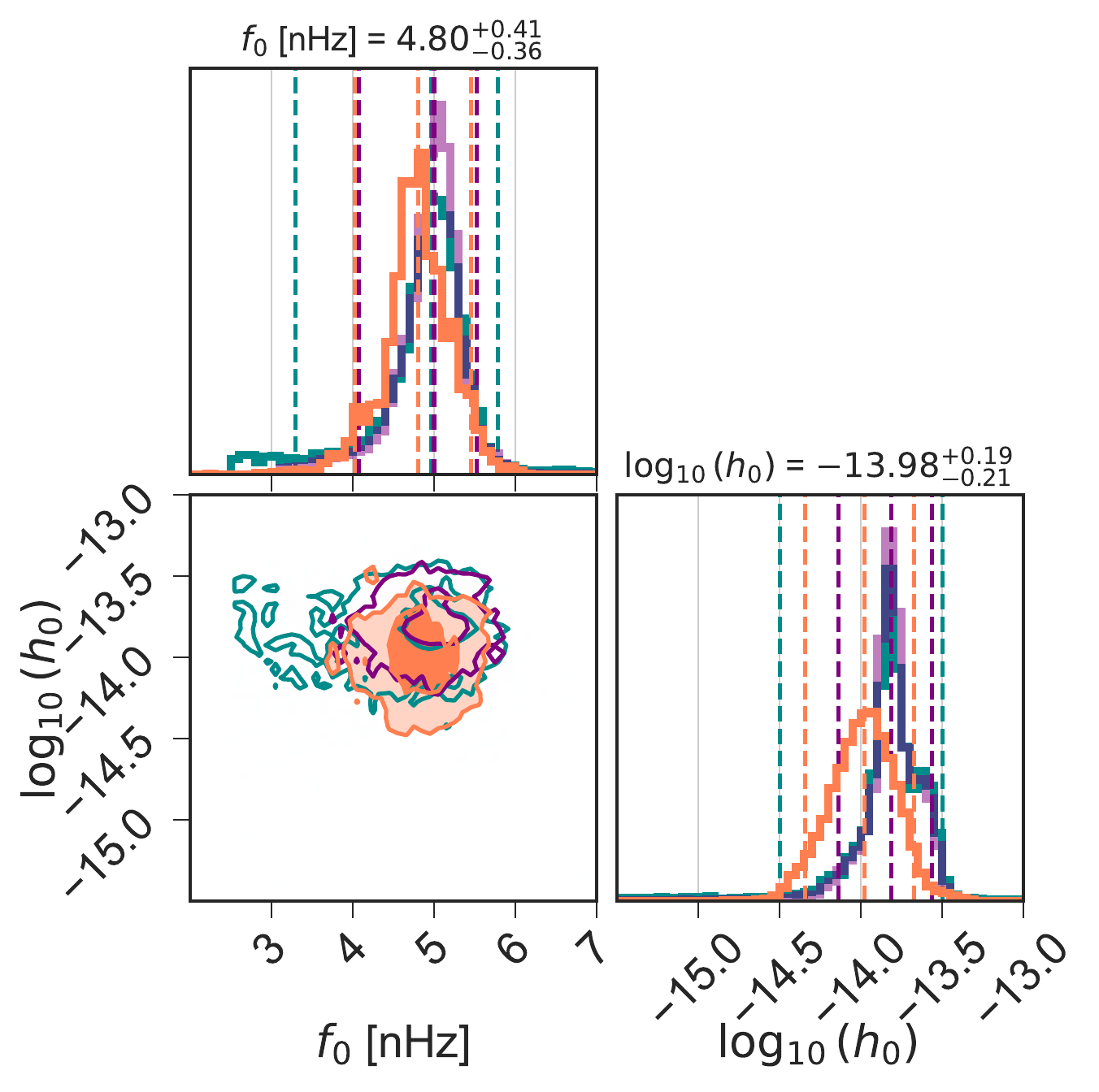}
\caption{Inference of the frequency and amplitude of a putative CGW in the CGW+CURN model. Plotted are the 50\% and 90\% credible regions. Dark-cyan contours are obtained using the Earth term only and \eprise{} with \ptmc{}, purple contours with \texttt{Eryn} sampler, while coral contours are produced using a sampler described in \cite{Becsy:2022zbu} and also include the pulsar term.}
\label{fig:CGW_CURN}
\end{figure}

Including the CURN in the model does not change these results significantly: we can still identify the CGW candidate with $BF^{\textrm{CGW+CURN}}_{\textrm{CURN}} = 7-20$, with the exact value depending on whether one includes the pulsar term ($BF=7$) or only the Earth term, as well as on the sampling method used ($BF= 12, 20$). Interestingly, the CURN parameters are much less constrained in the  CGW+CURN model. We show partial results of the Bayesian parameter inference in Figure \ref{fig:CGW_CURN}.  

Finally, we have considered a model containing a GWB plus a CGW and found that the GWB `absorbs' most of the CGW, that is, the CGW becomes poorly constrained. 
At the same time, our results indicate that we cannot exclude the presence of the CGW, since the BF is not informative ($BF^{\textrm{GWB}}_{\textrm{CGW}} = 1.2$), while the CGW model has a larger parameter space.
We note that the identified CGW candidate frequency of $\approx 5$~nHz is between the two lowest frequency bins that dominate the HD signal, as shown in Figure~\ref{fig:fs_pl}.
The rest of the bins do not contribute much to the GWB signal, due to the relatively short observation span of the \texttt{DR2new} data set and the high level of white noise. To summarise, we find that the observed data is equally well explained by either a GWB or a single CGW. However, given the additional number of parameters for the CGW model and in the absence of additional data we favour the simpler GWB model.
A detailed analysis including extensive simulations will be presented in a forthcoming publication.

\begin{figure*}[htp]
\centering
     \begin{subfigure}[b]{0.45\textwidth}
         \centering
         \includegraphics[width=\textwidth]{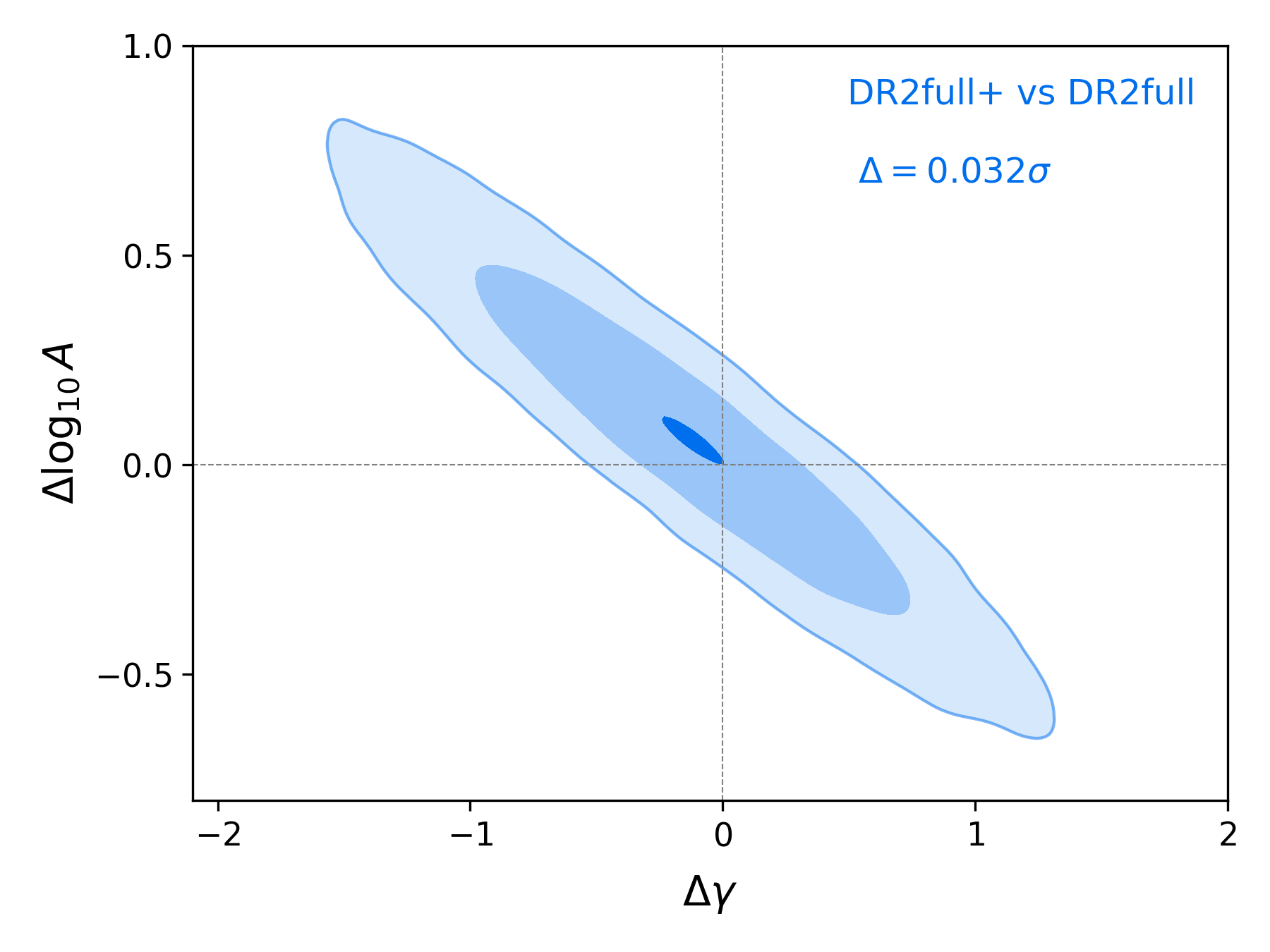}
         \end{subfigure}
         \hspace{0.05cm}
        \begin{subfigure}[b]{0.45\textwidth}
         \centering
         \includegraphics[width=\textwidth]{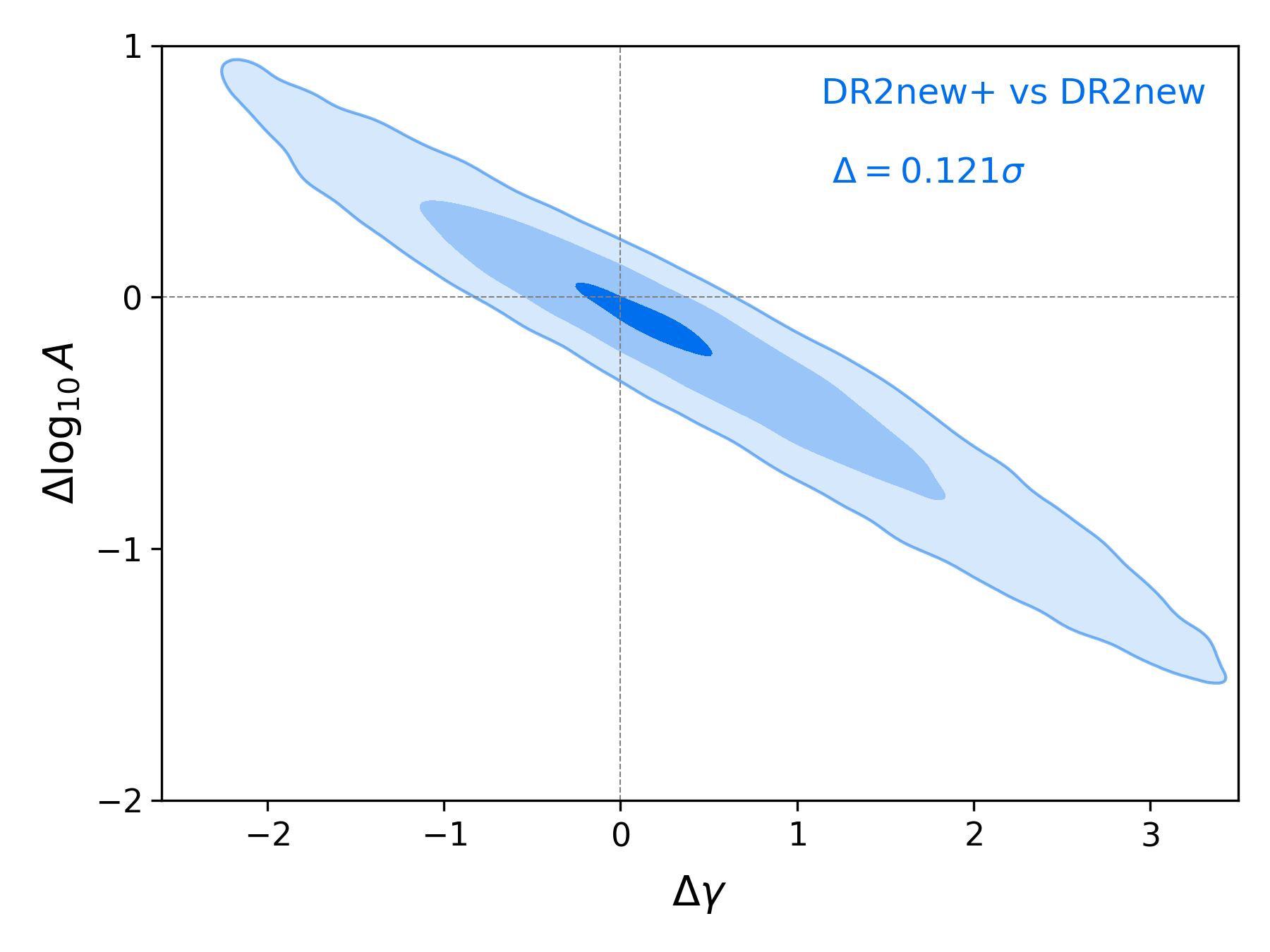}
     \end{subfigure}
        \caption{
        Difference distributions of posteriors while including the InPTA data.
        The left panel is associated with posteriors from
          \texttt{DR2full+} and \texttt{DR2full}. In contrast, the right panel involves the comparison between \texttt{DR2new+} and \texttt{DR2new}. Both panels show the tension for the GWB model.
        These plots provide three contours: $1\sigma$, $2\sigma$, and the $\Delta$ that represent the computed tension value.}
        \label{tension-inpta-plots}
\end{figure*}

\begin{table*}
\caption{90\% credible regions for the constraints of power law parameters in the different Bayesian analyses with DE440 for both \texttt{DR2full+} and \texttt{DR2new+}, which include the addition of InPTA data.}
\begin{center}
\def\arraystretch{1.5}
\begin{tabular*}{0.6\textwidth}{c||cc|cc}
\hline
 & \multicolumn{2}{c}{\texttt{DR2full+}} & \multicolumn{2}{c}{\texttt{DR2new+}} \\
Software + Model & $\log_{10} A_{\rm CRS}$ & $\gamma_{\rm CRS}$ & $\log_{10} A_{\rm CRS}$ & $\gamma_{\rm CRS}$ \\
\hline \hline
\eprise{} + CURN             & $-14.44_{-0.20}^{+0.17}$ & $3.98_{-0.37}^{+0.40}$ & $-14.30_{-0.52}^{+0.33}$ & $3.53_{-0.84}^{+1.12}$  \\
\hline
\eprise{} + GWB              & $-14.48_{-0.20}^{+0.18}$ & $4.06_{-0.40}^{+0.39}$ & $-14.10_{-0.44}^{+0.25}$ & $3.03_{-0.67}^{+1.02}$  \\
\hline
\end{tabular*}
\label{tab:pl_inpta}
\end{center}
\end{table*}

\begin{table}
\caption{
\texttt{Tensiometer} package based 
discrepancies between GWB posteriors that arise 
from \texttt{DR2full+}, \texttt{DR2full}, \texttt{DR2new+}, and \texttt{DR2new}. The entries provide the Z-score (in number of $\sigma$).
}
\def\arraystretch{1.5}
\centering
\begin{tabular}{c||c|c} 
 \hline
    & {\small \texttt{DR2full+} vs \texttt{DR2full}} & 
    {\small \texttt{DR2new+} vs \texttt{DR2new}} \\
 \hline \hline
 CURN   &    0.037 &   0.229	\\
 GWB    &	0.032 &   0.121 \\	 
 \hline
\end{tabular}
\label{tension-table-inpta}
\end{table}

\section{Search results on the EPTA$+$InPTA data sets \texttt{DR2full+} and \texttt{DR2new+}}
\label{sec:InPTA}

Analyses of \texttt{DR2full+} and \texttt{DR2new+} data sets including the InPTA data indicate general consistency of the results with the \texttt{DR2full} and \texttt{DR2new} data sets, respectively. We provide a comparison of posteriors for the CURN and GWB between the data sets, with and without InPTA data, in Tables~\ref{tab:pl_inpta} and \ref{tension-table-inpta}, as well as in Figure \ref{tension-inpta-plots}; see also Appendix \ref{App:B}. While the \texttt{DR2full+} and \texttt{DR2full} produce very consistent posteriors, a $\sim 0.2 \sigma$ difference can be found between \texttt{DR2new+} and \texttt{DR2new}. The impact of the InPTA data is more noticeable in the shorter data set.

The results for the OS shown in Table \ref{tab:os} are consistent with the EPTA-only data sets. Both \texttt{DR2full+} and \texttt{DR2full} give similar amplitudes and S/Ns for the three correlation models. An increase in the S/Ns of about $0.5$ for monopole, dipole and HD correlations can be found with the \texttt{DR2new+} against the S/Ns in the \texttt{DR2new}.

The corresponding BFs are in general agreement with the EPTA-only results (cf. Table~\ref{tab:bf_dr2}). The BF between GWB and CURN in the \texttt{DR2new+} of $65$ is comparable to the $60$ from \texttt{DR2new}. However, when testing for additional processes, we find significantly larger BFs for PSRN+GWB+CLK (row ID 8) and PSRN+GWB+EPH (row ID 9): $57$ versus $28$ and $43$ versus $33$, \texttt{DR2new+} versus \texttt{DR2new}, respectively. The small BF difference between the models including the CLK or EPH error terms and the GWB-only model further supports the assumption that the signal could be a GWB.

Finally, we investigate the effect of the additional InPTA on the contribution of the individual pulsars to the CURN via the dropout factor in Figure \ref{fig:cvsqc}. As with the previous results, most dropout factors are consistent between the \texttt{DR2full+} and \texttt{DR2full} data sets. J1600$-$3053 shows an increase in the dropout factor, possibly due to better single pulsar noise modelling with the addition of the InPTA data. For the new generation, EPTA-only and EPTA+InPTA data sets the differences are more pronounced. Most pulsars have dropout factors that are in agreement. Two pulsars, J1713$+$0747 and J0613$-$0200, have reduced dropout factors, whereas J1909$-$3744 shows an increase.

In summary, the results from the EPTA and InPTA combination are in broad agreement with the EPTA-only data set. The consistency between \texttt{DR2full+} and \texttt{DR2full} shows the robustness of the results from the full EPTA+InPTA combination. When comparing \texttt{DR2new+} and \texttt{DR2new} the effects of the InPTA data are more visible with differences in the power law posteriors, increased BFs and higher OS S/Ns for the GWB, but also for other possible noises. Further investigation is needed to assess and improve the combined sensitivity for GW searches.

\section{Discussions and conclusions}
\label{sec:conclusions}

The EPTA with its six telescopes and multiple observing systems has collected PTA observations for almost 25 years and using a new generation of observing systems for the last decade. For the DR2 we have increased the number of pulsars combined with the most recent observations from 6 to 25. A selection scheme has been applied to find the 25 pulsars that are sufficient to contribute about 95\% of the expected sensitivity of the full array with 42 pulsars from DR1 \citep{2023MNRAS.518.1802S}. Here, we used the optimal timing and noise models obtained in \cite{wm1,wm2} to search for a stochastic GWB. In addition, we combined data for ten common pulsars between InPTA DR1 \citep{2022PASA...39...53T} and EPTA DR2 and conducted GWB searches also on those extended EPTA$+$InPTA data sets. 

We present the main results of the GWB search using two versions of the EPTA 25-pulsar data set, the full data set (\texttt{DR2full}) with a time span of 24.7\,yr and a new backends-only data set (\texttt{DR2new}) with a time span of 10.3\, yr. Both data sets measure a common red signal. By virtue of its length, the full data set yields a better constraint on the spectral properties of the GWB. However, the new backends-only data set provides a better measurement of the cross-correlated power. In the following, we give a summary of the results of our analysis and discuss possible reasons for the discrepancies between the two data sets.

The power law amplitude for an HD-correlated process using \texttt{DR2full} is $\log_{10} A = -14.54_{-0.41}^{+0.28}$, with a spectral index $\gamma = 4.19_{-0.63}^{+0.73}$ that is close to the expected value of $13/3$ for a GWB from circular SMBHBs driven by GW emission. With \texttt{DR2new} we obtain a flatter spectrum with $\log_{10} A = -13.94_{-0.48}^{+0.23}$ and $\gamma = 2.71_{-0.73}^{+1.18}$.
Fixing the spectral index to $13/3$, the amplitudes for the two data sets are consistent, with values around $\log_{10}A=-14.61_{-0.15}^{+0.11}$. This indicates that the two data sets constrain the same amount of power in the GWB, although the detailed spectral shape appears to be different. 

The free spectrum analysis provides the possibility to directly compare the results from different data sets in the frequency domain. Ignoring the lowest 1/24.7\,yr$^{-1}$ frequency bin of \texttt{DR2full}, the remaining bins show good consistency with those of \texttt{DR2new}. Including the lowest frequency bin in the power law fitting may lead to a steepening of the power law. With a shorter data span \texttt{DR2new} probes a different frequency range starting at about two times higher than 1/24.7\,yr$^{-1}$. Power law fitting could also be affected by the frequency bins just above $10^{-8}\,$Hz, resulting in a flatter spectrum for \texttt{DR2new}. This could either indicate that a power law model does not provide a good fit to the common signal, or there is additional noise or signal around $10^{-8}\,$Hz or $10^{-9}\,$Hz.

The differences between the Bayesian correlation curves observed in the two data sets are most obvious around angular separations of 60\,deg, 80\,deg and 135\,deg. The correlation curve produced by \texttt{DR2full} shows a prominent monopolar-like structure, with a central part shifted upward compared to the HD curve, while the correlation curve produced by \texttt{DR2new} follows the HD correlation much more closely. These results are in agreement with the measured BFs (PSRN+GWB vs PSRN+CURN), which are four and 60 for \texttt{DR2full} and \texttt{DR2new}, respectively. From the null hypothesis distributions of BFs (PSRN+GWB vs PSRN+CURN) constructed with phase shifts, we can infer a p-value of 0.04 for \texttt{DR2full} and 0.001 for \texttt{DR2new}. There is essentially no evidence for other correlation patterns or additional common processes in either data set, with perhaps the exception of a tentative hint of an extra monopole in the \texttt{DR2full} data set, which will be the subject of further studies.

Optimal statistic analysis shows even more significant discrepancies between the two data sets. For \texttt{DR2full} the squared amplitude is only $2.7\times10^{-30}$, which is lower than the amplitude from the Bayesian analysis ($A^2=6.02\times 10^{-30}$). There is a large scatter in the correlation coefficients, giving a similar S/N of around $1.2$ for both the HD and the monopole correlations. The p-value for the S/N of the HD correlation is $0.07$ from phase shifts and $0.08$ from sky scrambles. For \texttt{DR2new} the squared amplitude is $1.0\times10^{-29}$, which is more consistent with (yet higher than) the Bayesian result. The correlation coefficients also match the HD curve much better. The S/N for the HD correlation increases to $3.5$, while the S/N for the monopole correlation drops to almost zero. Sky scrambles give a p-value of $0.002$ for HD correlation, while phase shifts yield a p-value $<0.0001$. This corresponds to $>3\sigma$ significance of the GWB signal.

Preliminary analysis including a CGW suggests that its contribution to the observed HD-correlated power cannot be ruled out. The presence of a CGW is not supported in the \texttt{DR2full} data set; its presence is preferred over CURN only in \texttt{DR2new}. The source amplitude and frequency are well-constrained. The candidate is also localised in the sky, but its position and error region depends on whether we include the pulsar term.   However, adding a GWB to the analysis absorbs most of the power of the CGW, preventing any strong claim about its actual presence in the data. A more thorough analysis involving the CGW model will be presented in a separate paper.

The analysis of the combined EPTA DR2 and InPTA DR1 shows broadly consistent results with the EPTA DR2 alone. The power law parameter constraints with \texttt{DR2full+} show little difference to those without the InPTA data. For \texttt{DR2new+}, the effect of the additional InPTA data is more pronounced. The power law parameters experience a small shift of 0.17$\sigma$ towards a steeper spectral index. The BFs and OS S/Ns are also in general agreement with the EPTA-only data sets. Increases in the evidence for additional monopole and dipole correlated signals of about 0.5 can be found in \texttt{DR2new+}. A larger impact on the shorter data set can be expected, since the InPTA, with three years of time span, is a more significant fraction of \texttt{DR2new} (10.3 years) compared to \texttt{DR2full} (24.7 years).

With the high amplitude and large uncertainty in the spectral index, the observed HD correlated signal is broadly consistent with the expectation from a cosmic population of SMBHBs. In particular, as shown by \cite{msc+2021} the high amplitude of ${\rm log}A=-14.61$ inferred when fixing $\gamma=13/3$ is consistent with the recent discovery of over-massive black holes \citep[e.g.][]{2011Natur.480..215M} and the upward revision of the normalisation of the SMBH-host galaxy relations \citep[see, e.g.][]{2013ApJ...764..184M,2013ARA&A..51..511K}. It is not straightforward, however, to construct a self-consistent SMBH and host galaxy cosmic evolution model that results in such a high GWB signal fulfilling other observational constraints on the SMBH mass function and on the evolution of the bolometric quasar luminosity function with redshift \citep{2022MNRAS.509.3488I}. A spectrum significantly flatter than $\gamma=13/3$ can arise for a number of different reasons, including strong coupling with the environment, the predominance of highly eccentric SMBHBs \citep[see e.g.][]{2013CQGra..30v4014S} or simply by the presence of extra power at high frequencies due to sparse and loud marginally resolvable individual binaries \citep{msc+2021}. Besides a cosmic population of SMBHBs, the detected signal can be generated by processes occurring in the early Universe \citep{2018CQGra..35p3001C} as well as specific models of dark matter \citep{2018PhRvD..98j2002P}. We plan to investigate the implications of this signal for all these formation scenarios in follow-up papers.

Our results seem to indicate that \texttt{DR2new} provides better constraints on the cross-correlated power than \texttt{DR2full}. 
It would normally be expected that the addition of more data would lead to a more significant detection of a stationary process. 
There are a few possible factors that could be contributing to this discrepancy that needs to be investigated in more detail.

\begin{enumerate}

    \item Lower quality of the early data, which lacks multi-radio frequency coverage and polarisation calibration, may have allowed for residual unmodelled noise. This can lead to different noises and signals being recovered with the early and new generation observations. Investigating better noise modelling can help to increase the sensitivity and reliability of the early data.

    \item Improper weights for the power law fitting in different frequency bins could introduce bias in the recovery of the spectral properties. In particular, the lowest frequency bins only have contributions from a few pulsars with the longest time spans, but their weights are the highest since the largest amount of power in a common red signal is at the lowest frequencies. Considering a weighting scheme for the different frequency components for the power law fitting could produce an unbiased result.

    \item The presence of excess power at low frequencies can lead to a steepening of the power law. While excess noise at high frequencies can make the spectrum appear shallower. In both cases, noise leaks into the GW signal giving erroneous power law constraints.

    \item Non-stationarity of the pulsar noise or of the putative GW signal can cause the measured spectral properties to be different between the early and late part of a data set as the properties of the noise and signal could evolve over time.
    
\end{enumerate}

Some of these differences between pulsars are smoothed out in the \texttt{DR2new} data set, as all pulsars have roughly the same time span $\sim 10\,$yr. This may help to measure the cross-correlations between pulsar pairs more robustly.

An extensive simulation campaign is ongoing to help to better understand the features of our data sets and to build more confidence in the internal consistency of our findings. Verification of the analysis algorithms and their performance on a realistic PTA data set is needed to set our expectations. These can then be compared against the real data set to test the effects that data quality and the noise/signal properties have on the results of the final analysis. Several simulation projects tackling different questions will be published in separate works.

Concurrent efforts from the NANOGrav collaboration on their 15-year data set (Arzoumanian et al., 2023) and the PPTA on their DR3 (Reardon et al., 2023) provide independent results on the search for a gravitational wave background. These will be compared in the IPTA framework to increase our confidence and prepare for the next IPTA data set. Additionally, the CPTA is preparing its first data release and analysis for a GWB signal (Xu et al., 2023).

Moving forward, we plan to add more pulsars timed with the new backends to \texttt{DR2new}. The EPTA data set is unique in its combination of time span, cadence, number of pulsars and, when combined with the InPTA data set, in DM monitoring. Indeed,
among the pulsars timed by EPTA, there are more than 30 sources observed with new backends for a time span $>8$ years displaying RMS$<2\,\mu$s, which can add significant value to the EPTA data set. A combination of the resulting data set with NANOGrav15, PPTADR3 and MeerKAT DR1 under the aegis of the third data release (DR3) of the IPTA, will produce a data set of unprecedented sensitivity that will help to pin down the nature of the signal presented here and to constrain its properties.

\begin{acknowledgements}
The European Pulsar Timing Array (EPTA) is a collaboration between
European and partner institutes, namely ASTRON (NL), INAF/Osservatorio
di Cagliari (IT), Max-Planck-Institut f\"{u}r Radioastronomie (GER),
Nan\c{c}ay/Paris Observatory (FRA), the University of Manchester (UK),
the University of Birmingham (UK), the University of East Anglia (UK),
the University of Bielefeld (GER), the University of Paris (FRA), the
University of Milan-Bicocca (IT), the Foundation for Research and 
Technology, Hellas (GR), and Peking University (CHN), with the
aim to provide high-precision pulsar timing to work towards the direct
detection of low-frequency gravitational waves. An Advanced Grant of
the European Research Council allowed to implement the Large European Array
for Pulsars (LEAP) under Grant Agreement Number 227947 (PI M. Kramer). 
The EPTA is part of the
International Pulsar Timing Array (IPTA); we thank our
IPTA colleagues for their support and help with this paper and the external Detection Committee members for their work on the Detection Checklist.

Part of this work is based on observations with the 100-m telescope of
the Max-Planck-Institut f\"{u}r Radioastronomie (MPIfR) at Effelsberg
in Germany. Pulsar research at the Jodrell Bank Centre for
Astrophysics and the observations using the Lovell Telescope are
supported by a Consolidated Grant (ST/T000414/1) from the UK's Science
and Technology Facilities Council (STFC). ICN is also supported by the
STFC doctoral training grant ST/T506291/1. The Nan{\c c}ay radio
Observatory is operated by the Paris Observatory, associated with the
French Centre National de la Recherche Scientifique (CNRS), and
partially supported by the Region Centre in France. We acknowledge
financial support from ``Programme National de Cosmologie and
Galaxies'' (PNCG), and ``Programme National Hautes Energies'' (PNHE)
funded by CNRS/INSU-IN2P3-INP, CEA and CNES, France. We acknowledge
financial support from Agence Nationale de la Recherche
(ANR-18-CE31-0015), France. The Westerbork Synthesis Radio Telescope
is operated by the Netherlands Institute for Radio Astronomy (ASTRON)
with support from the Netherlands Foundation for Scientific Research
(NWO). The Sardinia Radio Telescope (SRT) is funded by the Department
of University and Research (MIUR), the Italian Space Agency (ASI), and
the Autonomous Region of Sardinia (RAS) and is operated as a National
Facility by the National Institute for Astrophysics (INAF).

The work is supported by the National SKA programme of China
(2020SKA0120100), Max-Planck Partner Group, NSFC 11690024, CAS
Cultivation Project for FAST Scientific. This work is also supported
as part of the ``LEGACY'' MPG-CAS collaboration on low-frequency
gravitational wave astronomy. JA acknowledges support from the
European Commission (Grant Agreement number: 101094354). JA and SCha 
were partially supported by the Stavros
Niarchos Foundation (SNF) and the Hellenic Foundation for Research and
Innovation (H.F.R.I.) under the 2nd Call of the ``Science and Society --
Action Always strive for excellence -- Theodoros Papazoglou''
(Project Number: 01431). AC acknowledges support from the Paris
\^{I}le-de-France Region. AC, AF, ASe, ASa, EB, DI, GMS, MBo acknowledge
financial support provided under the European Union's H2020 ERC
Consolidator Grant ``Binary Massive Black Hole Astrophysics'' (B
Massive, Grant Agreement: 818691). GD, KLi, RK and MK acknowledge support
from European Research Council (ERC) Synergy Grant ``BlackHoleCam'', 
Grant Agreement Number 610058. This work is supported by the ERC 
Advanced Grant ``LEAP'', Grant Agreement Number 227947 (PI M. Kramer). 
AV and PRB are supported by the UK's Science
and Technology Facilities Council (STFC; grant ST/W000946/1). AV also acknowledges
the support of the Royal Society and Wolfson Foundation. JPWV acknowledges
support by the Deutsche Forschungsgemeinschaft (DFG) through thew
Heisenberg programme (Project No. 433075039) and by the NSF through
AccelNet award \#2114721. NKP is funded by the Deutsche
Forschungsgemeinschaft (DFG, German Research Foundation) --
Projektnummer PO 2758/1--1, through the Walter--Benjamin
programme. ASa thanks the Alexander von Humboldt foundation in
Germany for a Humboldt fellowship for postdoctoral researchers. APo, DP
and MBu acknowledge support from the research grant “iPeska”
(P.I. Andrea Possenti) funded under the INAF national call
Prin-SKA/CTA approved with the Presidential Decree 70/2016
(Italy). RNC acknowledges financial support from the Special Account
for Research Funds of the Hellenic Open University (ELKE-HOU) under
the research programme ``GRAVPUL'' (grant agreement 319/10-10-2022).
EvdW, CGB and GHJ acknowledge support from the Dutch National Science
Agenda, NWA Startimpuls – 400.17.608.
BG is supported by the Italian Ministry of Education, University and 
Research within the PRIN 2017 Research Program Framework, n. 2017SYRTCN. LS acknowledges the use of the HPC system Cobra at the Max Planck Computing and Data Facility.

\ifnum\wm>1 The Indian Pulsar Timing Array (InPTA) is an Indo-Japanese
collaboration that routinely employs TIFR's upgraded Giant Metrewave
Radio Telescope for monitoring a set of IPTA pulsars.  BCJ, YG, YM,
SD, AG and PR acknowledge the support of the Department of Atomic
Energy, Government of India, under Project Identification \# RTI 4002.
BCJ, YG and YM acknowledge support of the Department of Atomic Energy,
Government of India, under project No. 12-R\&D-TFR-5.02-0700 while SD,
AG and PR acknowledge support of the Department of Atomic Energy,
Government of India, under project no. 12-R\&D-TFR-5.02-0200.  KT is
partially supported by JSPS KAKENHI Grant Numbers 20H00180, 21H01130,
and 21H04467, Bilateral Joint Research Projects of JSPS, and the ISM
Cooperative Research Program (2021-ISMCRP-2017). AS is supported by
the NANOGrav NSF Physics Frontiers Center (awards \#1430284 and
2020265).  AKP is supported by CSIR fellowship Grant number
09/0079(15784)/2022-EMR-I.  SH is supported by JSPS KAKENHI Grant
Number 20J20509.  KN is supported by the Birla Institute of Technology
\& Science Institute fellowship.  AmS is supported by CSIR fellowship
Grant number 09/1001(12656)/2021-EMR-I and T-641 (DST-ICPS).  TK is
partially supported by the JSPS Overseas Challenge Program for Young
Researchers.  We acknowledge the National Supercomputing Mission (NSM)
for providing computing resources of ‘PARAM Ganga’ at the Indian
Institute of Technology Roorkee as well as `PARAM Seva' at IIT
Hyderabad, which is implemented by C-DAC and supported by the Ministry
of Electronics and Information Technology (MeitY) and Department of
Science and Technology (DST), Government of India. DD acknowledges the 
support from the Department of Atomic Energy, Government of India 
through Apex Project - Advance Research and Education in Mathematical 
Sciences at IMSc. \fi

The work presented here is a culmination of many years of data
analysis as well as software and instrument development. In particular,
we thank Drs. N.~D'Amico, P.~C.~C.~Freire, R.~van Haasteren, 
C.~Jordan, K.~Lazaridis, P.~Lazarus, L.~Lentati, O.~L\"{o}hmer and 
R.~Smits for their past contributions. We also
thank Dr. N. Wex for supporting the calculations of the
galactic acceleration as well as the related discussions.
The EPTA is also grateful
to staff at its observatories and telescopes who have made the
continued observations possible. 
\linebreak\linebreak\textit{Author contributions.}
The EPTA is a multi-decade effort and all authors have
contributed through conceptualisation, funding acquisition,
data-curation, methodology, software and hardware
 developments as well as (aspects of) the continued running of
the observational campaigns, which includes writing and
proofreading observing proposals, evaluating observations
and observing systems, mentoring students, developing
science cases. All authors also helped in (aspects of)
verification of the data, analysis and results as well as
in finalising the paper draft. Specific contributions from individual 
EPTA members are listed in the CRediT\footnote{\url{https://credit.niso.org/}} format below.

InPTA members contributed in uGMRT observations and data reduction to
create the InPTA data set which is employed while assembling the
\texttt{DR2full+} and \texttt{DR2new+} data sets. 

\ifnum\wm=1

JJan, KLi, GMS equally share the correspondence of the paper.

\linebreak\linebreak\textit{CRediT statement:}\newline
Conceptualisation: APa, APo, AV, BWS, CGB, CT, GHJ, GMS, GT, IC, JA, JJan, JPWV, JW, JWM, KJL, KLi, MK.\\
Methodology: APa, AV, DJC, GMS, IC, JA, JJan, JPWV, JWM, KJL, KLi, LG, MK.\\
Software: AC, AJ, APa, CGB, DJC, GMS, IC, JA, JJan, JJaw, JPWV, KJL, KLi, LG, MJK, RK.\\
Validation: AC, APa, CGB, CT, GMS, GT, IC, JA, JJan, JPWV, JWM, KLi, LG.\\
Formal Analysis: APa, CGB, DJC, DP, EvdW, GHJ, GMS, JA, JJan, JPWV, JWM, KLi.\\
Investigation: APa, APo, BWS, CGB, DJC, DP, GMS, GT, IC, JA, JJan, JPWV, JWM, KLi, LG, MBM, MBu, MJK, RK.\\
Resources: APa, APe, APo, BWS, GHJ, GMS, GT, HH, IC, JA, JJan, JPWV, JWM, KJL, KLi, LG, MJK, MK, RK.\\
Data Curation: AC, AJ, APa, BWS, CGB, DJC, DP, EG, EvdW, GHJ, GMS, GT, HH, IC, JA, JJan, JPWV, JWM, KLi, LG, MBM, MBu, MJK, MK, NKP, RK, SChe, YJG.\\
Writing – Original Draft: APa, GMS, JA, JJan, KLi, LG.\\
Writing – Review \& Editing: AC, AF, APa, APo, DJC, EB, EFK, GHJ, GMS, GT, JA, JJan, JPWV, JWM, KLi, MK, SChe, VVK.\\
Visualisation: APa, GMS, JA, JJan, KLi.\\
Supervision: APo, ASe, AV, BWS, CGB, DJC, EFK, GHJ, GMS, GT, IC, JA, JPWV, KJL, KLi, LG, MJK, MK, VVK.\\
Project Administration: APo, ASe, AV, BWS, CGB, CT, GHJ, GMS, GT, IC, JJan, JPWV, JWM, KLi, LG, MK.\\
Funding Acquisition: APe, APo, ASe, BWS, GHJ, GT, IC, JA, JJan, LG, MJK, MK.\\

\fi

\ifnum\wm=2
InPTA members contributed to the discussions that probed the impact of 
including InPTA data on single pulsar noise analysis. Furthermore, they 
provided quantitative comparisons of various noise models, wrote a brief 
description of the underlying \texttt{Tensiometer} package, and helped 
with the related interpretations.

APa, AC, MJK equally share the correspondence of the paper. 

\linebreak\linebreak\textit{CRediT statement:}\newline
Conceptualisation: AC, APa, APo, AV, BWS, CT, GMS, GT, JPWV, JWM, KJL, KLi, MJK, MK.\\
Methodology: AC, APa, AV, DJC, GMS, IC, JWM, KJL, KLi, LG, MJK, MK, SB, SChe, VVK.\\
Software: AC, AJ, APa, APe, GD, GMS, KJL, KLi, MJK, RK, SChe, VVK.\\
Validation: AC, APa, BG, GMS, IC, JPWV, JWM, KLi, LG, MJK.\\
Formal Analysis: AC, APa, BG, EvdW, GHJ, GMS, JWM, KLi, MJK.\\
Investigation: AC, APa, APo, BWS, CGB, DJC, DP, GMS, IC, JPWV, JWM, KLi, LG, MBM, MBu, MJK, RK, VVK.\\
Resources: AC, APa, APe, APo, BWS, GHJ, GMS, GT, IC, JPWV, JWM, KJL, KLi, LG, MJK, MK, RK.\\
Data Curation: AC, AJ, APa, BWS, CGB, DJC, DP, EvdW, GHJ, GMS, JA, JWM, KLi, MBM, MJK, MK, NKP, RK, SChe.\\
Writing – Original Draft: AC, APa, GMS, MJK.\\
Writing – Review \& Editing: AC, AF, APa, APo, BG, EB, EFK, GMS, GT, JA, JPWV, JWM, KLi, MJK, MK, SChe, VVK.\\
Visualisation: AC, APa, GMS, KLi, MJK.\\
Supervision: AC, APo, ASe, AV, BWS, CGB, DJC, EFK, GHJ, GT, JPWV, KJL, LG, MJK, MK, VVK.\\
Project Administration: AC, APo, ASe, AV, BWS, CGB, CT, GHJ, GMS, GT, JPWV, JWM, LG, MJK, MK.\\
Funding Acquisition: APe, APo, ASe, BWS, GHJ, GT, IC, LG, MJK, MK.\\
\fi

\ifnum\wm=3

Additionally, InPTA members contributed to GWB search efforts with 
\texttt{DR2full+} and \texttt{DR2new+} data sets and their interpretations. 
Further, they provided quantitative comparisons of GWB posteriors that 
arise from these data sets and multiple pipelines.

For this work specifically, SChen and YJG equally share the 
correspondence of the paper. 

\linebreak\linebreak\textit{CRediT statement:}\newline
Conceptualisation: AC, APa, APe, APo, ASe, AV, BG, CT, GMS, GT, IC, JA, JPWV, JWM, KJL, KLi, MK.\\
Methodology: AC, APa, ASe, AV, DJC, GMS, JWM, KJL, KLi, LS, MK, SChe.\\
Software: AC, AJ, APa, APe, GD, GMS, KJL, KLi, MJK, RK, SChe, VVK.\\
Validation: AC, APa, ASe, AV, BG, GMS, HQL, JPWV, JWM, KLi, LS, SChe, YJG.\\
Formal Analysis: AC, APa, ASe, AV, BG, EvdW, GMS, HQL, JWM, KLi, LS, MF, NKP, PRB, SChe, YJG.\\
Investigation: APa, APo, ASe, AV, BWS, CGB, DJC, DP, GMS, JWM, KLi, LS, MBM, MBu, MF, PRB, RK, SB, SChe, YJG.\\
Resources: AC, APa, APe, APo, ASe, AV, BWS, GHJ, GMS, GT, IC, JPWV, JWM, KJL, KLi, LG, LS, MJK, MK, RK.\\
Data Curation: AC, AJ, APa, BWS, CGB, DJC, DP, EvdW, GMS, JA, JWM, KLi, MBM, MJK, MK, RK, SChe.\\
Writing – Original Draft: AC, APa, BG, DJC, GMS, JA, KLi, SB, SChe, YJG.\\
Writing – Review \& Editing: AC, AF, APa, APo, ASe, AV, BG, DJC, EB, EFK, GMS, GT, JA, JPWV, JWM, KLi, LS, MBo, MK, NKP, PRB, SChe, VVK, YJG.\\
Visualisation: APa, BG, GMS, KLi, MF, PRB, SChe.\\
Supervision: APo, ASe, AV, BWS, CGB, DJC, EFK, GHJ, GMS, GT, JPWV, KJL, MK, SB.\\
Project Administration: APo, ASe, AV, BWS, CGB, CT, GHJ, GMS, GT, JPWV, JWM, LG, MK, SChe.\\
Funding Acquisition: APe, APo, ASe, AV, BWS, GHJ, GT, IC, JA, LG, MJK, MK, SB.\\
\fi

\end{acknowledgements}

%
%

\include{journals}
\bibliographystyle{aa}
\bibliography{psrrefs,references}

\begin{thebibliography}{89}
\expandafter\ifx\csname natexlab\endcsname\relax\def\natexlab#1{#1}\fi

\bibitem[{{Abbott} {et~al.}(2016){Abbott}, {Abbott}, {Abbott}, {Abernathy},
  {Acernese}, {Ackley}, {Adams}, {Adams}, {Addesso}, {Adhikari}, \&
  et~al.}]{abb2016}
{Abbott}, B.~P., {Abbott}, R., {Abbott}, T.~D., {et~al.} 2016, Physical Review
  Letters, 116, 061102

\bibitem[{{Allen} \& {Romano}(2022)}]{2022arXiv220807230A}
{Allen}, B. \& {Romano}, J.~D. 2022, arXiv e-prints, arXiv:2208.07230

\bibitem[{Anholm {et~al.}(2009)Anholm, Ballmer, Creighton, Price, \&
  Siemens}]{abc+2009}
Anholm, M., Ballmer, S., Creighton, J. D.~E., Price, L.~R., \& Siemens, X.
  2009, Phys. Rev. D, 79, 084030

\bibitem[{{Antoniadis} {et~al.}(2022){Antoniadis}, {Arzoumanian}, {Babak},
  {Bailes}, {Bak Nielsen}, {Baker}, {Bassa}, {B{\'e}csy}, {Berthereau},
  {Bonetti}, {Brazier}, {Brook}, {Burgay}, {Burke-Spolaor}, {Caballero},
  {Casey-Clyde}, {Chalumeau}, {Champion}, {Charisi}, {Chatterjee}, {Chen},
  {Cognard}, {Cordes}, {Cornish}, {Crawford}, {Cromartie}, {Crowter}, {Dai},
  {DeCesar}, {Demorest}, {Desvignes}, {Dolch}, {Drachler}, {Falxa}, {Ferrara},
  {Fiore}, {Fonseca}, {Gair}, {Garver-Daniels}, {Goncharov}, {Good}, {Graikou},
  {Guillemot}, {Guo}, {Hazboun}, {Hobbs}, {Hu}, {Islo}, {Janssen}, {Jennings},
  {Johnson}, {Jones}, {Kaiser}, {Kaplan}, {Karuppusamy}, {Keith}, {Kelley},
  {Kerr}, {Key}, {Kramer}, {Lam}, {Lamb}, {Lazio}, {Lee}, {Lentati}, {Liu},
  {Luo}, {Lynch}, {Lyne}, {Madison}, {Main}, {Manchester}, {McEwen}, {McKee},
  {McLaughlin}, {Mickaliger}, {Mingarelli}, {Ng}, {Nice}, {Os{\l}owski},
  {Parthasarathy}, {Pennucci}, {Perera}, {Perrodin}, {Petiteau}, {Pol},
  {Porayko}, {Possenti}, {Ransom}, {Ray}, {Reardon}, {Russell}, {Samajdar},
  {Sampson}, {Sanidas}, {Sarkissian}, {Schmitz}, {Schult}, {Sesana},
  {Shaifullah}, {Shannon}, {Shapiro-Albert}, {Siemens}, {Simon}, {Smith},
  {Speri}, {Spiewak}, {Stairs}, {Stappers}, {Stinebring}, {Swiggum}, {Taylor},
  {Theureau}, {Tiburzi}, {Vallisneri}, {van der Wateren}, {Vecchio},
  {Verbiest}, {Vigeland}, {Wahl}, {Wang}, {Wang}, {Wang}, {Witt}, {Zhang}, \&
  {Zhu}}]{ipta+22}
{Antoniadis}, J., {Arzoumanian}, Z., {Babak}, S., {et~al.} 2022, \mnras, 510,
  4873

\bibitem[{{Arzoumanian} {et~al.}(2020){Arzoumanian}, {Baker}, {Blumer},
  {B{\'e}csy}, {Brazier}, {Brook}, {Burke-Spolaor}, {Chatterjee}, {Chen},
  {Cordes}, {Cornish}, {Crawford}, {Cromartie}, {Decesar}, {Demorest}, {Dolch},
  {Ellis}, {Ferrara}, {Fiore}, {Fonseca}, {Garver-Daniels}, {Gentile}, {Good},
  {Hazboun}, {Holgado}, {Islo}, {Jennings}, {Jones}, {Kaiser}, {Kaplan},
  {Kelley}, {Key}, {Laal}, {Lam}, {Lazio}, {Lorimer}, {Luo}, {Lynch},
  {Madison}, {McLaughlin}, {Mingarelli}, {Ng}, {Nice}, {Pennucci}, {Pol},
  {Ransom}, {Ray}, {Shapiro-Albert}, {Siemens}, {Simon}, {Spiewak}, {Stairs},
  {Stinebring}, {Stovall}, {Sun}, {Swiggum}, {Taylor}, {Turner}, {Vallisneri},
  {Vigeland}, {Witt}, \& {Nanograv Collaboration}}]{abb+2020}
{Arzoumanian}, Z., {Baker}, P.~T., {Blumer}, H., {et~al.} 2020, \apjl, 905, L34

\bibitem[{{Arzoumanian} {et~al.}(2018){Arzoumanian}, {Baker}, {Brazier},
  {Burke-Spolaor}, {Chamberlin}, {Chatterjee}, {Christy}, {Cordes}, {Cornish},
  {Crawford}, {Thankful Cromartie}, {Crowter}, {DeCesar}, {Demorest}, {Dolch},
  {Ellis}, {Ferdman}, {Ferrara}, {Folkner}, {Fonseca}, {Garver-Daniels},
  {Gentile}, {Haas}, {Hazboun}, {Huerta}, {Islo}, {Jenet}, {Jones}, {Jones},
  {Kaplan}, {Kaspi}, {Lam}, {Lazio}, {Levin}, {Lommen}, {Lorimer}, {Luo},
  {Lynch}, {Madison}, {McLaughlin}, {McWilliams}, {Mingarelli}, {Ng}, {Nice},
  {Park}, {Pennucci}, {Pol}, {Ransom}, {Ray}, {Rasskazov}, {Siemens}, {Simon},
  {Spiewak}, {Stairs}, {Stinebring}, {Stovall}, {Swiggum}, {Taylor},
  {Vallisneri}, {Vigeland}, \& {Zhu}}]{abb+2018}
{Arzoumanian}, Z., {Baker}, P.~T., {Brazier}, A., {et~al.} 2018, ArXiv e-prints
  [\eprint[arXiv]{1801.02617}]

\bibitem[{{Arzoumanian} {et~al.}(2015){Arzoumanian}, {Brazier},
  {Burke-Spolaor}, {Chamberlin}, {Chatterjee}, {Christy}, {Cordes}, {Cornish},
  {Crowter}, {Demorest}, {Dolch}, {Ellis}, {Ferdman}, {Fonseca},
  {Garver-Daniels}, {Gonzalez}, {Jenet}, {Jones}, {Jones}, {Kaspi}, {Koop},
  {Lam}, {Lazio}, {Levin}, {Lommen}, {Lorimer}, {Luo}, {Lynch}, {Madison},
  {McLaughlin}, {McWilliams}, {Nice}, {Palliyaguru}, {Pennucci}, {Ransom},
  {Siemens}, {Stairs}, {Stinebring}, {Stovall}, {Swiggum}, {Vallisneri}, {van
  Haasteren}, {Wang}, \& {Zhu}}]{abb+2015a}
{Arzoumanian}, Z., {Brazier}, A., {Burke-Spolaor}, S., {et~al.} 2015, \apj,
  813, 65

\bibitem[{Babak \& Sesana(2012)}]{Babak:2011mr}
Babak, S. \& Sesana, A. 2012, Phys. Rev. D, 85, 044034

\bibitem[{{Bailes} {et~al.}(2020){Bailes}, {Jameson}, {Abbate}, {Barr}, {Bhat},
  {Bondonneau}, {Burgay}, {Buchner}, {Camilo}, {Champion}, {Cognard},
  {Demorest}, {Freire}, {Gautam}, {Geyer}, {Griessmeier}, {Guillemot}, {Hu},
  {Jankowski}, {Johnston}, {Karastergiou}, {Karuppusamy}, {Kaur}, {Keith},
  {Kramer}, {van Leeuwen}, {Lower}, {Maan}, {McLaughlin}, {Meyers},
  {Os{\l}owski}, {Oswald}, {Parthasarathy}, {Pennucci}, {Posselt}, {Possenti},
  {Ransom}, {Reardon}, {Ridolfi}, {Schollar}, {Serylak}, {Shaifullah},
  {Shamohammadi}, {Shannon}, {Sobey}, {Song}, {Spiewak}, {Stairs}, {Stappers},
  {van Straten}, {Szary}, {Theureau}, {Venkatraman Krishnan}, {Weltevrede},
  {Wex}, {Abbott}, {Adams}, {Burger}, {Gamatham}, {Gouws}, {Horn}, {Hugo},
  {Joubert}, {Manley}, {McAlpine}, {Passmoor}, {Peens-Hough}, {Ramudzuli},
  {Rust}, {Salie}, {Schwardt}, {Siebrits}, {Van Tonder}, {Van Tonder}, \&
  {Welz}}]{bja+2020}
{Bailes}, M., {Jameson}, A., {Abbate}, F., {et~al.} 2020, \pasa, 37, e028

\bibitem[{{Bassa} {et~al.}(2016){Bassa}, {Janssen}, {Karuppusamy}, {Kramer},
  {Lee}, {Liu}, {McKee}, {Perrodin}, {Purver}, {Sanidas}, {Smits}, \&
  {Stappers}}]{bjk+16}
{Bassa}, C.~G., {Janssen}, G.~H., {Karuppusamy}, R., {et~al.} 2016, 456, 2196

\bibitem[{B\'ecsy {et~al.}(2022)B\'ecsy, Cornish, \& Digman}]{Becsy:2022zbu}
B\'ecsy, B., Cornish, N.~J., \& Digman, M.~C. 2022, Phys. Rev. D, 105, 122003

\bibitem[{{Begelman} {et~al.}(1980){Begelman}, {Blandford}, \&
  {Rees}}]{1980Natur.287..307B}
{Begelman}, M.~C., {Blandford}, R.~D., \& {Rees}, M.~J. 1980, \nat, 287, 307

\bibitem[{{Caballero} {et~al.}(2016){Caballero}, {Lee}, {Lentati}, {Desvignes},
  {Champion}, {Verbiest}, {Janssen}, {Stappers}, {Kramer}, {Lazarus},
  {Possenti}, {Tiburzi}, {Perrodin}, {Os{\l}owski}, {Babak}, {Bassa}, {Brem},
  {Burgay}, {Cognard}, {Gair}, {Graikou}, {Guillemot}, {Hessels},
  {Karuppusamy}, {Lassus}, {Liu}, {McKee}, {Mingarelli}, {Petiteau}, {Purver},
  {Rosado}, {Sanidas}, {Sesana}, {Shaifullah}, {Smits}, {Taylor}, {Theureau},
  {van Haasteren}, \& {Vecchio}}]{cll+2016}
{Caballero}, R.~N., {Lee}, K.~J., {Lentati}, L., {et~al.} 2016, \mnras, 457,
  4421

\bibitem[{{Caprini} {et~al.}(2010){Caprini}, {Durrer}, \&
  {Siemens}}]{2010PhRvD..82f3511C}
{Caprini}, C., {Durrer}, R., \& {Siemens}, X. 2010, \prd, 82, 063511

\bibitem[{{Caprini} \& {Figueroa}(2018)}]{2018CQGra..35p3001C}
{Caprini}, C. \& {Figueroa}, D.~G. 2018, Classical and Quantum Gravity, 35,
  163001

\bibitem[{Carlin \& Chib(1995)}]{cc1995}
Carlin, B.~P. \& Chib, S. 1995, Journal of the royal statistical society series
  b-methodological, 57, 473

\bibitem[{{Chalumeau} {et~al.}(2022){Chalumeau}, {Babak}, {Petiteau}, {Chen},
  {Samajdar}, {Caballero}, {Theureau}, {Guillemot}, {Desvignes},
  {Parthasarathy}, {Liu}, {Shaifullah}, {Hu}, {van der Wateren}, {Antoniadis},
  {Bak Nielsen}, {Bassa}, {Berthereau}, {Burgay}, {Champion}, {Cognard},
  {Falxa}, {Ferdman}, {Freire}, {Gair}, {Graikou}, {Guo}, {Jang}, {Janssen},
  {Karuppusamy}, {Keith}, {Kramer}, {Lee}, {Liu}, {Lyne}, {Main}, {McKee},
  {Mickaliger}, {Perera}, {Perrodin}, {Porayko}, {Possenti}, {Sanidas},
  {Sesana}, {Speri}, {Stappers}, {Tiburzi}, {Vecchio}, {Verbiest}, {Wang},
  {Wang}, \& {Xu}}]{2022MNRAS.509.5538C}
{Chalumeau}, A., {Babak}, S., {Petiteau}, A., {et~al.} 2022, \mnras, 509, 5538

\bibitem[{Chamberlin {et~al.}(2015)Chamberlin, Creighton, Siemens, Demorest,
  Ellis, Price, \& Romano}]{ccx+2015}
Chamberlin, S.~J., Creighton, J. D.~E., Siemens, X., {et~al.} 2015, Phys. Rev.
  D, 91, 044048

\bibitem[{{Chen} {et~al.}(2021){Chen}, {Caballero}, {Guo}, {Chalumeau}, {Liu},
  {Shaifullah}, {Lee}, {Babak}, {Desvignes}, {Parthasarathy}, {Hu}, {van der
  Wateren}, {Antoniadis}, {Bak Nielsen}, {Bassa}, {Berthereau}, {Burgay},
  {Champion}, {Cognard}, {Falxa}, {Ferdman}, {Freire}, {Gair}, {Graikou},
  {Guillemot}, {Jang}, {Janssen}, {Karuppusamy}, {Keith}, {Kramer}, {Liu},
  {Lyne}, {Main}, {McKee}, {Mickaliger}, {Perera}, {Perrodin}, {Petiteau},
  {Porayko}, {Possenti}, {Samajdar}, {Sanidas}, {Sesana}, {Speri}, {Stappers},
  {Theureau}, {Tiburzi}, {Vecchio}, {Verbiest}, {Wang}, {Wang}, \&
  {Xu}}]{ccg+21}
{Chen}, S., {Caballero}, R.~N., {Guo}, Y.~J., {et~al.} 2021, \mnras, 508, 4970

\bibitem[{{Cognard} {et~al.}(2013){Cognard}, {Theureau}, {Guillemot}, {Liu},
  {Lassus}, \& {Desvignes}}]{ctg+13}
{Cognard}, I., {Theureau}, G., {Guillemot}, L., {et~al.} 2013, in SF2A-2013:
  Proceedings of the Annual meeting of the French Society of Astronomy and
  Astrophysics, ed. L.~{Cambresy}, F.~{Martins}, E.~{Nuss}, \& A.~{Palacios},
  327--330

\bibitem[{{Coles} {et~al.}(2015){Coles}, {Kerr}, {Shannon}, {Hobbs},
  {Manchester}, {You}, {Bailes}, {Bhat}, {Burke-Spolaor}, {Dai}, {Keith},
  {Levin}, {Os{\l}owski}, {Ravi}, {Reardon}, {Toomey}, {van Straten}, {Wang},
  {Wen}, \& {Zhu}}]{cks+2015}
{Coles}, W.~A., {Kerr}, M., {Shannon}, R.~M., {et~al.} 2015, \apj, 808, 113

\bibitem[{{Cornish} \& {Sampson}(2016)}]{2016PhRvD..93j4047C}
{Cornish}, N.~J. \& {Sampson}, L. 2016, \prd, 93, 104047

\bibitem[{{Damour} \& {Vilenkin}(2000)}]{2000PhRvL..85.3761D}
{Damour}, T. \& {Vilenkin}, A. 2000, \prl, 85, 3761

\bibitem[{{Demorest} {et~al.}(2013){Demorest}, {Ferdman}, {Gonzalez}, {Nice},
  {Ransom}, {Stairs}, {Arzoumanian}, {Brazier}, {Burke-Spolaor}, {Chamberlin},
  {Cordes}, {Ellis}, {Finn}, {Freire}, {Giampanis}, {Jenet}, {Kaspi}, {Lazio},
  {Lommen}, {McLaughlin}, {Palliyaguru}, {Perrodin}, {Shannon}, {Siemens},
  {Stinebring}, {Swiggum}, \& {Zhu}}]{dfg+2013}
{Demorest}, P.~B., {Ferdman}, R.~D., {Gonzalez}, M.~E., {et~al.} 2013, \apj,
  762, 94

\bibitem[{{Desvignes} {et~al.}(2016){Desvignes}, {Caballero}, {Lentati},
  {Verbiest}, {Champion}, {Stappers}, {Janssen}, {Lazarus}, {Os{\l}owski},
  {Babak}, {Bassa}, {Brem}, {Burgay}, {Cognard}, {Gair}, {Graikou},
  {Guillemot}, {Hessels}, {Jessner}, {Jordan}, {Karuppusamy}, {Kramer},
  {Lassus}, {Lazaridis}, {Lee}, {Liu}, {Lyne}, {McKee}, {Mingarelli},
  {Perrodin}, {Petiteau}, {Possenti}, {Purver}, {Rosado}, {Sanidas}, {Sesana},
  {Shaifullah}, {Smits}, {Taylor}, {Theureau}, {Tiburzi}, {van Haasteren}, \&
  {Vecchio}}]{dcl+2016}
{Desvignes}, G., {Caballero}, R.~N., {Lentati}, L., {et~al.} 2016, \mnras, 458,
  3341

\bibitem[{Detweiler(1979)}]{det79}
Detweiler, S. 1979, \apj, 234, 1100

\bibitem[{{Di Marco} {et~al.}(2023){Di Marco}, {Zic}, {Miles}, {Reardon},
  {Thrane}, \& {Shannon}}]{2023arXiv230504464D}
{Di Marco}, V., {Zic}, A., {Miles}, M.~T., {et~al.} 2023, arXiv e-prints,
  arXiv:2305.04464

\bibitem[{Ellis {et~al.}(2012)Ellis, Siemens, \& Creighton}]{Ellis:2012zv}
Ellis, J.~A., Siemens, X., \& Creighton, J. D.~E. 2012, Astrophys. J., 756, 175

\bibitem[{Ellis {et~al.}(2020)Ellis, Vallisneri, Taylor, \& Baker}]{enterprise}
Ellis, J.~A., Vallisneri, M., Taylor, S.~R., \& Baker, P.~T. 2020, ENTERPRISE:
  Enhanced Numerical Toolbox Enabling a Robust PulsaR Inference SuitE, Zenodo

\bibitem[{Falxa {et~al.}(2023)Falxa, Babak, \& Le~Jeune}]{Falxa:2022yrm}
Falxa, M., Babak, S., \& Le~Jeune, M. 2023, Phys. Rev. D, 107, 022008

\bibitem[{{Foster} \& {Backer}(1990)}]{fb1990}
{Foster}, R.~S. \& {Backer}, D.~C. 1990, \apj, 361, 300

\bibitem[{Foster \& Backer(1990)}]{fb90}
Foster, R.~S. \& Backer, D.~C. 1990, \apj, 361, 300

\bibitem[{{Gair} {et~al.}(2014){Gair}, {Romano}, {Taylor}, \&
  {Mingarelli}}]{grt+2014}
{Gair}, J., {Romano}, J.~D., {Taylor}, S., \& {Mingarelli}, C. M.~F. 2014,
  \prd, 90, 082001

\bibitem[{{Goncharov} {et~al.}(2021{\natexlab{a}}){Goncharov}, {Reardon},
  {Shannon}, {Zhu}, {Thrane}, {Bailes}, {Bhat}, {Dai}, {Hobbs}, {Kerr},
  {Manchester}, {Os{\l}owski}, {Parthasarathy}, {Russell}, {Spiewak},
  {Thyagarajan}, \& {Wang}}]{grs+2021}
{Goncharov}, B., {Reardon}, D.~J., {Shannon}, R.~M., {et~al.}
  2021{\natexlab{a}}, \mnras, 502, 478

\bibitem[{{Goncharov} {et~al.}(2021{\natexlab{b}}){Goncharov}, {Shannon},
  {Reardon}, {Hobbs}, {Zic}, {Bailes}, {Cury{\l}o}, {Dai}, {Kerr}, {Lower},
  {Manchester}, {Mandow}, {Middleton}, {Miles}, {Parthasarathy}, {Thrane},
  {Thyagarajan}, {Xue}, {Zhu}, {Cameron}, {Feng}, {Luo}, {Russell},
  {Sarkissian}, {Spiewak}, {Wang}, {Wang}, {Zhang}, \& {Zhang}}]{gsr+2021}
{Goncharov}, B., {Shannon}, R.~M., {Reardon}, D.~J., {et~al.}
  2021{\natexlab{b}}, {On the Evidence for a Common-spectrum Process in the
  Search for the Nanohertz Gravitational-wave Background with the Parkes Pulsar
  Timing Array}

\bibitem[{{Goncharov} {et~al.}(2022{\natexlab{a}}){Goncharov}, {Thrane},
  {Shannon}, {Harms}, {Bhat}, {Hobbs}, {Kerr}, {Manchester}, {Reardon},
  {Russell}, {Zhu}, \& {Zic}}]{ppta+22}
{Goncharov}, B., {Thrane}, E., {Shannon}, R.~M., {et~al.} 2022{\natexlab{a}},
  \apjl, 932, L22

\bibitem[{{Goncharov} {et~al.}(2022{\natexlab{b}}){Goncharov}, {Thrane},
  {Shannon}, {Harms}, {Bhat}, {Hobbs}, {Kerr}, {Manchester}, {Reardon},
  {Russell}, {Zhu}, \& {Zic}}]{gts+2022}
{Goncharov}, B., {Thrane}, E., {Shannon}, R.~M., {et~al.} 2022{\natexlab{b}},
  {Consistency of the Parkes Pulsar Timing Array Signal with a Nanohertz
  Gravitational-wave Background}

\bibitem[{{Guo} {et~al.}(2019){Guo}, {Li}, {Lee}, \& {Caballero}}]{gll+2019}
{Guo}, Y.~J., {Li}, G.~Y., {Lee}, K.~J., \& {Caballero}, R.~N. 2019, \mnras,
  489, 5573

\bibitem[{{Guzzetti} {et~al.}(2016){Guzzetti}, {Bartolo}, {Liguori}, \&
  {Matarrese}}]{2016NCimR..39..399G}
{Guzzetti}, M.~C., {Bartolo}, N., {Liguori}, M., \& {Matarrese}, S. 2016, Nuovo
  Cimento Rivista Serie, 39, 399

\bibitem[{{Hazboun} {et~al.}(2023){Hazboun}, {Meyers}, {Romano}, {Siemens}, \&
  {Archibald}}]{2023arXiv230501116H}
{Hazboun}, J.~S., {Meyers}, P.~M., {Romano}, J.~D., {Siemens}, X., \&
  {Archibald}, A.~M. 2023, arXiv e-prints, arXiv:2305.01116

\bibitem[{{Hee} {et~al.}(2016){Hee}, {Handley}, {Hobson}, \&
  {Lasenby}}]{hhh+2016}
{Hee}, S., {Handley}, W.~J., {Hobson}, M.~P., \& {Lasenby}, A.~N. 2016, \mnras,
  455, 2461

\bibitem[{Hellings \& Downs(1983)}]{hd83}
Hellings, R.~W. \& Downs, G.~S. 1983, \apjl, 265, L39

\bibitem[{{Hobbs}(2013)}]{hob2013}
{Hobbs}, G. 2013, Classical and Quantum Gravity, 30, 224007

\bibitem[{Hourihane {et~al.}(2022)Hourihane, Meyers, Johnson, Chatziioannou, \&
  Vallisneri}]{Hourihane:2022ner}
Hourihane, S., Meyers, P., Johnson, A., Chatziioannou, K., \& Vallisneri, M.
  2022 [\eprint[arXiv]{2212.06276}]

\bibitem[{{Izquierdo-Villalba} {et~al.}(2022){Izquierdo-Villalba}, {Sesana},
  {Bonoli}, \& {Colpi}}]{2022MNRAS.509.3488I}
{Izquierdo-Villalba}, D., {Sesana}, A., {Bonoli}, S., \& {Colpi}, M. 2022,
  \mnras, 509, 3488

\bibitem[{Jaffe \& Backer(2003)}]{jb03}
Jaffe, A.~H. \& Backer, D.~C. 2003, \apj, 583, 616

\bibitem[{{Jenet} {et~al.}(2009){Jenet}, {Finn}, {Lazio}, {Lommen},
  {McLaughlin}, {Stairs}, {Stinebring}, {Verbiest}, {Archibald}, {Arzoumanian},
  {Backer}, {Cordes}, {Demorest}, {Ferdman}, {Freire}, {Gonzalez}, {Kaspi},
  {Kondratiev}, {Lorimer}, {Lynch}, {Nice}, {Ransom}, {Shannon}, \&
  {Siemens}}]{jenet09}
{Jenet}, F., {Finn}, L.~S., {Lazio}, J., {et~al.} 2009, arXiv e-prints,
  arXiv:0909.1058

\bibitem[{{Jiang} {et~al.}(2019){Jiang}, {Yue}, {Gan}, {Yao}, {Li}, {Pan},
  {Sun}, {Yu}, {Liu}, {Tang}, {Qian}, {Lu}, {Yan}, {Peng}, {Zhang}, {Wang},
  {Li}, \& {Li}}]{jyg+19}
{Jiang}, P., {Yue}, Y., {Gan}, H., {et~al.} 2019, Science China Physics,
  Mechanics, and Astronomy, 62, 959502

\bibitem[{{Joshi} {et~al.}(2018){Joshi}, {Arumugasamy}, {Bagchi},
  {Bandyopadhyay}, {Basu}, {Dhanda Batra}, {Bethapudi}, {Choudhary}, {De},
  {Dey}, {Gopakumar}, {Gupta}, {Krishnakumar}, {Maan}, {Manoharan}, {Naidu},
  {Nandi}, {Pathak}, {Surnis}, \& {Susobhanan}}]{2018JApA...39...51J}
{Joshi}, B.~C., {Arumugasamy}, P., {Bagchi}, M., {et~al.} 2018, Journal of
  Astrophysics and Astronomy, 39, 51

\bibitem[{Karnesis {et~al.}(2023)Karnesis, Katz, Korsakova, Gair, \&
  Stergioulas}]{Karnesis:2023ras}
Karnesis, N., Katz, M.~L., Korsakova, N., Gair, J.~R., \& Stergioulas, N. 2023
  [\eprint[arXiv]{2303.02164}]

\bibitem[{{Kormendy} \& {Ho}(2013)}]{2013ARA&A..51..511K}
{Kormendy}, J. \& {Ho}, L.~C. 2013, \araa, 51, 511

\bibitem[{{Kramer} \& {Champion}(2013)}]{kc2013}
{Kramer}, M. \& {Champion}, D.~J. 2013, Classical and Quantum Gravity, 30,
  224009

\bibitem[{{Lam} {et~al.}(2018){Lam}, {Ellis}, {Grillo}, {Jones}, {Hazboun},
  {Brook}, {Turner}, {Chatterjee}, {Cordes}, {Lazio}, {DeCesar}, {Arzoumanian},
  {Blumer}, {Cromartie}, {Demorest}, {Dolch}, {Ferdman}, {Ferrara}, {Fonseca},
  {Garver-Daniels}, {Gentile}, {Gupta}, {Lorimer}, {Lynch}, {Madison},
  {McLaughlin}, {Ng}, {Nice}, {Pennucci}, {Ransom}, {Spiewak}, {Stairs},
  {Stinebring}, {Stovall}, {Swiggum}, {Vigeland}, \& {Zhu}}]{leg2018}
{Lam}, M.~T., {Ellis}, J.~A., {Grillo}, G., {et~al.} 2018, \apj, 861, 132

\bibitem[{{Lentati} {et~al.}(2013){Lentati}, {Alexander}, {Hobson}, {Taylor},
  {Gair}, {Balan}, \& {van Haasteren}}]{lah+2013}
{Lentati}, L., {Alexander}, P., {Hobson}, M.~P., {et~al.} 2013, \prd, 87,
  104021

\bibitem[{{Lentati} {et~al.}(2015){Lentati}, {Taylor}, {Mingarelli}, {Sesana},
  {Sanidas}, {Vecchio}, {Caballero}, {Lee}, {van Haasteren}, {Babak}, {Bassa},
  {Brem}, {Burgay}, {Champion}, {Cognard}, {Desvignes}, {Gair}, {Guillemot},
  {Hessels}, {Janssen}, {Karuppusamy}, {Kramer}, {Lassus}, {Lazarus}, {Liu},
  {Os{\l}owski}, {Perrodin}, {Petiteau}, {Possenti}, {Purver}, {Rosado},
  {Smits}, {Stappers}, {Theureau}, {Tiburzi}, \& {Verbiest}}]{ltm+2015}
{Lentati}, L., {Taylor}, S.~R., {Mingarelli}, C.~M.~F., {et~al.} 2015, \mnras,
  453, 2576

\bibitem[{{Lorimer} \& {Kramer}(2004)}]{lk04}
{Lorimer}, D.~R. \& {Kramer}, M. 2004, {Handbook of Pulsar Astronomy}, Vol.~4

\bibitem[{{Magorrian} {et~al.}(1998){Magorrian}, {Tremaine}, {Richstone},
  {Bender}, {Bower}, {Dressler}, {Faber}, {Gebhardt}, {Green}, {Grillmair},
  {Kormendy}, \& {Lauer}}]{1998AJ....115.2285M}
{Magorrian}, J., {Tremaine}, S., {Richstone}, D., {et~al.} 1998, \aj, 115, 2285

\bibitem[{{Manchester}(2006)}]{man06}
{Manchester}, R.~N. 2006, Nomenclature, Precession and New Models in
  Fundamental Astronomy, 26th meeting of the IAU, Joint Discussion 16, 22-23
  August 2006, Prague, Czech Republic, JD16, \#66, 16
  [\eprint{astro-ph/0604288}]

\bibitem[{{McConnell} \& {Ma}(2013)}]{2013ApJ...764..184M}
{McConnell}, N.~J. \& {Ma}, C.-P. 2013, \apj, 764, 184

\bibitem[{{McConnell} {et~al.}(2011){McConnell}, {Ma}, {Gebhardt}, {Wright},
  {Murphy}, {Lauer}, {Graham}, \& {Richstone}}]{2011Natur.480..215M}
{McConnell}, N.~J., {Ma}, C.-P., {Gebhardt}, K., {et~al.} 2011, \nat, 480, 215

\bibitem[{{Middleton} {et~al.}(2021){Middleton}, {Sesana}, {Chen}, {Vecchio},
  {Del Pozzo}, \& {Rosado}}]{msc+2021}
{Middleton}, H., {Sesana}, A., {Chen}, S., {et~al.} 2021, \mnras, 502, L99

\bibitem[{{Park} {et~al.}(2021){Park}, {Folkner}, {Williams}, \&
  {Boggs}}]{2021AJ....161..105P}
{Park}, R.~S., {Folkner}, W.~M., {Williams}, J.~G., \& {Boggs}, D.~H. 2021,
  \aj, 161, 105

\bibitem[{{Porayko} {et~al.}(2018){Porayko}, {Zhu}, {Levin}, {Hui}, {Hobbs},
  {Grudskaya}, {Postnov}, {Bailes}, {Bhat}, {Coles}, {Dai}, {Dempsey}, {Keith},
  {Kerr}, {Kramer}, {Lasky}, {Manchester}, {Os{\l}owski}, {Parthasarathy},
  {Ravi}, {Reardon}, {Rosado}, {Russell}, {Shannon}, {Spiewak}, {van Straten},
  {Toomey}, {Wang}, {Wen}, {You}, \& {PPTA
  Collaboration}}]{2018PhRvD..98j2002P}
{Porayko}, N.~K., {Zhu}, X., {Levin}, Y., {et~al.} 2018, \prd, 98, 102002

\bibitem[{Raveri \& Doux(2021)}]{raveri2021non}
Raveri, M. \& Doux, C. 2021, Physical Review D, 104, 043504

\bibitem[{{Raveri} \& {Hu}(2019)}]{RaveriHu}
{Raveri}, M. \& {Hu}, W. 2019, \prd, 99, 043506

\bibitem[{Sazhin(1978)}]{saz78}
Sazhin, M.~V. 1978, 22, 36

\bibitem[{{Sesana}(2013)}]{2013CQGra..30v4014S}
{Sesana}, A. 2013, Classical and Quantum Gravity, 30, 224014

\bibitem[{{Sesana} {et~al.}(2008){Sesana}, {Vecchio}, \& {Colacino}}]{svc08}
{Sesana}, A., {Vecchio}, A., \& {Colacino}, C.~N. 2008, \mnras, 390, 192

\bibitem[{{Speri} {et~al.}(2023){Speri}, {Porayko}, {Falxa}, {Chen}, {Gair},
  {Sesana}, \& {Taylor}}]{2023MNRAS.518.1802S}
{Speri}, L., {Porayko}, N.~K., {Falxa}, M., {et~al.} 2023, \mnras, 518, 1802

\bibitem[{{Tarafdar} {et~al.}(2022){Tarafdar}, {Nobleson}, {Rana}, {Singha},
  {Krishnakumar}, {Joshi}, {Paladi}, {Kolhe}, {Batra}, {Agarwal}, {Bathula},
  {Dandapat}, {Desai}, {Dey}, {Hisano}, {Ingale}, {Kato}, {Kharbanda},
  {Kikunaga}, {Marmat}, {Pandian}, {Prabu}, {Srivastava}, {Surnis}, {Susarla},
  {Susobhanan}, {Takahashi}, {Arumugam}, {Bagchi}, {Banik}, {De}, {Girgaonkar},
  {Gopakumar}, {Gupta}, {Maan}, {Manoharan}, {Naidu}, \&
  {Pathak}}]{2022PASA...39...53T}
{Tarafdar}, P., {Nobleson}, K., {Rana}, P., {et~al.} 2022, \pasa, 39, e053

\bibitem[{Taylor \& Weisberg(1989)}]{tw89}
Taylor, J.~H. \& Weisberg, J.~M. 1989, \apj, 345, 434

\bibitem[{Taylor {et~al.}(2021)Taylor, Baker, Hazboun, Simon, \&
  Vigeland}]{enterprise_extensions}
Taylor, S.~R., Baker, P.~T., Hazboun, J.~S., Simon, J., \& Vigeland, S.~J.
  2021, enterprise\_extensions

\bibitem[{{Taylor} \& {Gair}(2013)}]{tg2013}
{Taylor}, S.~R. \& {Gair}, J.~R. 2013, \prd, 88, 084001

\bibitem[{{Taylor} {et~al.}(2017){Taylor}, {Lentati}, {Babak}, {Brem}, {Gair},
  {Sesana}, \& {Vecchio}}]{tlb+2017}
{Taylor}, S.~R., {Lentati}, L., {Babak}, S., {et~al.} 2017, \prd, 95, 042002

\bibitem[{{the EPTA and InPTA Collaborations}(2023)}]{wm2}
{the EPTA and InPTA Collaborations}. 2023, \aa

\bibitem[{{the EPTA Collaboration}(2023)}]{wm1}
{the EPTA Collaboration}. 2023, \aa

\bibitem[{{Tiburzi} {et~al.}(2016){Tiburzi}, {Hobbs}, {Kerr}, {Coles}, {Dai},
  {Manchester}, {Possenti}, {Shannon}, \& {You}}]{thk+2016}
{Tiburzi}, C., {Hobbs}, G., {Kerr}, M., {et~al.} 2016, \mnras, 455, 4339

\bibitem[{{Vallisneri} {et~al.}(2020){Vallisneri}, {Taylor}, {Simon},
  {Folkner}, {Park}, {Cutler}, {Ellis}, {Lazio}, {Vigeland}, {Aggarwal},
  {Arzoumanian}, {Baker}, {Brazier}, {Brook}, {Burke-Spolaor}, {Chatterjee},
  {Cordes}, {Cornish}, {Crawford}, {Cromartie}, {Crowter}, {DeCesar},
  {Demorest}, {Dolch}, {Ferdman}, {Ferrara}, {Fonseca}, {Garver-Daniels},
  {Gentile}, {Good}, {Hazboun}, {Holgado}, {Huerta}, {Islo}, {Jennings},
  {Jones}, {Jones}, {Kaplan}, {Kelley}, {Key}, {Lam}, {Levin}, {Lorimer},
  {Luo}, {Lynch}, {Madison}, {McLaughlin}, {McWilliams}, {Mingarelli}, {Ng},
  {Nice}, {Pennucci}, {Pol}, {Ransom}, {Ray}, {Siemens}, {Spiewak}, {Stairs},
  {Stinebring}, {Stovall}, {Swiggum}, {van Haasteren}, {Witt}, \&
  {Zhu}}]{2020ApJ...893..112V}
{Vallisneri}, M., {Taylor}, S.~R., {Simon}, J., {et~al.} 2020, \apj, 893, 112

\bibitem[{{van Haasteren} \& {Levin}(2013)}]{vl2013}
{van Haasteren}, R. \& {Levin}, Y. 2013, \mnras, 428, 1147

\bibitem[{{van Haasteren} {et~al.}(2009){van Haasteren}, {Levin}, {McDonald},
  \& {Lu}}]{vlm+2009}
{van Haasteren}, R., {Levin}, Y., {McDonald}, P., \& {Lu}, T. 2009, \mnras,
  395, 1005

\bibitem[{{van Haasteren} \& {Vallisneri}(2014)}]{vv2014}
{van Haasteren}, R. \& {Vallisneri}, M. 2014, \prd, 90, 104012

\bibitem[{{Verbiest} {et~al.}(2009){Verbiest}, {Bailes}, {Coles}, {Hobbs}, {van
  Straten}, {Champion}, {Jenet}, {Manchester}, {Bhat}, {Sarkissian}, {Yardley},
  {Burke-Spolaor}, {Hotan}, \& {You}}]{vbc+2009}
{Verbiest}, J.~P.~W., {Bailes}, M., {Coles}, W.~A., {et~al.} 2009, \mnras, 400,
  951

\bibitem[{{Vigeland} {et~al.}(2018){Vigeland}, {Islo}, {Taylor}, \&
  {Ellis}}]{2018PhRvD..98d4003V}
{Vigeland}, S.~J., {Islo}, K., {Taylor}, S.~R., \& {Ellis}, J.~A. 2018, \prd,
  98, 044003

\bibitem[{{Wang} {et~al.}(2021){Wang}, {Yang}, {Fan}, {Hennawi}, {Barth},
  {Banados}, {Bian}, {Boutsia}, {Connor}, {Davies}, {Decarli}, {Eilers},
  {Farina}, {Green}, {Jiang}, {Li}, {Mazzucchelli}, {Nanni}, {Schindler},
  {Venemans}, {Walter}, {Wu}, \& {Yue}}]{2021ApJ...907L...1W}
{Wang}, F., {Yang}, J., {Fan}, X., {et~al.} 2021, \apjl, 907, L1

\bibitem[{{Wang} {et~al.}(2019){Wang}, {Yang}, {Fan}, {Wu}, {Yue}, {Li},
  {Bian}, {Jiang}, {Ba{\~n}ados}, {Schindler}, {Findlay}, {Davies}, {Decarli},
  {Farina}, {Green}, {Hennawi}, {Huang}, {Mazzuccheli}, {McGreer}, {Venemans},
  {Walter}, {Dye}, {Lyke}, {Myers}, \& {Nunez}}]{2019ApJ...884...30W}
{Wang}, F., {Yang}, J., {Fan}, X., {et~al.} 2019, \apj, 884, 30

\bibitem[{{White} \& {Rees}(1978)}]{1978MNRAS.183..341W}
{White}, S.~D.~M. \& {Rees}, M.~J. 1978, \mnras, 183, 341

\bibitem[{{Williams} {et~al.}(2021){Williams}, {Veitch}, \&
  {Messenger}}]{nessai}
{Williams}, M.~J., {Veitch}, J., \& {Messenger}, C. 2021, \prd, 103, 103006

\bibitem[{{Zhu} {et~al.}(2015){Zhu}, {Stairs}, {Demorest}, {Nice}, {Ellis},
  {Ransom}, {Arzoumanian}, {Crowter}, {Dolch}, {Ferdman}, {Fonseca},
  {Gonzalez}, {Jones}, {Jones}, {Lam}, {Levin}, {McLaughlin}, {Pennucci},
  {Stovall}, \& {Swiggum}}]{zsd+2015}
{Zhu}, W.~W., {Stairs}, I.~H., {Demorest}, P.~B., {et~al.} 2015, \apj, 809, 41

\bibitem[{{Zic} {et~al.}(2022){Zic}, {Hobbs}, {Shannon}, {Reardon},
  {Goncharov}, {Bhat}, {Cameron}, {Dai}, {Dawson}, {Kerr}, {Manchester},
  {Mandow}, {Marshman}, {Russell}, {Thyagarajan}, \& {Zhu}}]{zhs+2022}
{Zic}, A., {Hobbs}, G., {Shannon}, R.~M., {et~al.} 2022, \mnras, 516, 410

\end{thebibliography}

\begin{appendix}

\section{Chebyshev and Legendre decomposition for the ORF}
\label{App:A}

\begin{figure*}
\centering
\includegraphics[width=\textwidth]{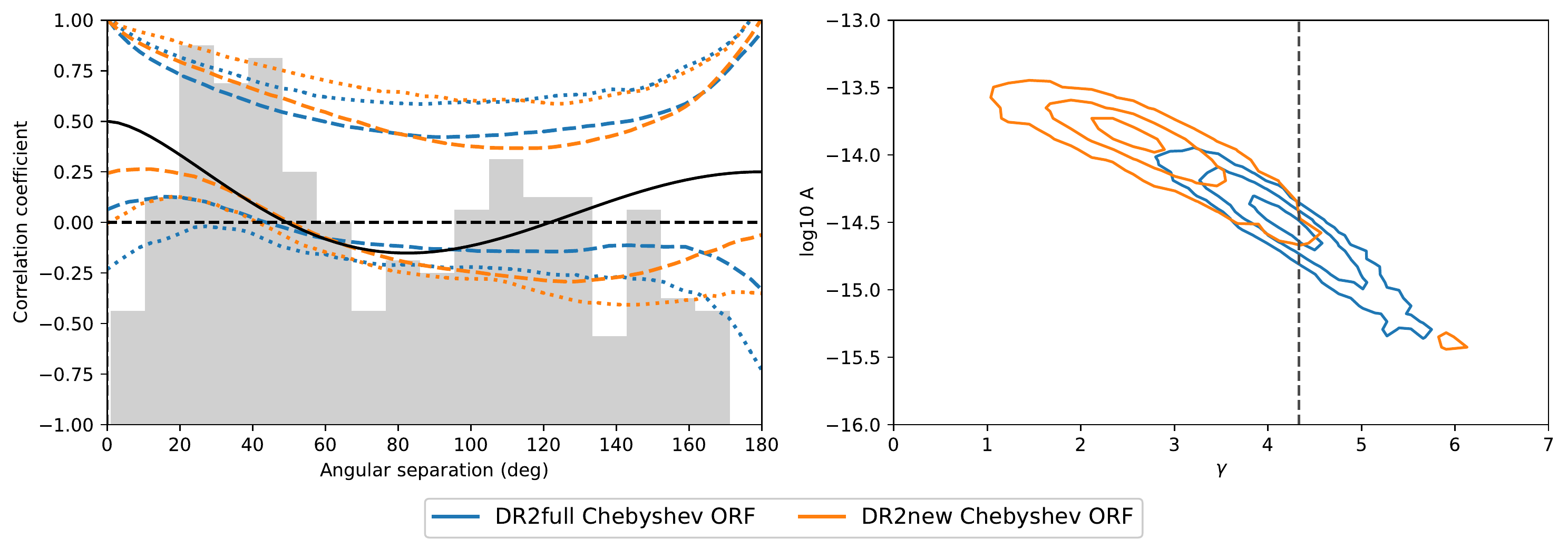}
\caption{
Overlap reduction function reconstructed using Chebyshev polynomial. The figure is in the style of Figure \ref{fig:bin_orf_pl}, with the difference in the left panel being that the dashed and dotted lines indicate the central 95 and 99.7 \% credible regions of the reconstructed spatial correlations.}
\label{fig:bin_orf_cheb}
\end{figure*}

\begin{figure*}
\centering
\includegraphics[width=\textwidth]{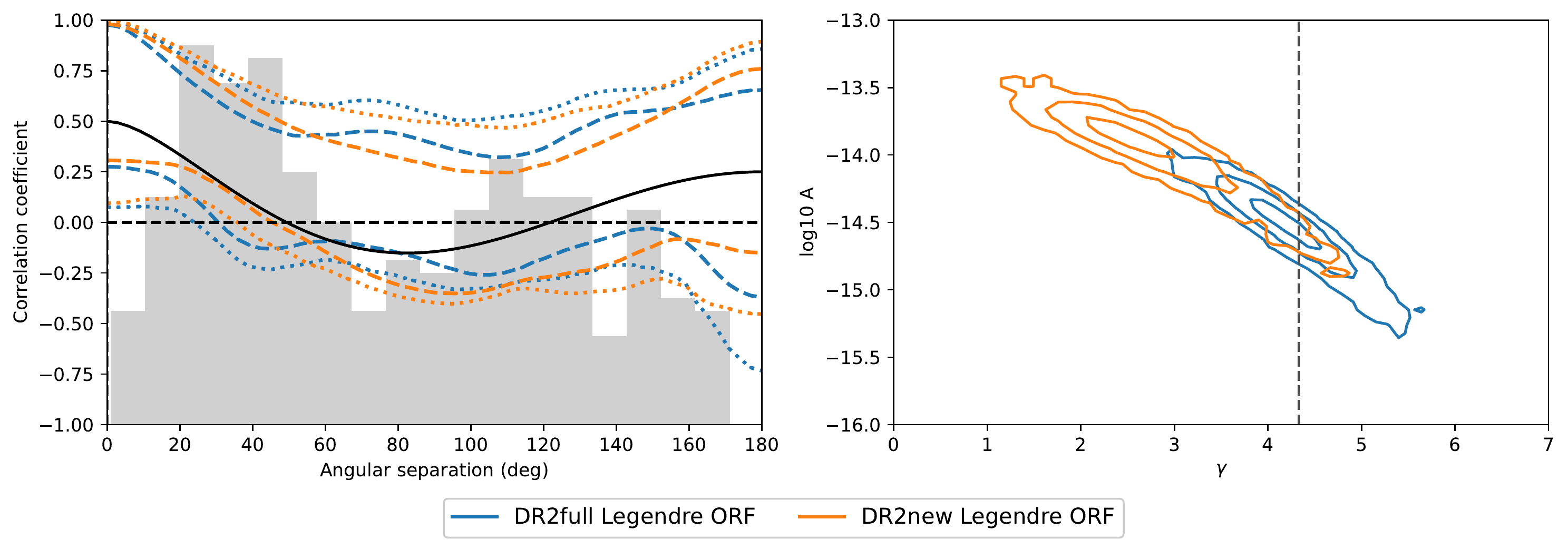}
\caption{
Overlap reduction function reconstructed using Legendre polynomial. The figure is in the style of Figure \ref{fig:bin_orf_pl}, with the difference in the left panel being that the dashed and dotted lines indicate the central 95 and 99.7 \% credible regions of the reconstructed spatial correlations.}
\label{fig:bin_orf_legendre}
\end{figure*}

In this Appendix, we present the spatial correlation measured among pulsar pairs as reconstructed from: i) a third order Chebishev polynomial decomposition and ii) a fifth order Legendre polynomial decomposition. Results are shown in Figures \ref{fig:bin_orf_cheb} and \ref{fig:bin_orf_legendre} respectively, which highlight the broad consistency with the average HD correlation expected for a GWB and with the binned Bayesian reconstruction shown in Figure~\ref{fig:bin_orf_pl} and discussed in Section~\ref{sec:spatial}.

\section{Comparison of the four data sets}
\label{App:B}

\begin{figure*}
\centering
\includegraphics[width=\textwidth]{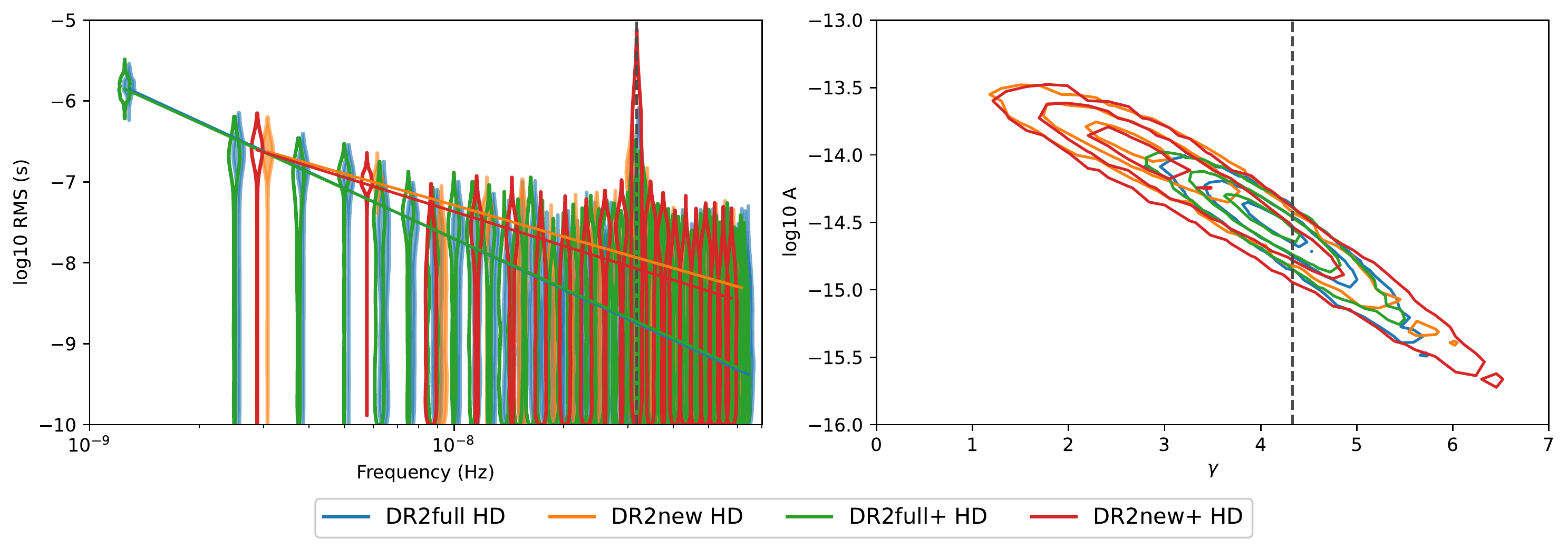}
\caption{Spectral properties of a CRS signal assuming HD correlations in the style of Figure \ref{fig:fs_pl} for all four data sets.}
\label{fig:fs_pl_all}
\end{figure*}

\begin{figure*}
\centering
\includegraphics[width=\textwidth]{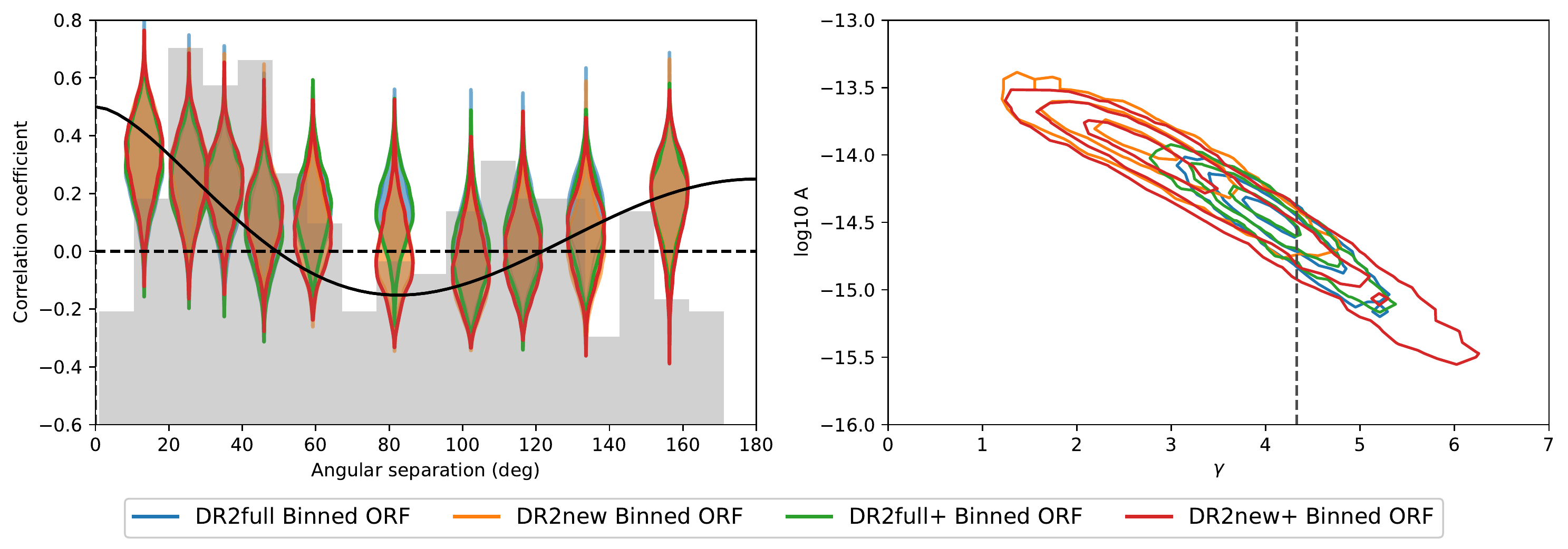}
\caption{Binned overlap reduction function in the style of Figure \ref{fig:bin_orf_pl} for all four data sets.}
\label{fig:orf_all}
\end{figure*}

In this Appendix, we present the constraints for the free spectrum, power law and binned correlations in Figures \ref{fig:fs_pl_all} and \ref{fig:orf_all} for the four data sets used in this work: \texttt{DR2full}, \texttt{DR2new}, \texttt{DR2full+} and \texttt{DR2new+}.

\end{appendix}

\end{document}